\documentclass[fleqn,usenatbib]{mnras}

\usepackage{newtxtext,newtxmath}
\usepackage[T1]{fontenc}
\usepackage{ae,aecompl}
\usepackage{graphicx}	
\usepackage{amsmath}	
\usepackage{multirow}
\usepackage{xcolor}
\usepackage{lipsum}

\newcounter{defcounter}
\newenvironment{myequation}{%
\addtocounter{equation}{-1}
\refstepcounter{defcounter}

\begin{eqnarray}}
{\end{eqnarray}}

\title[PDR Diagnostics Across Galactic Environments]{Photodissociation Region Diagnostics Across Galactic Environments}

\author[T. G. Bisbas et al.]{
Thomas G. Bisbas,$^{1,2}$\thanks{E-mail: bisbas@ph1.uni-koeln.de (TGB)} Jonathan C. Tan,$^{3,4}$ and Kei E. I. Tanaka$^{5}$
\\
$^{1}$I. Physikalisches Institut, Universit\"at zu K\"oln, Z\"ulpicher Stra{\ss}e 77, D-50923, K\"oln, Germany\\
$^{2}$Department of Physics, Aristotle University of Thessaloniki, GR-54124 Thessaloniki, Greece\\
$^{3}$Department of Space, Earth \& Environment, Chalmers University of Technology, SE-412 93 Gothenburg, Sweden \\
$^{4}$Department of Astronomy, University of Virginia, Charlottesville, VA 22904, USA \\
$^{5}$ALMA Project, National Astronomical Observatory of Japan, Mitaka, Tokyo 181-8588, Japan
}

\date{Accepted XXX. Received YYY; in original form ZZZ}

\pubyear{2015}

\begin{document}
\label{firstpage}
\pagerange{\pageref{firstpage}--\pageref{lastpage}}
\maketitle

\begin{abstract}
We present three-dimensional astrochemical simulations and synthetic observations of magnetised, turbulent, self-gravitating molecular clouds. We explore various galactic interstellar medium environments, including cosmic-ray ionization rates in the range of $\zeta_{\rm CR}=10^{-17}$--$10^{-14}\,{\rm s}^{-1}$, far-UV intensities in the range of $G_0=1$--$10^3$ and metallicities in the range of $Z=0.1$--$2\,{\rm Z}_{\odot}$. The simulations also probe a range of densities and levels of turbulence, including cases where the gas has undergone recent compression due to cloud-cloud collisions. We examine: i) the column densities of carbon species across the cycle of C{\sc ii}, C{\sc i} and CO, along with O{\sc i}, in relation to the H{\sc i}-to-H$_2$ transition; ii) the velocity-integrated emission of [C{\sc ii}]~$158\mu$m, [$^{13}$C{\sc ii}]~$158\mu$m, [C{\sc i}]~$609\mu$m and $370\mu$m, [O{\sc i}]~$63\mu$m and $146\mu$m, and of the first ten $^{12}$CO rotational transitions; iii) the corresponding Spectral Line Energy Distributions; iv) the usage of [C{\sc ii}] and [O{\sc i}]~$63\mu$m to describe the dynamical state of the clouds; v) the behavior of the most commonly used ratios between transitions of CO and [C{\sc i}]; and vi) the conversion factors for using CO and C{\sc i} as H$_2$-gas tracers. We find that enhanced cosmic-ray energy densities enhance all aforementioned line intensities. At low metallicities, the emission of [C{\sc ii}] is well connected with the H$_2$ column, making it a promising new H$_2$ tracer in metal-poor environments. The conversion factors of $X_{\rm CO}$ and $X_{\rm CI}$ depend on metallicity and the cosmic-ray ionization rate, but not on FUV intensity. In the era of ALMA, SOFIA and the forthcoming CCAT-prime telescope, our results can be used to understand better the behaviour of systems in a wide range of galactic and extragalactic environments.
\end{abstract}




\section{Introduction}
\label{sec:intro}

To determine the physical properties and evolution of the interstellar medium (ISM), we need to model its chemical conditions, since these help set its heating/cooling rates and ionisation state, which mediates coupling to magnetic fields. Calculating the intensity of various emission lines is important not only for estimating cooling rates, but also for predicting the diagnostic information they carry, so we can assess how well ISM conditions can be inferred from a given set of observables. The emission lines of [$^{12}$C{\sc ii}]~$158\mu$m (hereafter `[C{\sc ii}]'), [O{\sc i}]~$63\mu$m, [C{\sc i}] ${^3}P_1\!\rightarrow\!{^3}P_0$ at $609\mu$m (hereafter `[C{\sc i}]~(1-0)') and $^{12}$CO rotational transitions from $J=1-0$ to $J=20-19$ or above (hereafter `CO~(1-0)' etc) are frequently used as diagnostics to reveal the parameters of the ISM in different objects \citep[e.g.,][]{Kram04,Roel06,Lang10,Pine13,Beut14,Mash15,Okad19}. The aforementioned lines are emitted from so-called Photodissociation Regions  \citep[`PDRs';][]{Stern95,Holl99}, which characterise the interface between ionized and molecular gas phases. Given the close association of young, hot O and B stars with the production of extreme-UV and far-UV (FUV) radiation fields, studies of PDRs also help us to understand the life cycle of star formation and the ISM in galaxies. Currently, [C{\sc ii}] and [O{\sc i}]~$63\mu$m can be observed in the local Universe with the Stratospheric Observatory for Infrared Astronomy (SOFIA). 
Recently, the SOFIA-upGREAT instrument has also observed the optically thin isotopic line of [$^{13}$C{\sc ii}] in various objects \citep{Okad19b, Guev20} which can be used to infer to the optical depth of [C{\sc ii}].
The Atacama Large Millimeter/submillimeter Array (ALMA) at Llano de Chajnantor plateau in Chile is favored for observing such high frequency lines at higher redshifts, whereas both [C{\sc i}] fine-structure lines and CO rotational transitions can be observed at lower redshifts as well. Along with the forthcoming 6-meter CCAT-prime telescope, these instruments offer already an increasingly vast amount of observational data from local clouds to distant galaxies, thereby allowing the community to reveal ISM conditions under which star-formation takes place across all epochs. 

For the past 30 years or so, the main focus of PDR studies has been the understanding of the carbon cycle phase (C{\sc ii}/C{\sc i}/CO) in relation to the atomic-to-molecular (H{\sc i}-to-H$_2$) transition \citep[e.g.,][]{vDish88,Papa04,Roll07,Glov10,Glov11,Offn13,Papa18}. In the standard PDR picture, it is far-UV, optical and IR photons that regulate the thermal state, chemistry, abundances and emissivities of the various atoms and molecules at low column densities. For higher column densities, it is cosmic-rays that penetrate much deeper and control the thermal balance of the ISM \citep[see][for a review]{Stro07,Gren15} leading to different initial conditions for star formation \citep{Papa10}. Recent studies \citep{Meij11,Bial15,Bisb15,Bisb17a,Gach19a} found that the carbon cycle transition is much more sensitive to the cosmic-ray ionization rate, $\zeta_{\rm CR}$, than the relative location to the local H{\sc i}-to-H$_2$ transition. In particular, cosmic-rays are able to destroy CO molecules indirectly (via He$^+$) while the clouds may maintain their $\rm H_2$ molecular phase \citep{Bisb15}. For intermediate $\zeta_{\rm CR}$, this `CO-poor' molecular gas \citep{vDish92} is C{\sc i}-rich, while for high $\zeta_{\rm CR}$ values it is C{\sc ii}-rich.

However, it is not only cosmic-rays that are able to affect so drastically the carbon cycle and change the chemistry of the gas at high column densities. Mechanical heating \citep{Meij11} and X-ray heating \citep{Malo96, Meij06, Mack19} may play central roles in more special scenarios such as environments with high star-formation activity, supernova remnants, and relativistic jets. Lower gas-phase metallicity observed in various evolutionary stages of galaxies across the epochs \citep[see][for a review]{Maio19}, dwarf galaxies \citep[e.g.,][]{Tafe10,Requ16,Pine17}, high redshift galaxies \citep[e.g.,][]{Cook15, Wang17, Stro18} and the outer parts of large galaxies affect the shielding of H$_2$ and CO columns, which in turn impacts the carbon phases in H$_2$-rich gas \citep{Schr18}. Turbulence also affects the chemistry of the evolving ISM clouds by mixing the different phases of the gas \citep[e.g.,][]{Xie95,Holl99,Bial17}. In this work we shall only consider the effects caused due to cosmic-rays, FUV radiation, and metallicity.

Many groups focus on algorithms constructing synthetic observations of density and velocity distributions under different ISM conditions \citep[see][for a review]{Hawo18}. The general goal is to understand how these conditions affect the trends of emissivities of the different coolants. This has been achieved already to a great degree by various one-dimensional approaches \citep[e.g.,][]{Kauf99,LePe06,Bell06,Roll07,Meij11,Bisb14}. However, early works by \citet{Stut88}, \citet{Burt90} and later by \citet{Andr17} found that the emissivities may strongly depend on the spatial distribution and the structure of PDRs, thus a three-dimensional approach needs to be taken into account. This has been considered in hydrodynamical models coupled with chemical networks by \citet{Nels97,Glov10,Grass14,Walc15,Seif16,Giri16,Seif17,Fran18,Inou20} and in static, but chemically more detailed, PDRs by \citet{Bisb12,Bisb15b,Bisb17a,Bisb17b,Lim20}. An interesting new approach has been presented by \citet{Bisb19} who used entire column-density probability density functions as inputs to examine, at a minimal computational cost, the PDR properties of large ISM regions under different conditions.

Understanding the carbon cycle in relation to the atomic-to-molecular transition is essential to better estimate the molecular gas content in the ISM. The so-called `$X_{\rm CO}$-factor' is a scaling factor connecting the observed emission of CO~(1-0) with the column density of H$_2$, as the latter molecule does not emit radiation readily captured by radiotelescopes \citep[see][for a review]{Bola13}. Its recommended value is $X_{\rm CO}=2\times10^{20}\,{\rm cm}^{-2}\,{\rm K}^{-1}\,{\rm km}^{-1}\,{\rm s}$ with a $\pm30\%$ variation \citep{Bola13}. The $X_{\rm CO}$-factor is widely used in Milky Way and extragalactic observations \citep[e.g.,][]{Tacc08,Genz12,Papa12,Chen15,Luo20}. In the era of ALMA and SOFIA, alternative tracers using C{\sc i} and C{\sc ii} have attracted the interest of the community, but, as the methods currently stand, they face strong biases due to lack of necessary sophisticated synthetic observations based on state-of-the-art numerical simulations. The two transitions of [C{\sc i}] at $609\mu$m and $370\mu$m have been proposed to probe the cold H$_2$ gas mass as well as, or perhaps even better than, the CO lines do \citep{Papa04,Offn14,Glov15,Glov16,Gach19b}. In this regard, observations have shown that C{\sc i} is as reliable as CO in local clouds \citep{Lo14} as well as in extragalactic systems \citep{Zhan14,Both17}. When it comes to the study of early evolutionary stages of the H{\sc i}-to-H$_2$ transition, as well as the high-redshift Universe, C{\sc ii} is likely to be one of the best tracers of H$_2$ gas \citep{Madd20}. Such studies are, however, currently underdeveloped. As we will see in this work, the [C{\sc ii}]~$158\mu$m line may indeed trace the H$_2$ column density as well as CO and C{\sc i} in environments inundated by strong cosmic-ray energy densities. 

The aim of this paper is to study how the varying FUV radiation, metallicity and cosmic-ray ionization rate impacts the emission of the different carbon phases (C{\sc ii}, C{\sc i} and CO), as well as of atomic oxygen (O{\sc i}) in three-dimensional density and velocity distributions representing molecular clouds. The density and velocity distributions considered here are sub-regions from the magnetohydrodynamical (MHD) simulations of \citet{Wu17}. The simulations include self-gravity of the gas and studied the evolution of turbulent molecular clouds, including cases of cloud-cloud collisions. Approximate PDR-based heating and cooling functions were used to calculate heating and cooling rates in the evolution of these structures. We post-process representative outputs from the simulations with different environmental parameters, mimicking the conditions in different parts of our Galaxy, including the Galactic Centre. This work extends two different series of papers; the works of \citet{Bisb15} and \citet{Bisb17a} in understanding how the lines of [C{\sc ii}]~$158\mu$m, [C{\sc i}]~(1-0) and CO $J=1-0$ can be used as tracers of H$_2$-rich gas in environments with high $\zeta_{\rm CR}$, and the works of \citet{Wu15,Wu17} and \citet{Bisb17b,Bisb18} in examining how the shape of the `bridge-effect', which acts as a signature of a cloud-cloud collision process \citep{Hawo15a,Hawo15b,Bisb17b}, holds under extreme ISM environmental conditions.

This paper is organized as follows. Section~\ref{sec:method} discusses the density and velocity distributions used and the environmental ISM conditions considered. Section~\ref{sec:cds} presents the results of our simulations for the distribution of abundances and gas temperatures. Section~\ref{sec:emission} presents the resultant velocity integrated emission maps. Section~\ref{ssec:sleds} discusses how the CO spectral line energy distributions change in each environmental scenario. Section~\ref{ssec:spectra} shows how dynamical diagnostics using the fine-structure lines of [C{\sc ii}] and [O{\sc i}]~$63\mu$m depend on the ISM parameter explored.  Section~\ref{sec:ratios} presents results for the most commonly used line ratios and how they compare with observations. Section~\ref{sec:conversion} analyzes the impact of the ISM conditions considered here on the $X_{\rm CO}$- and $X_{\rm CI}$-factors. We conclude in Section~\ref{sec:conclusions}.

\section{Numerical approach}
\label{sec:method}

\subsection{Density and velocity distributions}
\label{ssec:densities}
In this work, two different three-dimensional density and velocity distributions are considered. These are taken from the \citet{Wu17} MHD simulations which were performed using the adaptive-mesh refinement code {\sc enzo} \citep{Brya14}. The first one is the case when two giant molecular clouds undergo collision (hereafter `dense cloud'); the second one is the case when they do not collide (hereafter `diffuse cloud') but simply overlap each other along the line of sight. In the `dense cloud', the clouds have relative velocities of $10\,{\rm km}\,{\rm s}^{-1}$. The \citet{Wu17} simulations contained self-gravity, supersonic turbulence and magnetic fields (treated in the limit of ideal MHD). In terms of chemistry, a PDR-based scheme was included to calculate the heating and cooling processes corresponding to ISM conditions under an external isotropic FUV radiation field of $G_0=4$ \citep[normalized according to][]{Habi68} and a cosmic-ray ionization rate of $\zeta_{\rm CR}=10^{-16}\,{\rm s}^{-1}$.

In an earlier work, \citet{Bisb17b} post-processed two such snapshots using {\sc 3d-pdr} (see \S\ref{ssec:3dpdr}). To avoid the high computational expense demanded by the astrochemical calculations, \citet{Bisb17b} interpolated a large grid of pre-calculated one-dimensional PDR simulations using a relation connecting the gas density of each cell, $n_{\rm H}$, with a most probable value of visual extinction, $A_{\rm V}$, following \citet{Wu15}. This, in turn, allowed assignment in each cell of the abundances of species, the gas and dust temperatures, and the level populations of various coolants. This technique provides a fast and reasonable estimation of PDR properties \citep[see also][for an extension to column-density distributions]{Bisb19}. However, it neglects any three-dimensional escape-probability effects and considers the same $A_{\rm V}$ value for a given $n_{\rm H}$, which introduces errors, particularly for densities $<\!10^3\,{\rm cm}^{-3}$ (see \citealt{Glov10, VanL13, Safr17, Seif17} for distributions of $A_V$ vs $n_{\rm H}$). 

To increase the accuracy of astrochemical calculations while keeping the computational cost low, we select two sub-regions (one from the `dense' and one from the `diffuse' cloud) from snapshots that have been evolved for $t=4\,{\rm Myr}$. The sub-region corresponding to the `dense cloud' is centered in the area holding the highest column densities along the $z$-axis. The choice of this sub-region is made to represent a high-density, collisionally-compressed, turbulent star-forming region. On the other hand, the sub-region corresponding to the `diffuse cloud' is to represent a part of a lower density, turbulent molecular cloud. Both these sub-regions have been presented in an earlier work \citep{Bisb18} and we now continue with the modeling of these same structures to be able to make comparisons to this prior work. The adaptive grid of each sub-region has been converted to a uniform grid and each cell has a spatial resolution of $0.125\,{\rm pc}$. Each sub-region has a total number of $112^3$ cells and therefore $14^3\,{\rm pc^3}$ in volume. The `dense' sub-region has a total mass of $M_{\rm tot}=4.3\times10^4\,{\rm M}_{\odot}$, mean total H-nucleus number density of $\langle n_{\rm H}\rangle\sim640\,{\rm cm}^{-3}$ and mean total column density of $\langle N_{\rm tot}\rangle\sim2.91\times10^{22}\,{\rm cm}^{-2}$; the `diffuse' sub-region has a total mass of $M_{\rm tot}=1.4\times10^4\,{\rm M}_{\odot}$, mean number density of $\langle n_{\rm H}\rangle\sim210\,{\rm cm}^{-3}$ and mean total column density of $\langle N_{\rm tot}\rangle\sim9.21\times10^{21}\,{\rm cm}^{-2}$.

\begin{figure}
    \centering
    \includegraphics[width=\linewidth]{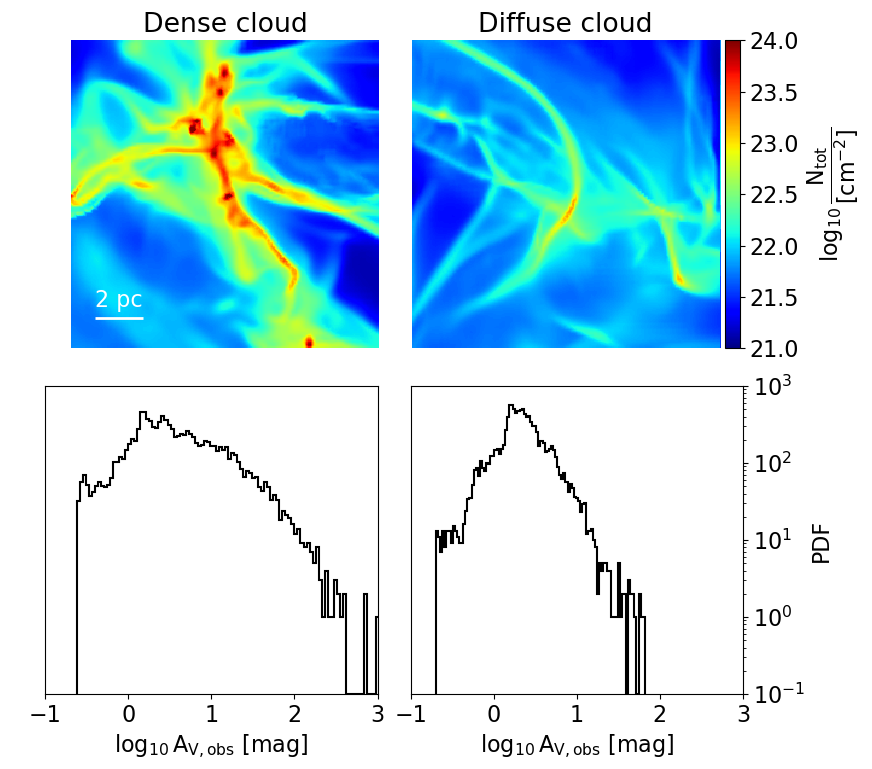}
    \caption{Top row: Column densities of the total H-nucleus number density, $N_{\rm tot}$, for the dense (left column) and the diffuse (right column) sub-regions considered in this work. In the dense cloud, the cloud-cloud collision occurs along the line-of-sight. Bottom row: the corresponding probability density functions of the observed visual extinction, $A_{\rm V,obs}$, of the top row. The metallicity here is taken to be $1\,{\rm Z}_{\odot}$.}
    \label{fig:pdfs}
\end{figure}

Figure~\ref{fig:pdfs} shows total H-nucleus column density maps ($N_{\rm tot}$) of the two sub-regions as well as their corresponding $A_{\rm V}$-PDF (probability density function) for the observed visual extinction, $A_{\rm V,obs}$. Throughout the paper, the collision in the `dense cloud' occurs along the line-of-sight of the observer. The histogram was generated using 100 bins. We convert from $N_{\rm tot}$ to $A_{\rm V}$ using the expression $A_{\rm V}=A_{\rm V,o}N_{\rm tot}(Z/{\rm Z}_{\odot})$ where $A_{\rm V,o}=6.3\times10^{-22}\,{\rm mag}\,{\rm cm}^{2}$ \citep{Wein01,Roll07} and $Z$ is the metallicity. The aforementioned relation of visual extinction with metallicity results from the $A_{\rm V}$ dependency on the column density of dust grains, the latter scaling linearly with metallicity for values down to 0.1 of the solar one \citep[][]{Lero11, Feld12}. Note that in the $A_{\rm V}$-PDF diagrams, the high column-density tail of the dense cloud extends more than that of the diffuse one as a result of the collision and the higher mass concentrated in dense structures.

\subsection{Photodissociation region calculations}
\label{ssec:3dpdr}

The publicly available code {\sc 3d-pdr}\footnote{https://uclchem.github.io/3dpdr.html} \citep{Bisb12} is used in this work to calculate the gas and dust temperatures, the abundances of species, as well as the level populations of the coolants examined. {\sc 3d-pdr} is able to treat the astrochemistry of photodissociation regions of an arbitrary three-dimensional density distribution. The code performs iterations over thermal balance and the calculations terminate once the total heating and the total cooling are equal to within $0.5\%$. Various heating and cooling processes are taken into account \citep[see][for full description]{Bisb12, Bisb14, Bisb17a}. The PDR calculations in this paper utilized a subset of the UMIST 2012 chemical network \citep{McEl13} consisting of 33 species (including e$^-$) and 330 reactions. Standard ISM abundances of $Z=1\,{\rm Z}_{\odot}$ metallicity have been adopted, i.e., [He]/[H]=0.1, [C]/[H]=$10^{-4}$ and [O]/[H]=$3\times10^{-4}$ \citep{Card96,Cart04,Roll07}. For environments with different metallicities, the aforementioned abundances are scaled accordingly.

The `ISM fiducial parameters' are those defined with a cosmic-ray ionization rate of $\zeta_{\rm CR}=10^{-16}\,{\rm s}^{-1}$, an external isotropic FUV intensity of $G_0=10$ and a metallicity of $Z=1\,{\rm Z}_{\odot}$. For each sub-region, a total of four cosmic-ray ionization rates are considered ($\zeta_{\rm CR}=10^{-17}-10^{-14}\,{\rm s}^{-1}$), four FUV intensities ($G_0=1-10^3$, normalized according to \citealt{Habi68}) and four metallicities ($Z=0.1-2\,{\rm Z}_{\odot}$). When a particular ISM parameter is explored, the other two remain those from the fiducial parameters. For each sub-region, ten simulations are thus performed, therefore twenty overall for this work. Table~\ref{tab:ics} shows a summary of all ISM environmental parameters used for each cloud. Furthermore, the microturbulent velocity (controlling the turbulent heating in {\sc 3d-pdr}) is set to $\varv_{\rm turb}=2\,{\rm km}\,{\rm s}^{-1}$, constant everywhere in both clouds, which approximates the local one-dimensional turbulent velocity dispersion of the MHD runs. 

\begin{figure}
    \centering
    \includegraphics[width=\linewidth]{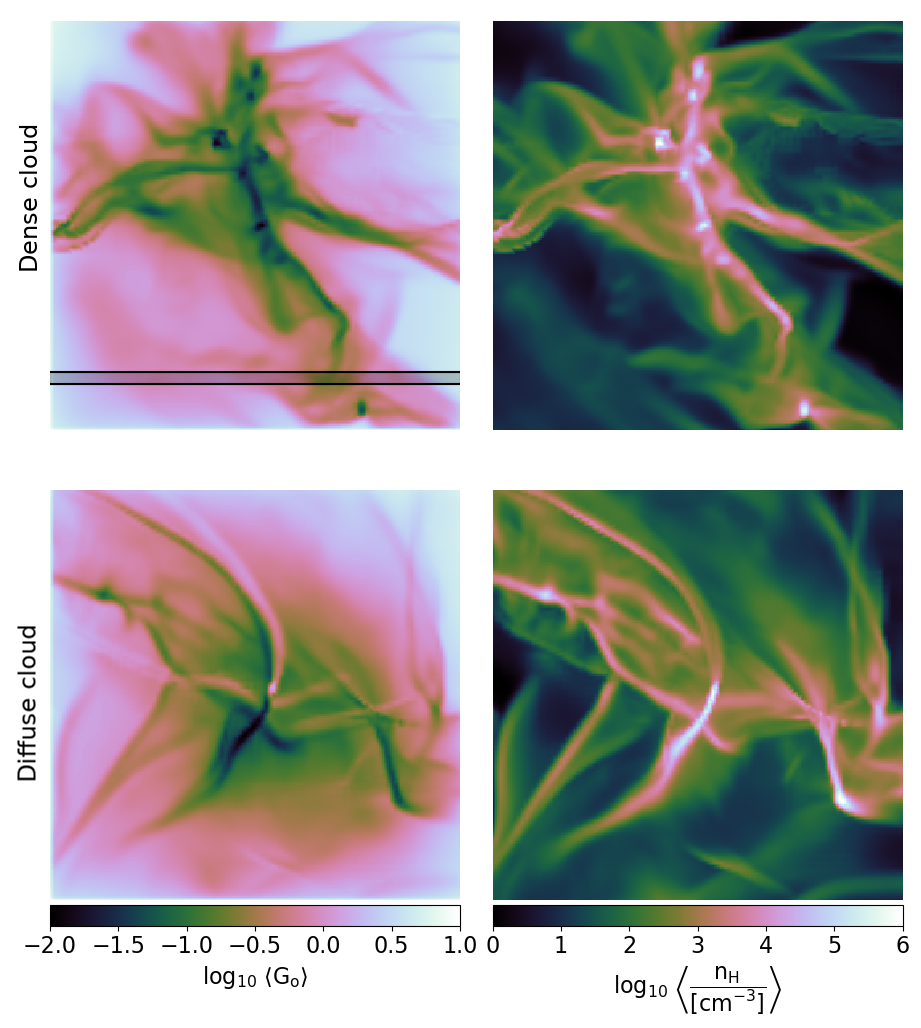}
    \caption{The distribution of radiation field (left column) and the average H-nucleus number density (right column) for the dense cloud (top row) and the diffuse cloud (bottom row). The shadowed stripe with thickness of $0.4\,{\rm pc}$ in the dense cloud shows the part of the cloud used to construct the spectra discussed in \S\ref{ssec:spectra}. Both density distributions are sub-regions from the \citet{Wu17} MHD simulations. The distribution of $\langle G_0\rangle$ shows the case for $G_0=10$. All other cases of FUV intensities obey the same qualitative distribution pattern.}
    \label{fig:GonH}
\end{figure}

Figure~\ref{fig:GonH} illustrates the local (and thus attenuated) density-weighted FUV radiation field ($\langle G_0\rangle$) as calculated by {\sc 3d-pdr}, as well as the density-weighted H-nucleus number density in each pixel. The distribution of $\langle G_0\rangle$ shows the case for an isotropic field of $G_0=10$ as the outer boundary condition (unattenuated). All the other FUV fields explored (not shown) obey the same qualitative distribution pattern. In each pixel, the above quantities ($Q$) are calculated as:
\begin{eqnarray}
\label{eqn:averages}
\langle Q\rangle = \frac{\int Q n_{\rm H} {\rm d}z}{\int n_{\rm H}{\rm d}z}.
\end{eqnarray}

\begin{table}
    \centering
    \begin{tabular}{l|c|c}
    \hline
    $\zeta_{\rm CR}\,({\rm s}^{-1})$    & $G_0$ & $Z\,({\rm Z}_{\odot})$\\\hline\hline
    ~~$10^{-17}$   & $10$ & $1$\\
    $\star10^{-16}$   & $10$ & $1$\\
    ~~$10^{-15}$   & $10$ & $1$\\
    ~~$10^{-14}$   & $10$ & $1$\\
    ~~$10^{-16}$   & $1$ & $1$\\
    ~~$10^{-16}$   & $10^2$ & $1$\\
    ~~$10^{-16}$   & $10^3$ & $1$\\
    ~~$10^{-16}$   & $10$ & $0.1$\\
    ~~$10^{-16}$   & $10$ & $0.5$\\
    ~~$10^{-16}$   & $10$ & $2$\\
    \hline
    \end{tabular}
    \caption{Summary of the ISM environmental parameters used for each cloud. The starred conditions correspond to the fiducial ISM parameters.}
    \label{tab:ics}
\end{table}

Cosmic-rays as low as $\zeta_{\rm CR}\sim10^{-17}\,{\rm s}^{-1}$ have been observed in the solar neighbourhood from the Voyager-1 spacecraft \citep{Cumm15,Cumm16}. However, a higher value of $\zeta_{\rm CR}\sim10^{-16}\,{\rm s}^{-1}$ has been suggested as the average one in the diffuse ISM of the Milky Way \citep{McCa03,Dalg06,Neuf10,Indr12,Indr15}. This will be the fiducial value in our simulations \citep[see also][]{Li18}. A much higher ionization rate of about $\zeta_{\rm CR}\sim10^{-14}\,{\rm s}^{-1}$ has been suggested to exist in regions close to the Galactic Centre, as revealed from H$_3^+$ \citep{Oka05,Goto08}, 6.4~keV Fe~K$\alpha$ line emission, gamma-rays, and synchrotron emission \citep{Yuse07,Yuse13}. Similar trends of increased FUV intensities in the Galactic Centre region have been found \citep[e.g.][]{Wolf03}. Metallicities as high as $Z\sim2\,{\rm Z}_{\odot}$ have been observed in the environment of inner Milky Way \citep{Give02} and known to decrease to the outer parts of our Galaxy \citep{Wolf03} with values close to $0.1\,{\rm Z}_{\odot}$ \citep[e.g.][]{Geno14,Inno20}. Therefore our choices of all the environmental parameters are to reflect the observed conditions of different parts in our Galaxy. However, these parameters are expected to be relevant much more broadly to conditions in a variety of galaxies.

As mentioned earlier, in this paper we post-process two static hydrodynamical snapshots under different ISM environmental parameters. However, different ISM conditions would inevitably result in different gas temperatures and, therefore, pressures leading to different density distributions. Solving simultaneously for such a detailed PDR chemistry as considered here with hydrodynamics is computationally very demanding. As an attempt to reduce the computational cost, various groups limit the chemical processes considered, consequently resulting in a limitation of their outcomes when it comes to ISM diagnostics. For the purposes of our work, we have adopted the limitation of removing the hydrodynamical calculations and focusing purely on the detailed PDR chemistry at equilibrium. This allows us to study simultaneously the behaviour and the trends of many different species and emission lines as a function of various ISM environmental parameters.

\subsection{Radiative transfer}
\label{ssec:rt}

In this work, we study the emission of the most important PDR coolants. These include the first ten $^{12}$CO transitions ($J=1-0,...,10-9$), [C{\sc i}]~(1-0) and (2-1), [O{\sc i}] at $63\mu$m and $146\mu$m, and [C{\sc ii}] at $158\mu$m. 
We solve the equation of radiative transfer for each line:
\begin{eqnarray}
\frac{{\rm d}I_{\nu}}{{\rm d}z}
=\alpha_{\nu{\rm,line}}(B_{\nu}(T_{\rm ex})-I_{\nu})
+\rho\kappa_{\nu{\rm,dust}}(B_{\nu}(T_{\rm dust})-I_{\nu}),
\label{eqn:rt}
\end{eqnarray}
where $z$ is the line-of-sight direction, $\nu$ is the frequency, $I_{\nu}$ is the intensity, $\alpha_{\nu{\rm,line}}$ is the absorption coefficient of the line, $B_{\nu}$ is the Planck function, $T_{\rm ex}$ is the excitation temperature of the line, $T_{\rm dust}$ is the dust temperature, $\rho$ is the density, and $\kappa_{\nu{\rm,dust}}$ is the dust opacity, respectively. We use the same radiative transfer algorithm developed in \citet{Bisb17b}, but have updated it to take into account the dust emission/absorption (the second term in the right side of Equation \ref{eqn:rt}). This is a necessary update in order to more accurately treat dense, high $A_V$ regions where the clouds become optically thick at shorter wavelengths. Therefore, the line intensity is evaluated by subtracting the dust continuum from the total intensity, i.e., $I_{\nu,{\rm line}} = I_{\nu}-I_{\nu,{\rm dust}}$. We utilize the values of $T_{\rm ex}$ and $T_{\rm dust}$  obtained from our \textsc{3d-pdr} results, while the density $\rho$ is from the \citet{Wu17} MHD simulations. For the dust opacity at solar metallicity, we adopt an approximate power-law of the form $\kappa_{\nu{\rm, dust}}=0.1\times(Z/{\rm Z}_\odot)(\nu/1000\,{\rm GHz})^2\,{\rm cm}^{2}\,{\rm g}^{-1}$ per unit of the total density of gas and dust \citep[e.g.,][]{Arzo11}. For other metallicity cases, we assume that the opacity scales proportionally to the metallicity \citep[e.g.,][]{Bate14,Bate19,Tanaka14,Tanaka18}. This simple assumption might cause an overestimation of the dust opacity, because the dust-to-gas mass ratio is lower than a linear relation at low metallicity of $\la0.2{\rm Z_{\odot}}$ \citep{Remy14}. The gas velocity is taken directly from the MHD simulations to evaluate the Doppler shifting of the line,
\begin{eqnarray}
\nu=\nu_0\left(1-\frac{\varv_{\rm los}}{c}\right), 
\end{eqnarray}
where $\nu_0$ is the frequency in the lab frame, $\varv_{\rm los}$ is the gas velocity along the line-of-sight, and $c$ is the speed of light, respectively \citep[see][for further details]{Bisb17b}.

In each emission map, we estimate the integrated antenna temperature over velocity as
\begin{eqnarray}
\label{eqn:emission}
W=\int_{\varv_{\rm min}}^{\varv_{\rm max}} T_{\rm A}{\rm d}\varv_{\rm los},
\end{eqnarray}
where $T_{\rm A}$ is the antenna temperature of the line, $\varv_{\rm min}=-20\,{\rm km}\,{\rm s}^{-1}$, and $\varv_{\rm max}=+20\,{\rm km}\,{\rm s}^{-1}$. The antenna temperature is defined as
\begin{eqnarray}
T_{\rm A}=\frac{c^2 I_{\nu,{\rm line}}}{2 k_{\rm B}\nu^2},
\end{eqnarray}
where $k_{\rm B}$ is the Boltzmann constant.

We calculate the emission of the [$^{13}$C{\sc ii}] isotope following \citet{Kirs20}. The antenna temperature of this isotope at a given velocity channel is connected with the antenna temperature of [C{\sc ii}] through the expression:
\begin{eqnarray}
\frac{T_{\rm A}([\rm ^{13}CII])}{T_{\rm A}(\rm [CII])}=\frac{1-e^{-\tau_{\rm CII}/r}}{1-e^{-\tau_{\rm CII}}}\approx\frac{\tau_{\rm CII}/r}{1-e^{-\tau_{\rm CII}}}
\end{eqnarray}
where $\tau_{\rm CII}$ is the optical depth of [C{\sc ii}] and $r={\rm ^{12}C}/ ^{13}{\rm C}$ is the atomic carbon isotopic abundance ratio which we take to be $r=80$ \citep{Wils99}. By integrating $T_{\rm A}([\rm ^{13}C])$ over velocities using Eqn.(\ref{eqn:emission}), we obtain the velocity integrated emission of the [$^{13}$C{\sc ii}] isotope. We note that lower abundance ratios have been observed, e.g., in Orion \citep{Lang90,Lang93} at a value of $r=67$ \citep[see also][]{Wake08, Osse13}. Such lower ratios will simply increase the results of $W(^{13}{\rm CII})$ presented here by, e.g., a factor of $80/67\sim1.2$.

\section{Distribution of abundances and gas temperatures}
\label{sec:cds}

\begin{figure*}
    \centering
    \includegraphics[width=0.99\linewidth]{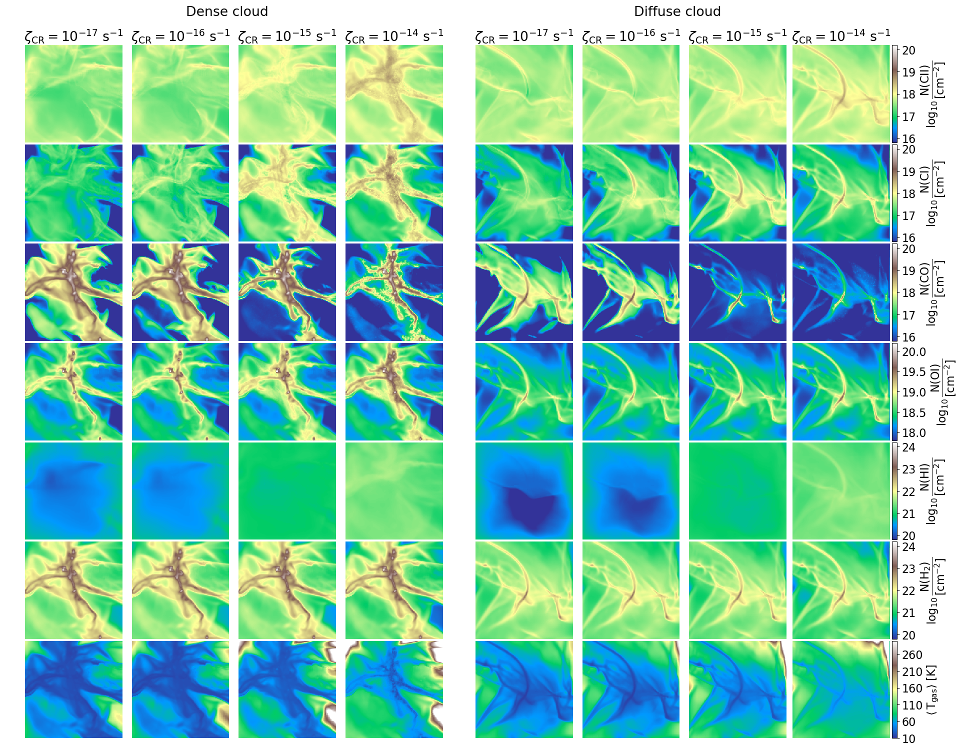}
	\caption{Distributions of column densities with $\zeta_{\rm CR}$ as the free parameter, density-weighted gas temperatures, density-weighted FUV radiation field and average total H-nucleus number densities for the dense cloud (left suite) and the diffuse cloud (right suite). From top-to-bottom: column densities of C{\sc ii}, C{\sc i}, CO, O{\sc i}, H{\sc i} and H$_2$ and density-weighted gas temperatures. The cosmic-ray ionization rate increases from left-to-right ($\zeta_{\rm CR}=10^{-17}\,{\rm s}^{-1}$ to $10^{-14}\,{\rm s}^{-1}$). As $\zeta_{\rm CR}$ increases, the abundances of C{\sc ii}, C{\sc i}, and O{\sc i} increase while CO significantly decreases everywhere except the dense filamentary structure in the left panel and the densemost part of the right panel. On the other hand, the abundance of H$_2$ remains remarkably unchanged as explained in \citet{Bisb15,Bisb17a}. Note that for $\zeta_{\rm CR}\sim10^{-14}\,{\rm s}^{-1}$, $T_{\rm gas}$ increases in the dense cloud to values $\gtrsim30\,{\rm K}$ everywhere in the filament and to $\gtrsim45\,{\rm K}$ in the diffuse cloud (see Tables~\ref{tab:dense}, \ref{tab:diffuse}). The colour bars are shown on the right and are common for each row. In all cases the intensity of the isotropic FUV radiation field is taken to be $G_0=10$.} 
    \label{fig:col_CRcd}
\end{figure*}

\begin{figure*}
    \centering
    \includegraphics[width=0.99\linewidth]{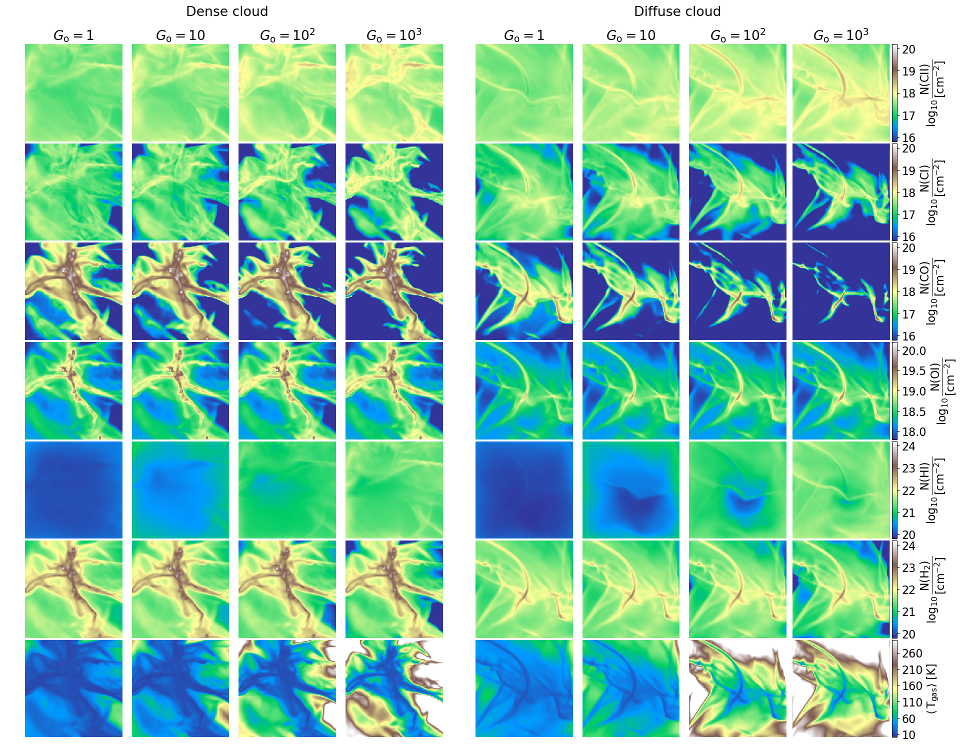}
	\caption{Same as Fig.~\ref{fig:col_CRcd} but for the FUV intensity as the free parameter, which increases from left-to-right ($G_0=1$ to $10^3$). As $G_0$ increases, the abundance of C{\sc ii} increases everywhere in both clouds and C{\sc i} increases in the dense and more shielded regions, while decreases in the diffuse outer gas as it converts to C{\sc ii}. The increase of both C{\sc ii} and C{\sc i} is as a result of the photodissociation of CO, the abundance of which consequently decreases. The effect is more prominent in the diffuse cloud. The abundance of O{\sc i} increases by a very small amount for higher radiation field intensities. In both clouds --and as expected-- the H$_2$ molecule also dissociates with increasing $G_0$ resulting in an increase of H{\sc i}. The gas temperature remains low at high column densities since the FUV photons extinguish rapidly. On the other hand the average gas temperature in the outer parts of both distributions and especially in the diffuse cloud increases in response to the increase of photoelectric heating for high $G_0$.}
    \label{fig:col_Gcd}
\end{figure*}

\begin{figure*}
    \centering
    \includegraphics[width=0.99\linewidth]{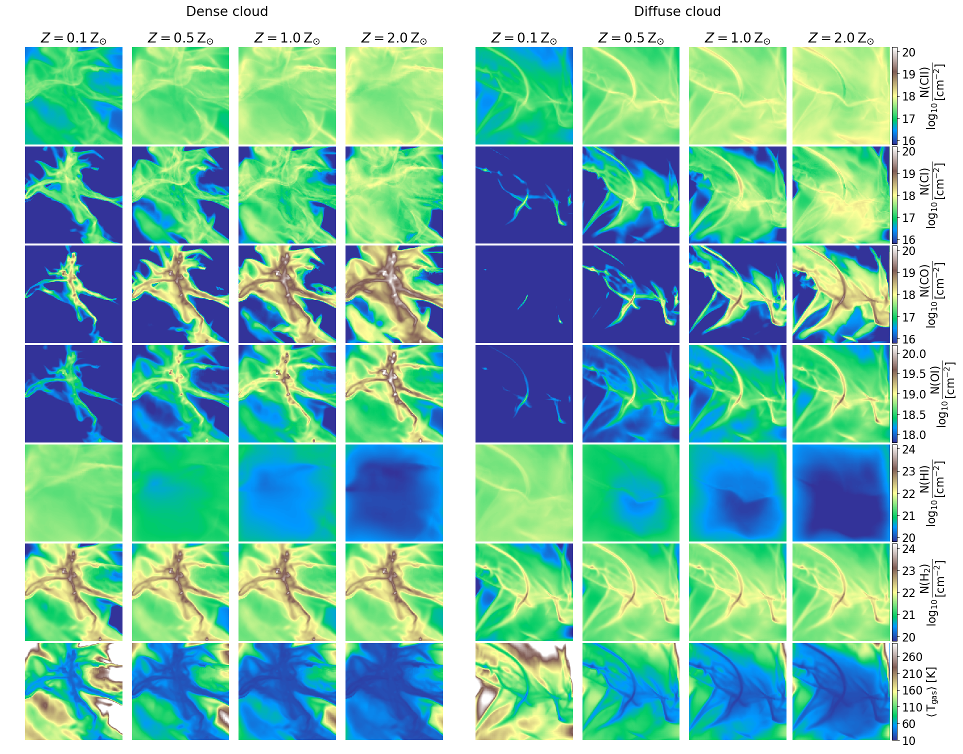}
	\caption{Same as Fig.~\ref{fig:col_CRcd} but for metallicity as the free parameter, which increases from left-to-right ($Z=0.1$ to $2\,{\rm Z_{\odot}}$). The size of both clouds is assumed to remain the same for all metallicities, implying that the visual extinction (both effective and observed) is being scaled accordingly. At low metallicities, the abundances of CO and H$_2$ decrease in both clouds with H$_2$ to decrease more mildly. In particular, at $Z=0.1\,{\rm Z}_{\odot}$ both clouds remain molecular, but the diffuse one has considerably high abundance of H{\sc i} as a result of the H$_2$ photodissociation. The clouds are also rich in C{\sc ii} and the gas temperature is increased since there is less shielding to cause extinction of FUV radiation.}
    \label{fig:col_Zcd}
\end{figure*}

Figures~\ref{fig:col_CRcd}, \ref{fig:col_Gcd} and \ref{fig:col_Zcd} show two suites of panels illustrating column density maps of C{\sc ii}, C{\sc i}, CO, O{\sc i}, H{\sc i} and H$_2$ species, as well as the average, density-weighted gas temperatures (using Eqn.~\ref{eqn:averages}) for the two clouds and for the cosmic-ray ionization rate ($\zeta_{\rm CR}$), the FUV intensity and the metallicity as the free ISM parameters, respectively. Note that the panels showing the carbon cycle column-densities share the same colour bar extent. Tables~\ref{tab:dense} and \ref{tab:diffuse} provide a summary of the average abundances and $\langle T_{\rm gas}\rangle$ for the dense and diffuse clouds respectively. In both tables we include additionally the results for an isotropic radiation field with strength $G_0=4$, which was adopted for the studies in our earlier works \citep{Bisb17b,Bisb18}. Such a radiation field is expected to appear in the inner part of Milky Way \citep{Wolf03}.

\subsection{Varying the cosmic-ray ionization rate}
\label{ssec:cdcr}

It is known \citep[e.g.,][]{Meij11,Bial15} that cosmic-rays ionize He creating He$^+$ which then reacts and destroys very effectively the molecule of CO following the reaction:
\begin{myequation}
\label{R1}
{\rm CO} + {\rm He}^+ \longrightarrow {\rm C}^+ + {\rm O} + {\rm He}
\end{myequation}
The above reaction creates and therefore increases the abundance of C{\sc ii} which then reacts quickly with free e$^-$ creating C{\sc i}-rich regions. On the other hand, although also destroyed by cosmic-rays, H$_2$ forms back to its molecular phase quickly following the ion-neutral reaction between H and H$_2^+$ for lower densities as well as between H$_3^+$ and O, C{\sc i}, or CO for higher densities \citep{Bisb15,Bisb17a}. All the above processes are needed to explain the trends observed in the panels of Fig.~\ref{fig:col_CRcd}. In particular, we see that $N({\rm H}_2)$ remains remarkably unchanged even in the extreme case of $\zeta_{\rm CR}=10^{-14}\,{\rm s}^{-1}$ and even when $N_{\rm tot}$ is as low as $N_{\rm tot}\sim 3\times10^{21}\,{\rm cm}^{-2}$. Any minor decrease in H$_2$ is reflected in a corresponding increase in H{\sc i}. However, $N({\rm CO})$ decreases in some H$_2$-rich regions by more than two orders of magnitude leading to extended CO-deficient regions. We also note that the cosmic-ray induced destruction of CO and H$_2$ are expected to be weaker in dense gas since the photodissociation by secondary UV photons due to cosmic-rays are compensated by the higher formation efficiency \citep{Bisb15,Wake17}.

\begin{table*}
    \centering
    \begin{tabular}{l|c|c|c|c|c|c|c}
         \hline
         Free parameter & $x$(H$_2$) & $x$(H{\sc i}) & $x$(C{\sc ii}) & $x$(C{\sc i}) & $x$(CO) & $x$(O{\sc i}) & $\langle\rm T_{\rm gas}\rangle$~[K] \\ \hline\hline
         ~~$\zeta_{\rm CR}=10^{-17}\,{\rm s}^{-1}$ & 0.494 & $1.20(-2)$ & $1.25(-5)$ & $3.61(-6)$ & $8.38(-5)$ & $1.44(-4)$ & 29.4 \\ 
         $\star\zeta_{\rm CR}=10^{-16}\,{\rm s}^{-1}$ & 0.493 & $1.40(-2)$ & $1.36(-5)$ & $7.42(-6)$ & $7.89(-5)$ & $1.44(-4)$ & 30.8 \\
         ~~$\zeta_{\rm CR}=10^{-15}\,{\rm s}^{-1}$ & 0.486 & $2.79(-2)$ & $1.96(-5)$ & $2.05(-5)$ & $5.99(-5)$& $1.77(-4)$ & 39.3 \\
         ~~$\zeta_{\rm CR}=10^{-14}\,{\rm s}^{-1}$ & 0.456 & $8.91(-2)$ & $3.49(-5)$ & $2.39(-5)$ & $4.11(-5)$&$2.22(-4)$ & 56.9 \\
         ~~$G_0=1$ & 0.498 & $4.20(-3)$ & $9.71(-6)$ & $7.67(-6)$ & $8.26(-5)$ & $1.33(-4)$ & 26.2 \\
         ~~$G_0=4$ & 0.496 & $8.35(-3)$ & $1.20(-5)$ & $7.42(-6)$ & $8.05(-5)$ & $1.40(-4)$ & 28.3 \\
         ~~$G_0=10^2$ & 0.481 & $3.81(-2)$ & $1.80(-5)$ & $7.85(-6)$ & $7.42(-5)$ & $1.56(-4)$ & 44.2 \\
         ~~$G_0=10^3$ & 0.460 & $7.90(-2)$ & $2.23(-5)$ & $8.85(-6)$ & $6.88(-5)$ & $1.68(-4)$ & 73.2 \\
         ~~$Z=0.1\,{\rm Z_{\odot}}$ & 0.450 & 0.100 & $3.64(-6)$ & $2.50(-6)$ & $3.82(-6)$ & $2.28(-5)$ & 70.8 \\
         ~~$Z=0.5\,{\rm Z_{\odot}}$ & 0.486 & $2.84(-2)$ & $1.00(-5)$ & $5.21(-6)$ & $3.47(-5)$ &$7.83(-5)$ & 37.4 \\
         ~~$Z=2.0\,{\rm Z_{\odot}}$ & 0.497 & $5.63(-3)$ & $1.70(-5)$ & $1.11(-5)$ & $1.72(-4)$ & $3.02(-4)$ & 26.2 \\
         \hline
    \end{tabular}
    \caption{Summary of the average abundances of species (H$_2$, H{\sc i}, C{\sc ii}, C{\sc i}, CO, O{\sc i}) and average gas temperatures for all ISM environmental parameters explored for the dense cloud. The starred simulation corresponds to the fiducial ISM environmental parameters ($\zeta_{\rm CR}=10^{-16}\,{\rm s}^{-1}$, $G_0=10$, $Z=1\,{\rm Z_{\odot}}$). The number in the brackets corresponds to the order of magnitude.}
    \label{tab:dense}
\end{table*}

\begin{table*}
    \centering
    \begin{tabular}{l|c|c|c|c|c|c|c}
         \hline
         Free parameter & $x$(H$_2$) & $x$(H{\sc i}) & $x$(C{\sc ii}) & $x$(C{\sc i}) & $x$(CO) & $x$(O{\sc i}) & $\langle\rm T_{\rm gas}\rangle$~[K] \\ \hline\hline
         ~~$\zeta_{\rm CR}=10^{-17}\,{\rm s}^{-1}$ & 0.483 & $3.27(-2)$ & $5.07(-5)$ & $1.34(-5)$ & $3.59(-5)$ & $2.68(-4)$ & 49.9\\
         $\star\zeta_{\rm CR}=10^{-16}\,{\rm s}^{-1}$ & 0.481 & $3.80(-2)$ & $5.31(-5)$ & $2.21(-5)$ & $2.50(-5)$ & $2.61(-4)$ & 52.7 \\
         ~~$\zeta_{\rm CR}=10^{-15}\,{\rm s}^{-1}$ & 0.457 & $8.69(-2)$ & $6.08(-5)$ & $3.13(-5)$ & $7.80(-6)$ & $2.87(-4)$ & 65.8 \\
         ~~$\zeta_{\rm CR}=10^{-14}\,{\rm s}^{-1}$ & 0.357 & 0.284 & $7.42(-5)$ & $2.22(-5)$ & $3.51(-6)$ & $2.95(-4)$ & 82.0 \\
         ~~$G_0=1$ & 0.494 & $1.16(-2)$ & $3.73(-5)$ & $2.64(-5)$ & $3.63(-5)$ & $2.49(-4)$ & 37.4 \\
         ~~$G_0=4$ & 0.489 & $2.20(-2)$ & $4.68(-5)$ & $2.35(-5)$ & $2.97(-5)$ & $2.61(-4)$ & 44.7 \\
         ~~$G_0=10^2$ & 0.433 & 0.134 & $6.80(-5)$ & $1.69(-5)$ & $1.51(-5)$ & $2.82(-4)$ & 154.3 \\
         ~~$G_0=10^3$ & 0.344 & 0.312 & $7.92(-5)$ & $1.22(-5)$ & $8.54(-6)$ & $2.90(-4)$ & 176.1 \\
         ~~$Z=0.1\,{\rm Z_{\odot}}$ & 0.300 & 0.400 & $9.29(-6)$ & $6.12(-7)$ & $9.40(-8)$ & $3.00(-5)$ & 135.7 \\
         ~~$Z=0.5\,{\rm Z_{\odot}}$ & 0.455 & $9.08(-2)$ & $3.63(-5)$ & $8.66(-6)$ & $5.03(-6)$ & $1.44(-4)$ & 70.2 \\
         ~~$Z=2.0\,{\rm Z_{\odot}}$ & 0.493 & $1.45(-2)$ & $6.65(-5)$ & $4.58(-5)$ & $8.76(-5)$ & $4.80(-4)$ & 46.7 \\
         \hline
    \end{tabular}
    \caption{As in Table~\ref{tab:dense} for the diffuse cloud.}
    \label{tab:diffuse}
\end{table*}

The above findings are in broad agreement with the 3D simulations discussed in \citet{Bisb17a}. In the present work, there are, however, some additional interesting features. In the dense cloud with $\zeta_{\rm CR}=10^{-17}\,{\rm s}^{-1}$, the column density of C{\sc ii} is almost entirely connected to the distribution of the rarefied medium with no obvious connection to the distribution of $N$(H$_2$). As $\zeta_{\rm CR}$ increases, and in particular when it reaches the maximum value of $10^{-14}\,{\rm s}^{-1}$, the abundance of C{\sc ii} at high densities increases to such a high level that the spatial distribution of C{\sc ii} column densities follows closely that of the molecular gas. A similar feature is observed with $N$(C{\sc i}), whereas $N$(CO) is progressively narrowed down to the densest parts of the filamentary structures. Note that at $\zeta_{\rm CR}=10^{-14}\,{\rm s}^{-1}$, both $N$(C{\sc ii}) and $N$(C{\sc i}) have very similar overall spatial distributions as $N$(H$_2$). On the other hand, the $N$(CO) and $\rm H_2$ correspondence is good when $\zeta_{\rm CR}=10^{-17}\,{\rm s}^{-1}$. The average gas temperature maps also show that cosmic-ray heating significantly increases the temperatures at high column densities, which has consequences for both the dynamics, i.e., by increasing thermal pressures, and for the emissivities of lines (see \S\ref{sec:emission}).

\subsection{Varying the FUV radiation field intensity}
\label{ssec:varyFUVcd}

As $G_0$ increases (Fig.~\ref{fig:col_Gcd}), it photodissociates the molecule of CO increasing the abundances of C{\sc i} and particularly C{\sc ii} \citep{vDish88,Stern95,Beut14}. Tables~\ref{tab:dense} and \ref{tab:diffuse} (rows with $G_0$ as the free parameter), show this decrease of CO abundance. In the dense cloud, the abundance of C{\sc i} slightly increases since the high-density filamentary structure (where the majority of the cloud mass is concentrated) shields the FUV photons preventing the ionization of C{\sc i} to C{\sc ii} to dominate over the photodissociation of CO to C{\sc i}. On the other hand, the diffuse cloud becomes quickly C{\sc ii}-dominated as FUV photons ionize C{\sc i}, the abundance of which is consequently decreasing. For the particular case of C{\sc i}, Fig.~\ref{fig:col_Gcd} shows additionally that this species is effectively destroyed in the outer gas as it is more translucent, thus N(C{\sc i}) decreases. However, in the high density regions N(C{\sc i}) increases mainly due to CO photodissociation. We find that $N$(C{\sc ii}) can increase up to one order of magnitude between the $G_0=1$ and $10^3$ cases. The photodissociation of CO creates also a surplus of O{\sc i} abundance, hence its column density also increases but by small amounts. For instance the average (density-weighted) abundance of O{\sc i} increases from $1.33\times10^{-4}$ to $1.68\times10^{-4}$ in the dense cloud and from $2.49\times10^{-4}$ to $2.90\times10^{-4}$ in the diffuse cloud (see Tables~\ref{tab:dense} and \ref{tab:diffuse}).

The FUV photons also dissociate the H$_2$ molecule producing H{\sc i} gas, especially in the shielded regions. This pushes the H{\sc i}-to-H$_2$ transition to higher column densities \citep{vDish86,Kauf99,Bial16}. As with C{\sc ii}, the column density of H{\sc i} also increases up to one order of magnitude on average in both clouds. Looking at the gas temperature, it is found that it remains always low in the parts with high densities, since the FUV photons extinguish rapidly. This is observed in both density distributions. However, the outermost parts become warmer as $G_0$ increases, reaching values of $\langle T_{\rm gas}\rangle\gtrsim500\,{\rm K}$.

The abundance ratios of CO/H$_2$, C{\sc i}/H$_2$ and C{\sc ii}/H$_2$ indicate that the dense cloud remains almost entirely molecular and rich in CO. In particular, the above ratios for the $G_0=10^3$ intensity are found to be $1.49\times10^{-4}$, $1.92\times10^{-5}$ and $4.84\times10^{-5}$ (Table~\ref{tab:dense}). For lower values of $G_0$, CO/H$_2$ increases (up to $1.66\times10^{-4}$ for $G_0=1$), while the other two decrease. On the other hand, the diffuse cloud is always C{\sc ii}-rich under all $G_0$ intensities, although it mostly remains molecular. Here, we found the ratios of CO/H$_2$ as $(7.35$--$2.48)\times10^{-5}$, C{\sc i}/H$_2$ as $(5.34$--$3.55)\times10^{-5}$, and C{\sc ii}/H$_2$ as $(7.55$--$23.03)\times10^{-5}$ for $G_0=1$--$10^3$, respectively (see Table~\ref{tab:diffuse}). Such values of abundance ratios are in agreement with the findings of \citet{Bisb19} in studying clouds with similar column density distribution.

\subsection{Varying the metallicity}
\label{ssec:Zvary}

\begin{figure*}
    \centering
    \includegraphics[width=\linewidth]{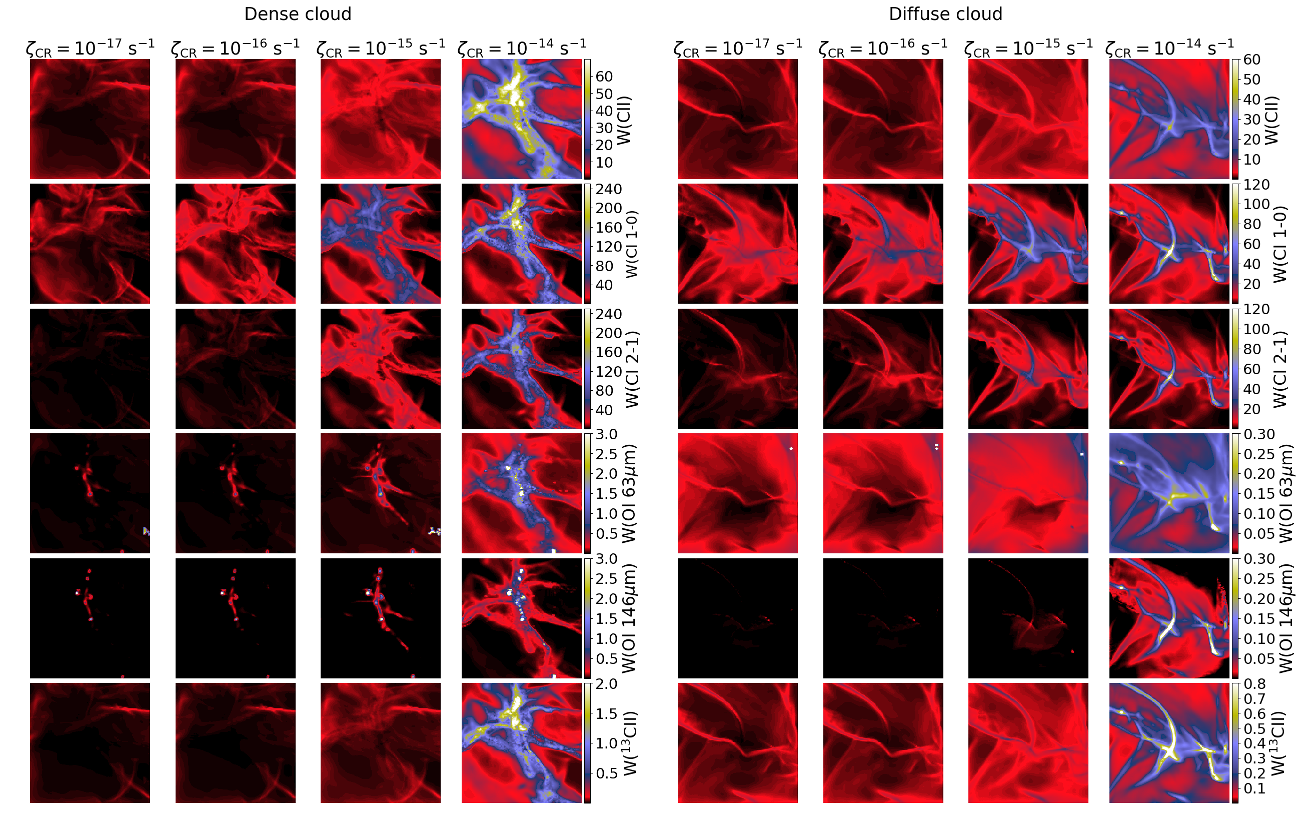}
	\caption{From top-to-bottom: emission maps (in units of [K~km/s]) for [C{\sc ii}]~$158\mu$m, [C{\sc i}]~$609\mu$m and $370\mu$m, [O{\sc i}]~$63\mu$m and $146\mu$m and [$^{13}$C{\sc ii}]. The left suite is for the dense cloud and the right for the diffuse cloud. In each suite, the cosmic-ray ionization rate increases from left-to-right. In both density distributions, the emission of the fine-structure lines of [C{\sc ii}], both [C{\sc i}], both [O{\sc i}] and [$^{13}$C{\sc ii}] increases and becomes more extended by increasing $\zeta_{\rm CR}$. Notably, at $\zeta_{\rm CR}=10^{-14}\,{\rm s}^{-1}$ the emissions of [C{\sc ii}] and [$^{13}$C{\sc ii}] originate from high (and therefore H$_2$-rich) column densities.} 
    \label{fig:col_CRemission}
\end{figure*}

Low gas-phase metallicity affects the dust-shielding of CO against FUV photons  as well as the formation of H$_2$ since the dust-to-gas ratio decreases. Consequently, the visual extinction scales linearly with $Z$. These lead to extended H$_2$-rich regions which are deficient in CO abundance and thus CO-dark \citep{Pak98,Bial15,Madd20}. On the other hand, high metallicities have high dust-to-gas ratios, therefore blocking the FUV photons penetrating the ISM; this creates H$_2$-rich clouds which are also CO-rich and have low gas temperature.

The above trends are seen in Fig.~\ref{fig:col_Zcd}. For $Z=2\,{\rm Z}_{\odot}$ both clouds are fully molecular with an average gas temperature of $\langle T_{\rm gas}\rangle{\simeq}26\,{\rm K}$ for the dense and $47\,{\rm K}$ for the diffuse cloud (Tables~\ref{tab:dense} \& \ref{tab:diffuse}, respectively). At these supersolar metallicities, CO/H$_2{\simeq}3.46\times10^{-4}$ for the dense and $1.78\times10^{-4}$ for the diffuse clouds. Similarly, C{\sc i}/H$_2{\simeq}2.23\times10^{-5}$ and $9.29\times10^{-5}$, and C{\sc ii}/H$_2{\simeq}3.42\times10^{-5}$ and $1.34\times10^{-4}$. Notably, the dense cloud has fifteen times higher CO abundance than C{\sc i} and ten times higher than C{\sc ii}, while the diffuse cloud has twice CO abundance than C{\sc i} and approximately equal abundance to C{\sc ii}.

On the other hand, for $Z=0.1\,{\rm Z}_{\odot}$ the dense cloud remains H$_2$-rich and has an average gas temperature of $71\,{\rm K}$. Its abundances in all carbon phases are quite low (since both C and O abundances are reduced) but similar to each other without significant differences. For instance the abundances of C{\sc ii} and CO are $3.7\times10^{-6}$ and that of C{\sc i} is slightly lower at $2.5\times10^{-6}$. Contrary to the dense cloud, the diffuse one is predominantly in atomic form (see Tables~\ref{tab:dense}, \ref{tab:diffuse}). It has a nearly double average gas temperature of $136\,{\rm K}$ and its carbon phase is found to be almost entirely in C{\sc ii} form. In both clouds, the increase in temperature is a result of the cooling reduction as well as the penetration of FUV photons to higher number densities, since the effective visual extinction is lower than that at higher $Z$.

\section{Velocity integrated emission maps}
\label{sec:emission}

\subsection{Varying the cosmic-ray ionization rate}
\label{ssec:emcr}

Figure~\ref{fig:col_CRemission} shows the emission maps of the [C{\sc ii}]~$158\mu$m, [C{\sc i}]~(1-0), (2-1), as well as [O{\sc i}]~$63\mu$m, $146\mu$m and [$^{13}$C{\sc ii}] lines as a function of $\zeta_{\rm CR}$, with an isotropic $G_0=10$ FUV field and $Z=1\,{\rm Z}_{\odot}$ at all times. As can be seen, for $\zeta_{\rm CR}\leq10^{-16}\,{\rm s}^{-1}$ the behaviour of the aforementioned lines in both clouds does not strongly vary with $\zeta_{\rm CR}$. The emission of [C{\sc ii}] is connected with the rarefied medium surrounding the denser one as a result of its interaction with the FUV field. In both clouds, the velocity integrated antenna temperature peaks at approximately $W({\rm C\textsc{ii}}){\sim}5-6\,{\rm K}\,{\rm km/s}$. On the other hand, both [C{\sc i}] fine structure lines are brighter than [C{\sc ii}]. Their emission is more extended for $\zeta_{\rm CR}=10^{-16}\,{\rm s}^{-1}$ than for $10^{-17}\,{\rm s}^{-1}$ due to the cosmic-ray induced destruction of CO which increases the abundance of C{\sc i}. This is demonstrated by the elevated brightness in the dense part of the diffuse cloud, which increases from $W({\rm C\textsc{i}\,1-0}){\sim}25$ to $40\,{\rm K}\,{\rm km/s}$ in the densest part at the centre, where the total column density is $N_{\rm tot}{\simeq}2\times10^{23}\,{\rm cm}^{-2}$. In both clouds and for $\zeta_{\rm CR}\leq10^{-16}\,{\rm s}^{-1}$, $W({\rm C\textsc{i}\,2-1})$ is quite low with values $\lesssim8\,{\rm K}\,{\rm km/s}$ for the dense cloud and $\lesssim14\,{\rm K}\,{\rm km/s}$ for the diffuse cloud. Note that the diffuse cloud is somewhat brighter in the [C{\sc i}]~(2-1) line for $\zeta_{\rm CR}=10^{-16}\,{\rm s}^{-1}$ than the corresponding one for the dense cloud, as a result of the optical thickness of the line; it is more optically thin in the diffuse cloud as the latter has overall lower $N_{\rm tot}$ values. Similarly to [C{\sc ii}], both [O{\sc i}] lines do not show any strong variation for $\zeta_{\rm CR}\leq10^{-16}\,{\rm s}^{-1}$. In the dense cloud, both [O{\sc i}] lines originate from the dense filamentary structure reaching values of as high as ${\sim}1\,{\rm K}\,{\rm km/s}$. In the diffuse cloud, the emission of the $63\mu$m line is ${\lesssim}0.5\,{\rm K}\,{\rm km/s}$ and it originates from the surrounding medium, whereas the emission of the $146\mu$m line is almost completely negligible. In summary, both [O{\sc i}] lines in the diffuse cloud are relatively faint.

For $\zeta_{\rm CR}{\geq}10^{-15}\,{\rm s}^{-1}$, there are interesting features observed in all aforementioned lines. In fact, the brightness of all of them increases revealing the inner and more dense parts of both clouds, which correspond to higher column densities. The most prominent feature is observed for the [C{\sc ii}] fine-structure line. For $\zeta_{\rm CR}=10^{-15}\,{\rm s}^{-1}$, its emission is more extended in both clouds, but its peak value remains at similar levels to that described earlier. However, a significant increase in [C{\sc ii}] brightness is seen once $\zeta_{\rm CR}=10^{-14}\,{\rm s}^{-1}$. There are two reasons that explain this elevation; the first one is connected with the increase in C{\sc ii} abundance due to the destruction of CO, and the second one is connected with the higher gas temperatures reached due to cosmic-ray heating. As described in \S\ref{sec:cds} and in Fig.~\ref{fig:col_CRcd}, the dense gas ($n_{\rm H}{\sim}10^5-10^6\,{\rm cm}^{-3}$) can be as warm as $40\,{\rm K}$ in the dense cloud and $50\,{\rm K}$ in the diffuse cloud. This increase excites [C{\sc ii}], given that the energy separation between the upper and lower states is $91.2\,{\rm K}$. For instance, the ratio of C{\sc ii} level populations in the excited state between a gas-temperature of $15\,{\rm K}$ and of $45\,{\rm K}$ (estimated for $\zeta_{\rm CR}=10^{-17}$ and $10^{-14}\,{\rm s}^{-1}$, respectively, where the gas remains very H$_2$-rich) is $e^{91.2/15}/e^{91.2/45}{\simeq}57$, resulting in the sudden brightness increase \citep[see also][]{Clar19}. At its peak, $W(\rm C\textsc{ii})\gtrsim60\,{\rm K}\,{\rm km/s}$ in the dense and ${\gtrsim}30\,{\rm K}\,{\rm km/s}$ in the diffuse cloud. Note also that the emission of [C{\sc ii}] is tightly connected with the distribution of the H$_2$ column density; this makes the $158\mu$m line a promising H$_2$ tracer in environments permeated with high cosmic-ray energy densities, particularly in the high-redshift Universe. Following this line, the emission of [$^{13}$C{\sc ii}] also increases with the cosmic-ray ionization rate, especially for $\zeta_{\rm CR}=10^{-14}\,{\rm s}^{-1}$. At this value of $\zeta_{\rm CR}$, the column density of C{\sc ii} is high enough to increase the optical depth in large parts of both clouds to values $\tau_{\rm CII}>0.6$ (see Fig.~\ref{fig:taucii}). In such cases, the emission of [$^{13}$C{\sc ii}], which also peaks where [C{\sc ii}]~$158\mu$m peaks, can be used to infer to the value of $\tau_{\rm CII}$ in observations \citep{Osse13,Guev20}.

In a similar way, the brightness temperatures of both transitions of [C{\sc i}] and [O{\sc i}] increase for higher $\zeta_{\rm CR}$. Since the energy separation of [C{\sc i}]~(1-0) is $23.6\,{\rm K}$, the emission of this line is already strong at $\zeta_{\rm CR}=10^{-15}\,{\rm s}^{-1}$, with peak values ${\sim}100\,{\rm K}\,{\rm km/s}$ for the dense and ${\sim}60\,{\rm K}\,{\rm km/s}$ for the diffuse clouds. At these levels of $\zeta_{\rm CR}$, [C{\sc i}]~(2-1) becomes bright with values ${\gtrsim}20\,{\rm K}\,{\rm km/s}$ in both clouds, while for $\zeta_{\rm CR}=10^{-14}\,{\rm s}^{-1}$ its emission may be much stronger. This is because the energy separation of [C{\sc i}]~(2-1) is $62.5\,{\rm K}$ and so it traces dense gas with temperatures ${\gtrsim}30\,{\rm K}$. On the other hand, the energy separations of the [O{\sc i}] lines are $227.7\,{\rm K}$ for the $63\mu$m and $326.6\,{\rm K}$ for the $146\mu$m. Therefore, it is difficult for the enhanced cosmic-ray heating to excite these lines and their emission is much fainter than the aforementioned ones. However, the elevated gas temperatures, in addition to the abundance increase of O{\sc i} for higher $\zeta_{\rm CR}$, increase the level populations in higher states causing an enhancement of [O{\sc i}] brightness temperatures. Interestingly, in certain high column density regions in both clouds, the emission of [O{\sc i}]~$146\mu$m may become stronger than that of [O{\sc i}]~$63\mu$m, as the second is much more optically thick than the first \citep{Gold19}.

\begin{figure*}
    \centering
    \includegraphics[width=\linewidth]{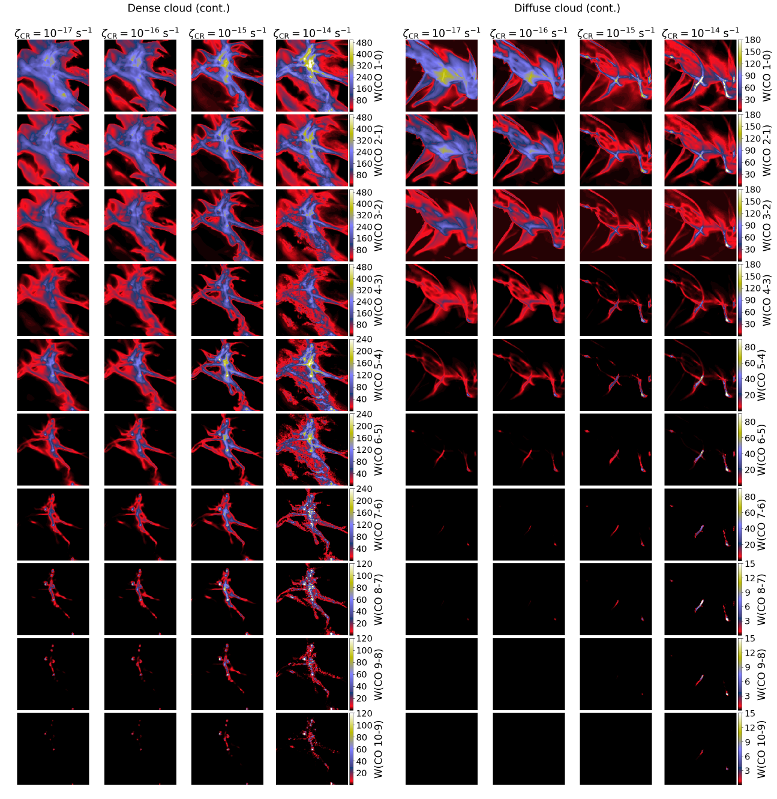}
	\caption{As in Fig.~\ref{fig:col_CRemission} but for CO and for all $J$-transitions examined here. The colour bar is in units of [K~km/s]. As $\zeta_{\rm CR}$ increases (from left-to-right), ${\rm W}({\rm CO}\,J=1-0)$ decreases in the more rarefied part but increases in the denser medium for higher $\zeta_{\rm CR}$. On the other hand, the abundance of CO is remarkably reduced due its cosmic-ray induced destruction (see Fig.~\ref{fig:col_CRcd} and \S\ref{sec:emission} for the relevant discussion). As the $J\rightarrow J-1$ transitions are progressively increased, regions with lower column densities become dark. For a constant $\zeta_{\rm CR}$, as the $J\rightarrow J-1$ transition increases, the emission becomes weaker and originates from higher column densities. Note the change in the colour bar range for mid- and higher transitions. At these higher transitions, both clouds and particularly the `diffuse cloud' become very faint.}
    \label{fig:col_CRemission2}
\end{figure*}

Figure~\ref{fig:col_CRemission2} shows the emission maps of all CO transitions, from $J=1-0$ to $J=10-9$. In general, low-$J$ CO lines are more extended, brighter and originate from intermediate column densities, whereas higher $J$-transitions are fainter and originate from the innermost parts of both clouds, where the column densities are high \citep[see also][]{Bisb14}. All CO transitions in each cloud show a very similar behaviour in the cases of $\zeta_{\rm CR}=10^{-17}$ and $10^{-16}\,{\rm s}^{-1}$; this is to be expected since the cosmic-ray ionization rate is not high enough to destroy CO or raise the gas temperature to a significant degree. Note, however, that low-$J$ CO lines are in general more extended for $\zeta_{\rm CR}=10^{-17}\,{\rm s}^{-1}$ than for $10^{-16}\,{\rm s}^{-1}$. For instance, there are low column-density regions where a decrease in $W(\rm CO\,1-0)$ can be seen, e.g., at the lower left of the dense cloud or the lower right of the diffuse one. In addition, the filamentary structures in the dense cloud remain bright for high-$J$ CO lines, whereas those of the diffuse one becomes very faint, especially for CO $J=7-6$ and above.

For $\zeta_{\rm CR}=10^{-15}\,{\rm s}^{-1}$, there are several interesting features seen in both clouds. First of all, for these cosmic-ray ionization rates and above, CO is destroyed very effectively as can be seen in the corresponding column-density panels of Fig.~\ref{fig:col_CRcd}. This is reflected in several parts of both clouds that have low total column densities and in general consist of gas with low number density. On the other hand, CO remains in regions with higher densities. This marks a special case in which only the densest parts of clouds are CO-rich and thus visible \citep{Bisb15}. This, in turn, can make large structures inundated by strong cosmic-ray energy densities to look more clumpy than they really are \citep[c.f.][]{Hodg12}. This effect is better demonstrated in the diffuse cloud for $\zeta_{\rm CR}=10^{-15}\,{\rm s}^{-1}$. Note that there, the emission maps of CO $J=1-0$ and also $J=2-1$ are different in terms of intensity distribution when compared to lower $\zeta_{\rm CR}$ but reminiscent to the distribution of H$_2$ column densities (see Fig.~\ref{fig:col_CRcd}). A remarkable consequence is also the increase in brightness of higher-$J$ CO lines that were previously very faint. This can be seen, for instance, in the CO $J=10-9$ emission of the dense cloud and the $J=7-6$ emission of the diffuse cloud.

\begin{figure*}
    \centering
    \includegraphics[width=\linewidth]{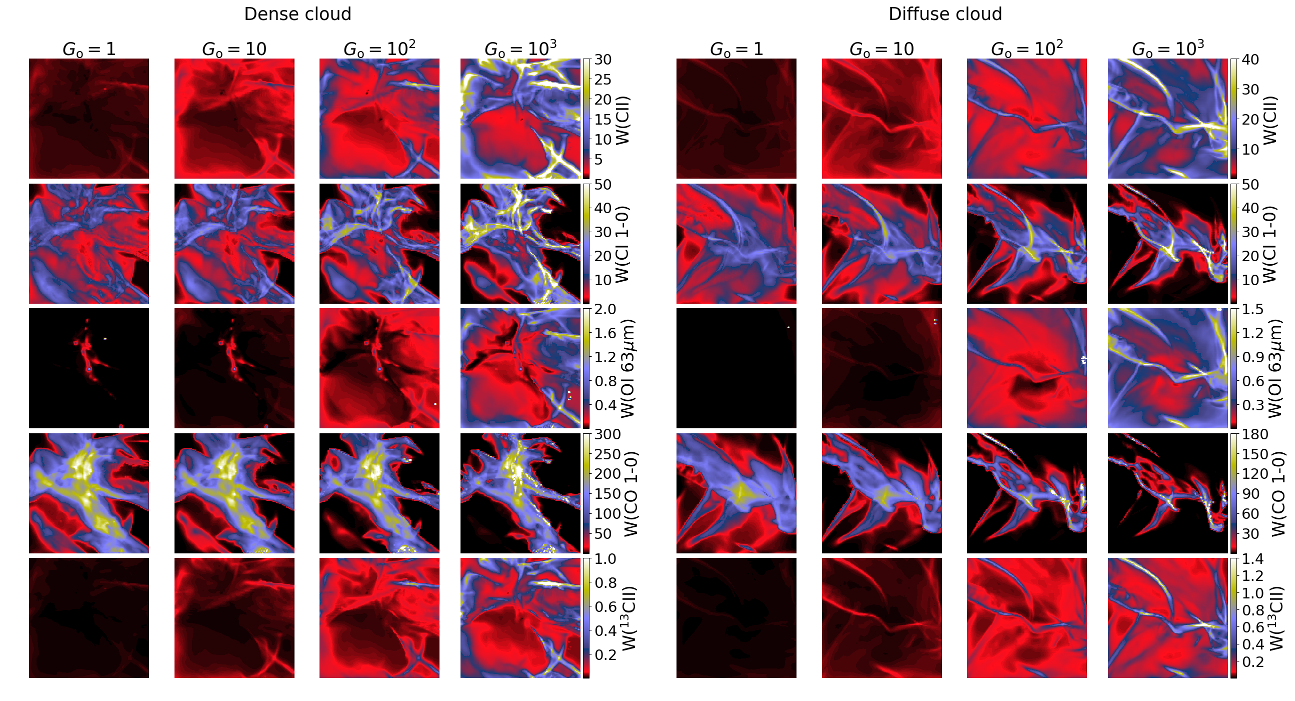}
	\caption{Emission maps of the most important emission lines with the intensity of FUV radiation ($G_0$) as the free parameter. The colour bar is in units of [K~km/s]. $G_0$ increases from left-to-right for each of the two clouds. From top-to-bottom the emission of [C{\sc ii}], [C{\sc i}]~(1-0), [O{\sc i}] at $63\mu$m, CO~(1-0) and [$^{13}$C{\sc ii}] are shown. High $G_0$ values increase overall the emission of [C{\sc ii}], [C{\sc i}]~(1-0), [O{\sc i}]~$63\mu$m and [$^{13}$C{\sc ii}]. Due to CO photodissociation, the emission of CO~(1-0) in the diffuse parts of both clouds decreases considerably. Figure~\ref{fig:Gapp} shows the rest of emission lines explored.}
    \label{fig:Ga_striking}
\end{figure*}

In the most extreme case of $\zeta_{\rm CR}=10^{-14}\,{\rm s}^{-1}$, cosmic-rays are expected to further destroy CO, leaving behind a very CO-poor H$_2$-rich ISM. However, as discussed in \citet{Bisb17a}, the increase in gas temperature initiates the formation of CO via the OH channel, resulting in a slight increase in CO abundance. The lower density gas remains very CO-poor, but H$_2$-rich. The emission of all CO transition lines increase remarkably. For instance, it is predicted that $W(\rm CO\,1-0)$ may obtain values ${\gtrsim}400\,{\rm K}\,{\rm km/s}$ and ${\gtrsim}150\,{\rm K}\,{\rm km/s}$ in structures similar to the dense and the diffuse clouds, respectively. The increase of the $J=1-0$ line emission is also supported by the decrease of the optical depth of this transition for high $\zeta_{\rm CR}$ (see Fig.~\ref{fig:tauco}). In both density distributions, higher-$J$ CO lines are also increased and parts of the diffuse cloud may now be detectable even to the $J=10-9$ transition. This is because the column density of CO does not vary significantly when compared to $\zeta_{\rm CR}=10^{-15}\,{\rm s}^{-1}$. However, the increase in gas temperature leads to lower optical depths overall, resulting in the increase in brightness temperatures. 

\subsection{Varying the FUV radiation field intensity}
\label{ssec:emfuv}

Figure~\ref{fig:Ga_striking} shows the emission maps of the most important cooling lines ([C{\sc ii}]~$158\mu$m, [C{\sc i}]~(1-0), [O{\sc i}]~$63\mu$m and CO $J=1-0$) with the FUV intensity of the radiation field as the free ISM parameter. The rest of emission maps of cooling lines are illustrated in Fig.~\ref{fig:Gapp}. In what follows, we define the average intensity as
\begin{eqnarray}
\label{eqn:avintensity}
\overline{W}=\frac{\int W{\rm d}S}{\int {\rm d}S},
\end{eqnarray}
where $W$ is the emission of the given line and $S$ the area of the whole map. As $G_0$ increases, the average emission of [C{\sc ii}] increases more than an order of magnitude. More specifically, it increases from $\overline{W}_{\rm CII}{\sim}0.45$ to ${\sim}10.8\,{\rm K}\,{\rm km/s}$ in the dense cloud and from ${\sim}0.51$ to ${\sim}12.0\,{\rm K}\,{\rm km/s}$ in the diffuse cloud. The emission of [$^{13}$C{\sc ii}] exceeds the value of $0.1\,{\rm K}\,{\rm km/s}$ (below which its detection becomes challenging) for $G_0\gtrsim10^2$ and only in small parts of both clouds. As with the case of cosmic-rays as the free ISM parameter, this increase in emission is a result of i) the photodissociation of CO which increases the abundance of C{\sc ii} and ii) the increase in gas temperature due to photoelectric heating. However, this effect is seen only in the outermost parts of both clouds --where the FUV intensity impinges from-- and not in the entire volume of each cloud as was the previous case with cosmic-rays. Considering the panels corresponding to $\langle G_0\rangle$ in Fig.~\ref{fig:Ga_striking}, it can be seen that $W({\rm CII})$ peaks in the ISM gas surrounding the higher density medium, particularly in the dense cloud.

The emission of [C{\sc i}]~(1-0) is not strongly affected by the increase of $G_0$. Comparing the emission maps of this line between the $G_0=1$ and $10^3$ cases, it is observed that $W({\rm CI\,1-0})$ substantially decreases in the regions with low column densities (the effect is better seen in the diffuse cloud) whereas it increases in the densemost parts. Overall, $\overline{W}_{\rm CI\,1-0}$ increases in the dense cloud (since it has high column densities along the filamentary structure, see also \S\ref{ssec:varyFUVcd}) from ${\sim}8.8$ to ${\sim}12.47\,{\rm K}\,{\rm km/s}$ but slightly decreases in the diffuse cloud from ${\sim}8.5$ to ${\sim}5.70\,{\rm K}\,{\rm km/s}$.

As explained in \S\ref{ssec:emcr}, the [O{\sc i}]~$63\mu$m cooling line becomes considerably bright when the gas has a temperature approaching the energy separation of this line. From the corresponding panels of Fig.~\ref{fig:col_Gcd} for $\langle T_{\rm gas}\rangle$, it can be seen these higher temperatures are achieved in the outer parts of the filamentary structures in both clouds; however, in these regions $N({\rm OI})$ is in general low. The overall result is an increase of the [O{\sc i}]~$63\mu$m emission, although it still remains quite weak. Notably, the central region of the dense cloud always has $N({\rm OI})\gtrsim5\times10^{19}\,{\rm cm}^{-2}$, which results in an emission of $W({\rm OI\,63\mu{\rm m}})\gtrsim0.1\,{\rm K}\,{\rm km/s}$ even for the lowest explored $G_0=1$.

\begin{figure*}
    \centering
    \includegraphics[width=\linewidth]{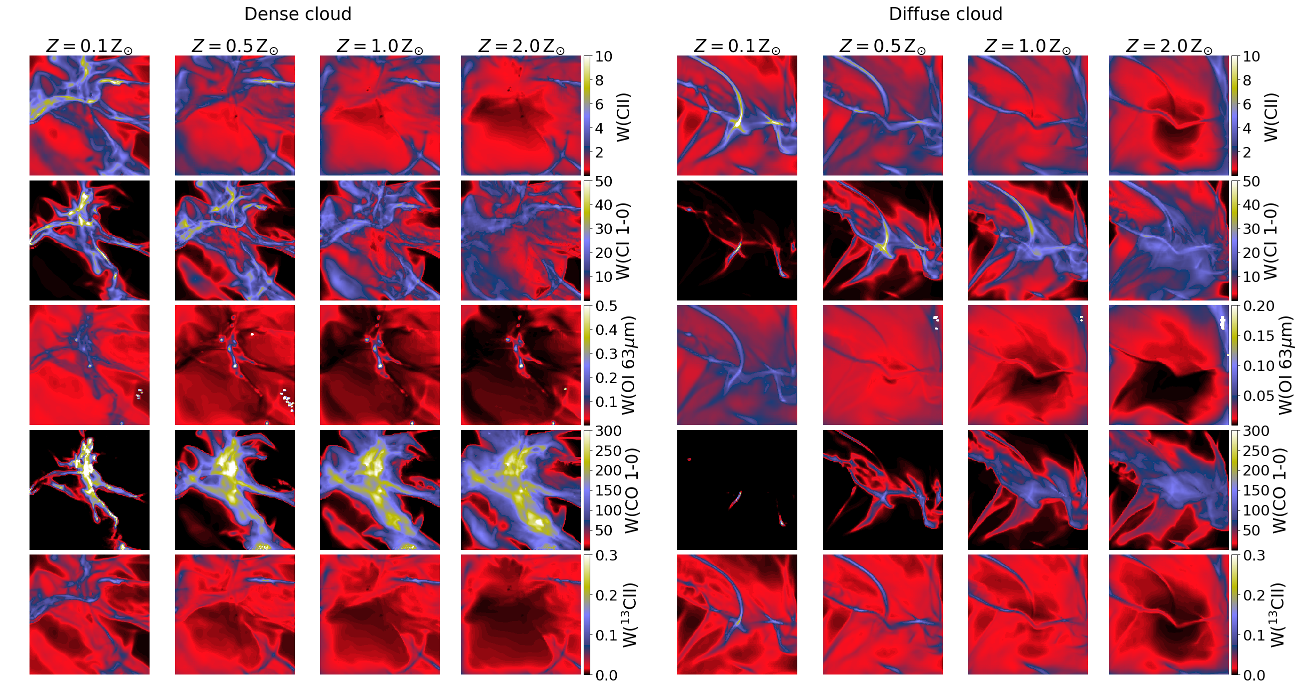}
	\caption{As in Fig.~\ref{fig:Ga_striking} but for the metallicity as the free parameter. The colour bar is in units of [K~km/s]. For low metallicities, the lines of [C{\sc ii}], [C{\sc i}]~(1-0) and [$^{13}$C{\sc ii}] are more strongly associated with the H$_2$ column density than in higher metallicities. Figure~\ref{fig:Zapp} shows the rest of emission lines explored.}
    \label{fig:Za_striking}
\end{figure*}

The intensity increase of the FUV radiation and the corresponding increase of photoelectric heating result in the destruction of CO through photodissociation. As described with C{\sc i}, the emission of CO decreases where the effective visual extinction and thus the observed column density is low. It is found that $\overline{W}_{\rm CO~1-0}$ decreases from ${\sim}120$ to ${\sim}90\,{\rm K}\,{\rm km/s}$ in the dense cloud and from ${\sim}66$ to ${\sim}41\,{\rm K}\,{\rm km/s}$ in the diffuse cloud. As an example, the small region located in the bottom left of the diffuse cloud is bright in CO $J=1-0$ for $G_0=1$, but progressively vanishes as $G_0$ increases. Similarly, the gas surrounding the dense structure in the diffuse cloud is also progressively weakening in the $W({\rm CO~1-0})$ emission line. Furthermore, the photodissociation of CO in these lower density regions is found to be responsible for the relative increase of the CO $J=1-0$ emission in the highest density parts of both clouds (the effect is more prominent in the diffuse cloud for $G_0=10^3$ when compared to $G_0=1$). This is better understood if we consider that these low density regions are where the CO $J=1-0$ optical depth starts to increase (see also \S\ref{ssec:emcr}) which results in ``hiding'' the denser structures. When the FUV photons destroy this outer layer, they shift this starting point to higher densities, thus revealing the more clumpy ISM. This effect is, however, very localized.

\subsection{Varying the metallicity}
\label{ssec:emz}

Figure~\ref{fig:Za_striking} shows how the most important cooling lines vary with metallicity as the free ISM parameter. As described in \S\ref{ssec:Zvary}, lower metallicities have high impact on the carbon phases in relation to the H{\sc i}-to-H$_2$ transition due to the low dust-to-gas ratio, thereby allowing the FUV photons to penetrate deeper inside the cloud increasing the overall gas temperature. This makes the emission of [C{\sc ii}] to increase from $\overline{W}_{\rm CII}{\sim}1\,{\rm K}\,{\rm km/s}$ at $Z=2\,{\rm Z}_{\odot}$ to ${\sim}2.1\,{\rm K}\,{\rm km/s}$ at $Z=0.1\,{\rm Z}_{\odot}$ for the dense cloud and from ${\sim}1.2$ to ${\sim}1.7\,{\rm K}\,{\rm km/s}$ for the diffuse cloud. Consequently, the emission of [$^{13}$C{\sc ii}] also increases, though remaining at levels that are challenging for easy detection. This is counter-intuitive considering that the C{\sc ii} abundance is almost five to seven times lower at $Z=0.1\,{\rm Z}_{\odot}$ than at $Z=2\,{\rm Z}_{\odot}$. However, the higher gas temperatures obtained at low metallicities increase $W({\rm CII})$ considerably. Furthermore, the spatial distribution of $W({\rm CII})$ as seen in the $Z=0.1\,{\rm Z}_{\odot}$ panels in both clouds (and especially in the diffuse cloud) is reminiscent to the $N({\rm H}_2)$ as seen in the corresponding panels in Fig.~\ref{fig:col_Zcd}. This suggests that [C{\sc ii}] may be an excellent alternative tracer of H$_2$-rich gas in low metallicity systems \citep[e.g.][]{Madd20}, since the CO $J=1-0$ emission is relatively weak, as described below. 

The emission line of [C{\sc i}]~(1-0) is more extended at $Z=2\,{\rm Z}_{\odot}$ and its average emission is $\overline{W}_{\rm CI\,1-0}{\sim}9\,{\rm K}\,{\rm km/s}$ for the dense cloud and ${\sim}10\,{\rm K}\,{\rm km/s}$ for the diffuse cloud. As we decrease in metallicity, the average emission of this line also decreases, i.e., at $Z=0.1\,{\rm Z}_{\odot}$ these values become ${\sim}5$ and ${\sim}0.5\,{\rm K}\,{\rm km/s}$, respectively. However, since C{\sc i} is associated with higher number densities in metal poor environments and given that the gas temperature increases, $W({\rm CI\,1-0})$ peaks at higher values than it does for metal rich environments. This effect is more prominent in the dense cloud, since in the diffuse one the carbon abundance is almost entirely in C{\sc ii} form. 

The emission line of the optically thick [O{\sc i}]~$63\mu$m is found to be brighter and more extended in the metal-poor runs. The increase of this line at low-Z is due to the increase of the gas temperature as explained above, thus revealing better the distribution of gas in both clouds. The average values of $W({\rm OI}\,63\mu{\rm m})$ are ${\sim}4.7\times10^{-2}\,{\rm K}\,{\rm km/s}$ for the dense cloud and ${\sim}3.6\times10^{-2}\,{\rm K}\,{\rm km/s}$ for the diffuse cloud, but still remain weak for such conditions. Additional heating that boosts the gas temperature to higher values than those estimated by the current ISM conditions would increase the emission of [O{\sc i}]. For instance, the emission of this line obtains its highest values for $\zeta_{\rm CR}=10^{-14}\,{\rm s}^{-1}$ (presented in \S\ref{ssec:emcr}) since it is in those runs that $T_{\rm gas}$ has been everywhere substantially increased. Note also that although high metallicities increase the abundance of O{\sc i} as discussed in \S\ref{ssec:Zvary}, its emission is weaker than the one at low-Z due to the decrease of gas temperature.

\begin{figure*}
    \centering
    \includegraphics[width=\linewidth]{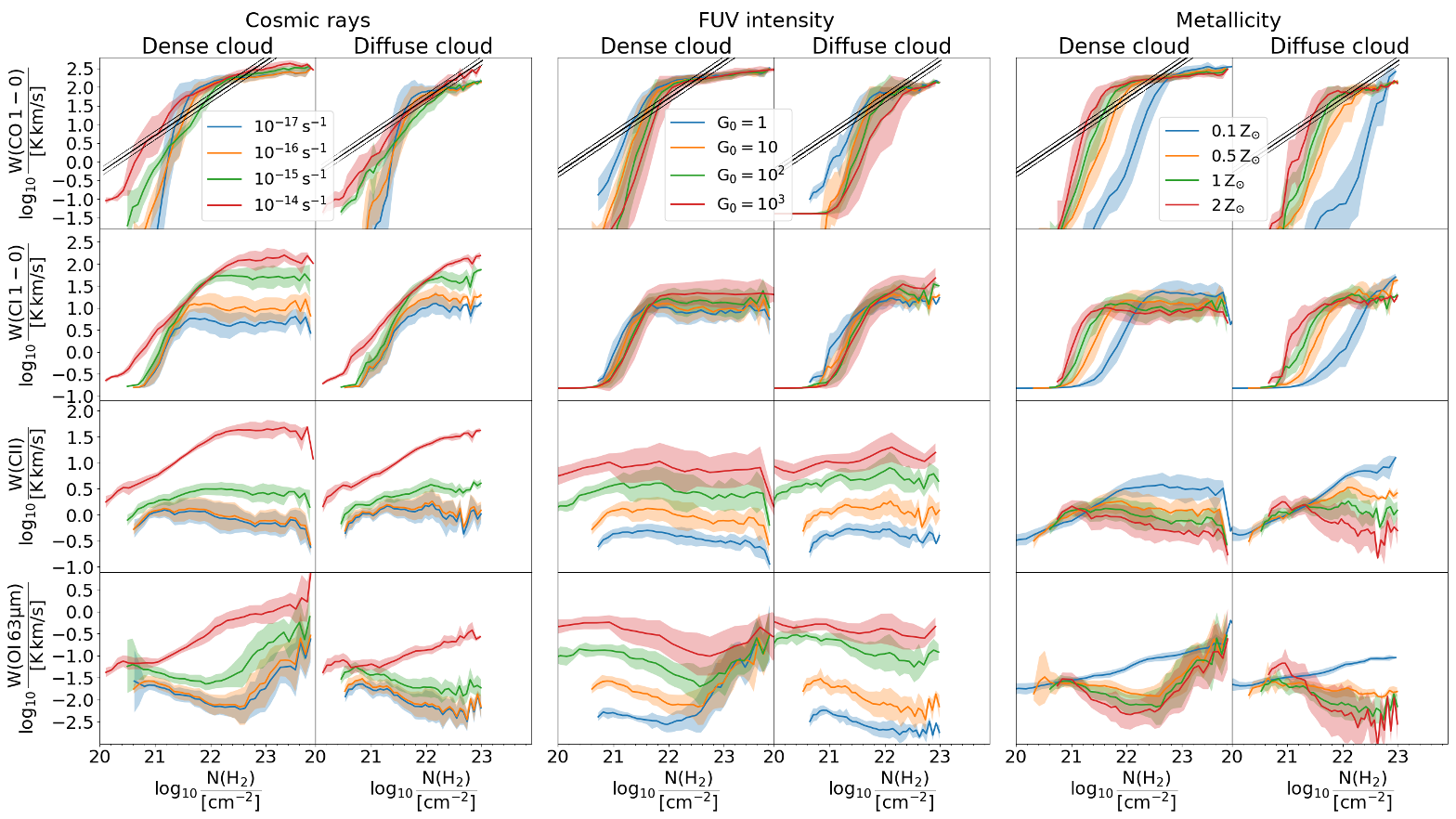}
	\caption{Relation of the velocity-integrated emission of (top-to-bottom) CO $J=1-0$, [C{\sc i}]~(1-0), [C{\sc ii}] and [O{\sc i}]~$63\,\mu$m versus $\rm N(H_2)$. There are three pairs of column corresponding to the three free parameters (left-to-right); the cosmic-ray ionization rate, the FUV intensity and the metallicity. Each solid line corresponds to the mean value of the velocity integrated emission for a given N(H$_2$). The shaded areas correspond to $1\sigma$ standard deviation. The left column of each pair corresponds to the `dense cloud' and the right one to the `diffuse cloud'. The thick solid diagonal lines in the CO $J=1-0$ panels show the correlation of $N(\rm H_2)$ and $W_{\rm CO,1-0}$ for a conversion factor $X_{\rm CO}=2\times10^{20}\,{\rm cm}^{-2}\,({\rm K}\,\,{\rm km}/{\rm s})^{-1}$ as well as the recommended $\pm30\%$ uncertainty \citep{Bola13}. See \S\ref{ssec:growth} for discussion.}
    \label{fig:satur}
\end{figure*}

As can be seen in the fourth row of Fig.~\ref{fig:Za_striking}, the emission of CO $J=1-0$ follows the trend of [C{\sc i}]~(1-0) in the outer gas and it is more extended in the metal-rich run. However, in the high-density regions, $W({\rm CO\,1-0})$ is much brighter than $W({\rm CI\,1-0})$. It is found that $\overline{W}_{\rm CO\,1-0}{\sim}127\,{\rm K}\,{\rm km/s}$ in the dense cloud and ${\sim}73\,{\rm K}\,{\rm km/s}$ in the diffuse cloud. The photodissociation of CO at low metallicities results in a negligible emission of that line in the ISM surrounding the dense material in both clouds, with the diffuse cloud being almost entirely CO-dark. Only the densest parts in the dense cloud remain CO-bright and that emission peaks at higher values than those observed in the metal-rich run. 

Figure~\ref{fig:Zapp} shows the rest of emission lines calculated for $Z$ as the free ISM environmental parameter.

\subsection{Relation of line intensities with $\rm H_2$ column density}
\label{ssec:growth}

Figure~\ref{fig:satur} (from top-to-bottom) shows the relation between the mean value of the velocity integrated emission of CO~(1-0), [C{\sc i}]~(1-0), [C{\sc ii}], and [O{\sc i}]~$63\mu$m for the dense and diffuse clouds versus the H$_2$ column density. The shaded areas correspond to $1\sigma$ standard deviation. There are three pairs of such relations (from left-to-right), each corresponding to a free ISM environmental parameter. Overall, the shape of CO~(1-0) does not strongly vary with $\zeta_{\rm CR}$ and $G_{0}$, but it builds at higher $N({\rm H}_2)$ for lower $Z$. Furthermore, for our fiducial ISM environmental parameters, this shape is in broad agreement with the findings of \citet{Seif20} who studied magnetised molecular clouds from the SILCC-Zoom project. For $N(\rm H_2)\gtrsim2\times10^{22}\,{\rm cm}^{-2}$, $W(\rm CO~1-0)$ remains remarkably unaffected for all ISM conditions explored, except for $Z=0.1\,{\rm Z}_{\odot}$, in both density distributions considered. In the CO~(1-0) panels, we additionally plot the correlation between $N(\rm H_2)$ and $W(\rm CO~1-0)$ for the \citet{Bola13} recommended value of $X_{\rm CO}=2\times10^{20}\,{\rm cm}^{-2}\,{\rm K}^{-1}\,{\rm km}^{-1}\,{\rm s}$ with the uncertainty of $\pm30\%$. As can be seen, both clouds have a $W(\rm CO~1-0)$-$N(\rm H_2)$ relation that varies substantially with respect to that of a constant $X_{\rm CO}$ for $N(\rm H_2)\lesssim3\times10^{21}\,{\rm cm}^{-2}$, for $N(\rm H_2)\gtrsim10^{23}\,{\rm cm}^{-2}$ and for different metallicities. In addition, for $N(\rm H_2)\sim3\times10^{21}\,{\rm cm}^{-2}$ at solar metallicity, the CO~(1-0) intensity saturates since it becomes optically thick; an effect that is seen more clearly in the dense cloud. This saturation has been also previously noticed in theoretical \citep[e.g.][]{Smit14,Gong18} and observational \citep[e.g.][]{Pine08} studies and it shifts to higher (lower) columns of H$_2$ gas as metallicity decreases (increases).

For $\zeta_{\rm CR}=10^{-17}\,{\rm s}^{-1}$, the emission of [C{\sc i}]~(1-0) increases for both clouds until $N(\rm H_2)\sim3-4\times10^{21}\,{\rm cm}^{-2}$, after which it saturates, as seen in the dense cloud \citep[see also][]{Glov15}. For $\zeta_{\rm CR}=10^{-16}\,{\rm s}^{-1}$, the emission of [C{\sc i}]~(1-0) follows the same correlation with $N(\rm H_2)$, although it saturates at an approximately 2-3 times higher intensity. Once the cosmic-ray ionization rate increases to $\zeta_{\rm CR}=10^{-15}\,{\rm s}^{-1}$, the intensity of the line increases and therefore saturates at slightly higher H$_2$ column densities, more specifically at ${\sim}10^{22}\,{\rm cm}^{-2}$. For the highest $\zeta_{\rm CR}$ of $10^{-14}\,{\rm s}^{-1}$, [C{\sc i}]~(1-0) becomes more optically thin (see Fig.\ref{fig:tauci}) and saturates at $N(\rm H_2)\sim3\times10^{22}\,{\rm cm}^{-2}$, which is approximately the highest column density found in the diffuse cloud. The fact that [C{\sc i}]~(1-0) becomes remarkably bright with increasing $\zeta_{\rm CR}$, in addition to being more optically thin, makes it a good tracer for H$_2$-rich clouds in cosmic-ray dominated regions \citep{Papa04,Bisb15} and especially for the diffuse ISM. In both clouds, the $W({\rm CI})-N({\rm H_2})$ relation does not substantially change for the $G_0$ range explored here. However, the observed trend is that $W({\rm CI})$ peaks at higher values for higher $G_0$, although it saturates at the same $N({\rm H}_2)$. On the other hand, since high metallicities increase the visual extinction, $W({\rm CI})$ becomes optically thick for lower $N({\rm H}_2)$ when compared to the metal-poor runs. The peak of [C{\sc i}]~(1-0) emission increases as $Z$ lowers, revealing the more clumpy and denser structures in response to the reduced C{\sc i} abundances in the lower density medium and the gas temperature increase (see \S\ref{ssec:Zvary} and \ref{ssec:emz}). It is further found that the gas is [C{\sc i}]-bright for $N({\rm H}_2)\gtrsim2\times10^{22}\,{\rm cm}^{-2}$ as evidenced from both clouds.

The ionic line of [C{\sc ii}]~$158\mu$m does not correlate with $N({\rm H}_2)$ except for the ISM parameters of $\zeta_{\rm CR}=10^{-14}\,{\rm s}^{-1}$ and $Z=0.1\,{\rm Z}_{\odot}$. This feature is observed in both clouds and it is a result of the origin of [C{\sc ii}] from low optical depths as a main PDR coolant \citep{Holl99}. In general, the $158\mu$m line increases its emission as $\zeta_{\rm CR}$ increases (valid for $\zeta_{\rm CR}\ge10^{-16}\,{\rm s}^{-1}$). The same is observed also either as $G_0$ increases or as $Z$ decreases. However, the $W({\rm CII})-N(\rm H_2)$ relation takes a completely different turn for the extreme cases of high cosmic-ray ionization rates and of low metallicities. In particular, [C{\sc ii}] becomes very bright for $\zeta_{\rm CR}=10^{-14}\,{\rm s}^{-1}$ and it shows a remarkably good correlation with the H$_2$ column density up to ${\sim}3\times10^{22}\,{\rm s}^{-1}$. This means that in cosmic-ray dominated regions [C{\sc ii}]~$158\mu$m may be considered as an alternative H$_2$ tracer, which is particularly interesting for studies of environments such as the Galactic Centre or extreme extragalactic objects \citep[see also][]{Lang14,Accu17a,Accu17b}. The same effect is observed for $Z=0.1\,{\rm Z}_{\odot}$ and particularly for the diffuse cloud, leading to suggestions of using this line to trace CO-dark H$_2$-rich gas in low metallicity systems, such as dwarf galaxies \citep{Madd20}.

\begin{figure*}
    \centering
    \includegraphics[width=\linewidth]{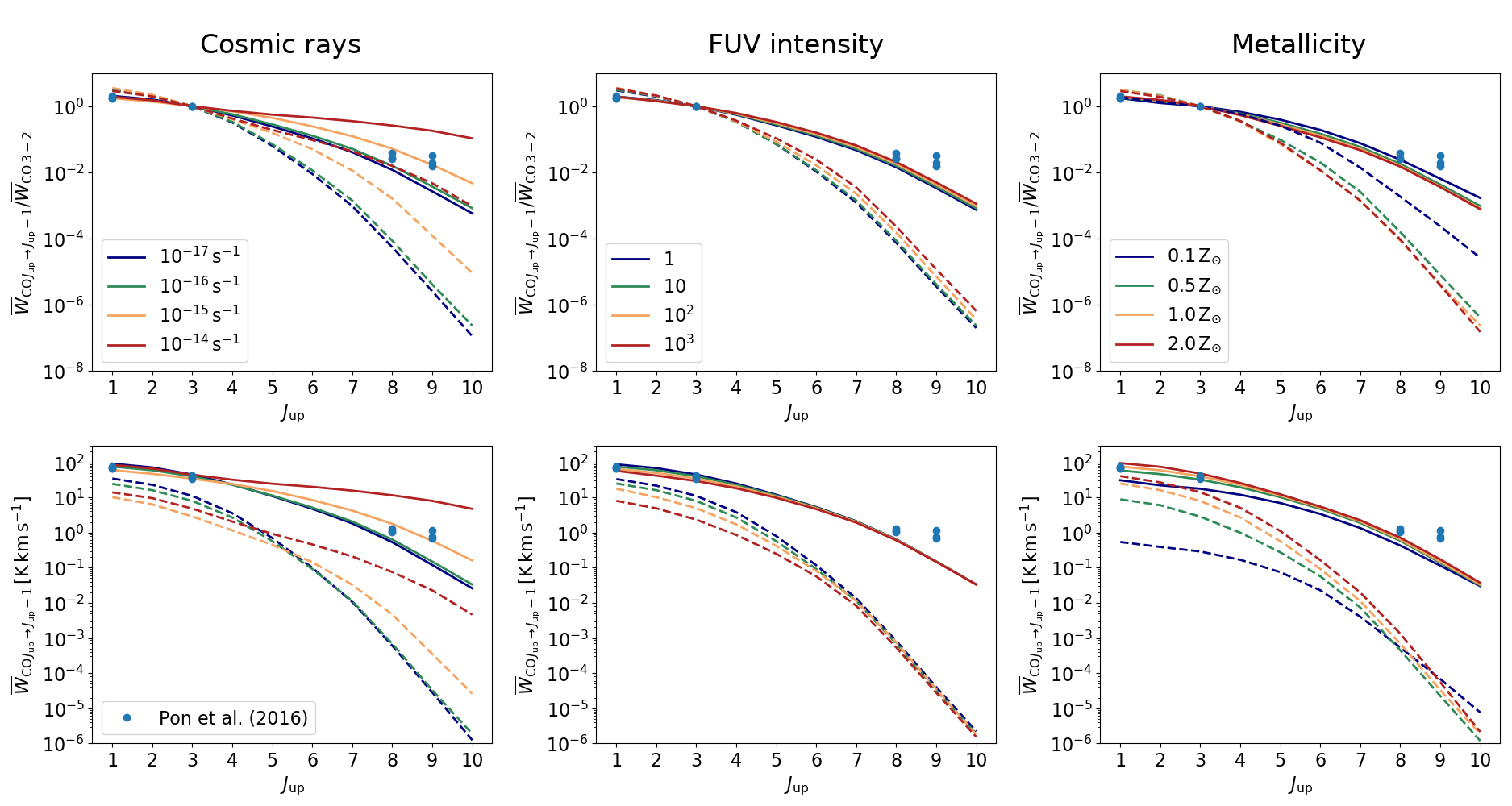}
    \caption{CO Spectral Line Energy Distributions (SLED) for the dense (solid) and diffuse (dashed) clouds for all ISM environmental parameters explored (from left-to-right, the cosmic-ray ionization rate, the FUV intensity and the metallicity as free parameters, respectively). The top panel shows the average emission of CO transitions normalized to the CO $J=3-2$ transition and the bottom panel actual, non-normalized average emission. High-$J$ CO lines are highly affected by the increase of the cosmic-ray ionization rate as the latter increase the gas temperature in high-density gas. Higher FUV intensities are decreasing the emission of low-$J$ CO transitions as a result of the CO photodissociation. In a similar way, but more significantly, lower metallicities decrease considerably the emission of the low-$J$ and mid-$J$ transitions. The blue points are observations of three different infrared dark clouds presented in \citet{Pon16}, for comparison.We find that these observations are best represented with the column density distribution of the `dense cloud' with $\zeta_{\rm CR}\simeq10^{-15}\,{\rm s}^{-1}$. See \S\ref{ssec:sleds} for further discussion.}
    \label{fig:sleds}
\end{figure*}

Finally, the fine-structure line of [O{\sc i}]~$63\mu$m, illustrated in the bottom row of Fig.~\ref{fig:satur}, correlates with $N(\rm H_2)$ in a more complicated way than [C{\sc ii}] does. In particular, as described above with [C{\sc ii}], [O{\sc i}] does not correlate with $\zeta_{\rm CR}$ for values $\le10^{-16}\,{\rm s}^{-1}$ and shows a stronger dependency with the FUV intensity. No strong correlation is observed for metallicities $Z\ge0.5\,{\rm Z}_{\odot}$. Higher FUV intensities increase the emission of the $63\mu$m line for H$_2$ columns $\lesssim2\times10^{22}\,{\rm cm}^{-2}$. In particular, for $N(\rm H_2)\lesssim2\times10^{22}\,{\rm cm}^{-2}$, [O{\sc i}] weakens with increasing H$_2$ column density as a result of the nature of the line being very optically thick and suffering from self-absorption \citep{Gold19}.  However, once $N({\rm H}_2){\gtrsim}2\times10^{22}\,{\rm cm}^{-2}$ (dense cloud), its emission becomes stronger with increasing $N(\rm H_2)$. The aforementioned behavior is also followed for all ISM environmental parameters explored, except for $\zeta_{\rm CR}=10^{-14}\,{\rm s}^{-1}$ and for $Z=0.1\,{\rm Z}_{\odot}$. As with [C{\sc ii}], once the cosmic-ray ionization rate increases to $\zeta_{\rm CR}=10^{-14}\,{\rm s}^{-1}$, [O{\sc i}]~$63\mu$m becomes much brighter and increases with increasing H$_2$ column density. This is a result of the abundance increase due to the destruction of CO by cosmic-rays as well as the gas-temperature increase due to cosmic-ray heating, which in turn affects the optical depth of this line. Similarly, for $Z=0.1\,{\rm Z}_{\odot}$ both clouds have higher gas temperatures, thereby increasing the $\log_{10} W({\rm OI}\,63\mu{\rm m})$ linearly with $\log_{10} N({\rm H}_2)$. For any of the above cases, however, the emission remains very faint.

\section{Spectral Line Energy Distributions}
\label{ssec:sleds}

Spectral Line Energy Distributions (SLEDs) of the rotational transitions of the CO molecule can become important diagnostics for the ISM properties as they can provide insights about its thermodynamics, molecular mass content and dynamical state \citep{Nara14}. Various observations focus on the construction of CO SLEDs \citep[e.g.,][]{Papa10a,Papa12b,Rose15,Mash15,Jobl18,Vall18,Vale20} from local clouds to extragalactic observations and to transitions up to $J=20-19$ or beyond. Elevated SLEDs of high-$J$ transitions are thought to be a result of various mechanisms including shocks, cosmic-rays and X-rays. In this Section, we investigate how our simulated SLEDs respond to the cosmic-ray ionization rate, the FUV intensity of the radiation field and the metallicity.

Figure~\ref{fig:sleds} shows the CO SLEDs for the dense (solid lines) and the diffuse (dashed lines) clouds against each free ISM environmental parameter. The average intensities were calculated using Eqn.~\ref{eqn:avintensity}. The upper panels of Fig.~\ref{fig:sleds} show the normalized average intensities (to the $J=3-2$ transition), while the lower panel shows the non-normalized SLEDs. In these results we do not consider any observational cut-off limit.

Cosmic-rays affect the CO SLEDs in both low-$J$ and high-$J$ transitions. For $\zeta_{\rm CR}\leq10^{-16}\,{\rm s}^{-1}$ no appreciable differences are observed. Higher $\zeta_{\rm CR}$ lowers the $\overline{W}_{\rm CO}$ emission for $J_{\rm up}\lesssim4$ in the dense cloud and $J_{\rm up}\lesssim5$. For the $J=1-0$ transition, we find that the emission decreases approximately four times in the diffuse cloud case and approximately two times in the dense cloud. This is due to the cosmic-ray induced CO destruction in the lower density medium which decreases the overall emission of CO at these low-$J$ transitions. Note, however, that the $\overline{W}_{\rm CO}$ for $\zeta_{\rm CR}=10^{-14}\,{\rm s}^{-1}$ is always higher than for $10^{-15}\,{\rm s}^{-1}$ (see \S\ref{ssec:cdcr} and \S\ref{ssec:emcr}). In both clouds, cosmic-rays excite the high-$J$ CO lines resulting in an increase of the SLEDs. We find that in this regime of $J_{\rm up}$ transitions, increases with $\zeta_{\rm CR}$. In the dense cloud and for $\zeta_{\rm CR}=10^{-14}\,{\rm s}^{-1}$, the intensity of the $J_{\rm up}=10$ transition is approximately two orders of magnitude greater when compared to $\zeta_{\rm CR}=10^{-17}\,{\rm s}^{-1}$, since the gas temperature of the high-density gas increases to values $T_{\rm gas}\sim50\,{\rm K}$ (see Figs.~\ref{fig:col_CRcd} and \ref{fig:col_CRemission2}) thus exciting high-$J$ transitions. This difference can be even higher for lower column densities. For instance, it is approximately four orders of magnitude in the diffuse cloud.

The range of FUV radiation field intensities explored does not drastically affect the high-$J$ CO SLEDs as can be seen in the middle column of Fig.~\ref{fig:sleds}. Only the low-$J$ CO lines are affected. In particular, the $\overline{W}_{\rm CO}$ of the $J=1-0$ transition decreases approximately $1.5$ times in the dense cloud and approximately $5$ times in the diffuse cloud. This decrease becomes more effective for stronger radiation field intensities, i.e. $G_0\gtrsim100$.

Since low-metallicity gas contains reduced abundances of C and O, the abundance of CO and the corresponding emission will be reduced. As can be seen from the third column of Fig.~\ref{fig:sleds}, the SLEDs of both clouds are progressively reduced as $Z$ lowers. As with the other two free ISM parameters, the effect is more prominent in the diffuse cloud. For the diffuse cloud, the $\overline{W}_{\rm CO}$ emission of the $J=1-0$ transition is reduced by approximately 70 times when comparing the $2\,{\rm Z}_{\odot}$ with the $0.1\,{\rm Z}_{\odot}$ metallicities. For the dense cloud, that emission is reduced by approximately three times. 

\begin{figure}
    \centering
    \includegraphics[width=0.95\linewidth]{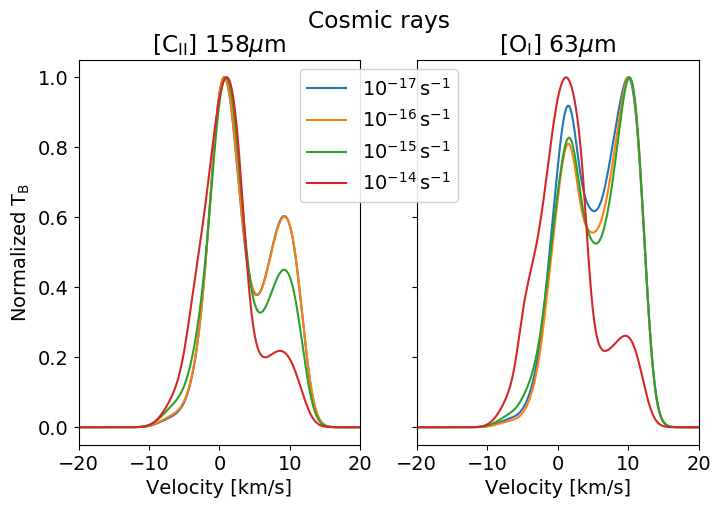}
    \includegraphics[width=0.95\linewidth]{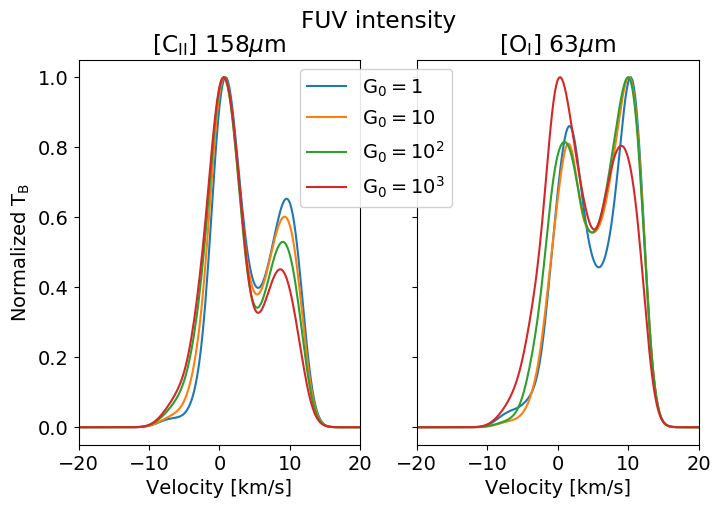} 
    \includegraphics[width=0.95\linewidth]{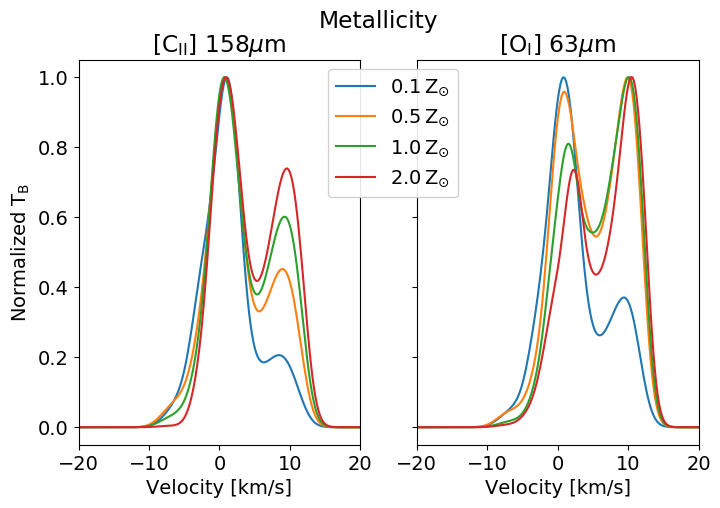} 
	\caption{Normalized brightness temperature ($\rm T_{\rm B}$)-velocity diagrams for the `dense cloud'. From top-to-bottom, the free ISM parameters are the cosmic-ray ionization rate, the FUV intensity and the metallicity. The left column corresponds to the spectra for [C{\sc ii}] and the right one for [O{\sc i}]~$63\mu$m fine-structure lines. Given that this cloud is a sub-region of the \citet{Wu17} cloud-cloud collision MHD simulation, we explore how the `bridge-effect' is affected with each free parameter. We find that this colliding signature holds for all environments except for the cases with $\zeta_{\rm CR}=10^{-14}\,{\rm s}^{-1}$ and $Z=2\,{\rm Z}_{\odot}$.}
    \label{fig:signatures}
\end{figure}

\citet{Pon16} reported observations of high-$J$ CO lines for three different infrared dark clouds (IRDCs), namely the IRDC C (G028.37+00.07), F (G034.43+00.24) and G (G034.77-0.55). Their observations in the CO lines with $J_{\rm up}=1,3,8$ and $9$ are presented in Fig.~\ref{fig:sleds} with blue thick dots. Although it is not our primary intention to model the above IRDCs in detail, it is interesting to see how our results compare to the \citet{Pon16} observations and fitting. These observations show an elevated emission of the high-$J$ CO lines. We find that they can be reproduced with the column density distribution of the dense cloud and for ISM conditions corresponding to $\zeta_{\rm CR}{\sim}10^{-15}\,{\rm s}^{-1}$, $G_0=10$ and $Z=1\,{\rm Z}_{\odot}$. In their paper, \citet{Pon16} perform PDR simulations including shock heating and they find that the elevated high-$J$ CO intensities may be explained with the presence of a mechanism heating high column densities. Since the intensity of the FUV photons extinguishes fast, they propose that shock heating may be responsible for the observed high-$J$ CO intensities. This is in agreement with our simulations in the sense that a heating mechanism, other than photoelectric that can penetrate the cloud, such as cosmic-rays, may be needed to excite these high transitions. This is additionally supported with the low dust temperatures of ${\sim}15\,{\rm K}$ observed in such IRDCs \citep{Pere10,Zhu14,Pon16,Lim16}; neither shocks nor high cosmic-ray ionization rates can significantly increase dust temperatures.

\section{Diagnostics of Collisions}
\label{ssec:spectra}

In a previous work, \citet{Bisb17b} examined how molecular, atomic and ionic line emissions can be used as diagnostics for collisions between giant molecular clouds (GMC). They found that fine-structure lines, such as [C{\sc ii}]~$158\mu$m and [O{\sc i}]~$63\mu$m, can be used as alternatives to the commonly used low$-J$ CO lines \citep[e.g.][]{Hawo15a,Hawo15b}. In particular, since these fine-structure lines are emitted from the rarefied ISM surrounding the denser and more compact gas, they can be used to identify more evolved collisions in which the dense parts of the clouds have already merged and the velocity information obtained from CO lines from these colliding parts no longer distinguished the two clouds.

As described in \S\ref{ssec:densities}, the two density distributions are sub-regions from evolved MHD simulations corresponding to a time of $t=4\,{\rm Myr}$ after the beginning of the simulation. At this particular time and as discussed in \citet{Bisb17b}, the low-$J$ CO and [C{\sc i}]~(1-0) position-velocity diagrams in the cloud-cloud collision case show that the colliding signature has disappeared. This is because the dense structures of the two GMCs have been merged to form the filament seen in Fig.~\ref{fig:pdfs} (`dense cloud'). However, the `bridge-effect' for the binary GMC collision can be seen in the [C{\sc ii}]~$158\mu$m and [O{\sc i}]~$63\mu$m lines \citep[see also][]{Bisb18}. It is, therefore, interesting to examine how the ISM environmental parameters considered here affect the shape of the `bridge' that connects two peaks in brightness temperatures for different velocities.

Figure~\ref{fig:signatures} shows normalized spectra to the peak value of [C{\sc ii}] and [O{\sc i}]~$63\mu$m lines for the dense cloud only and under all ISM environments explored. The spectra correspond to the slice from the highlighted region shown in the top left panel of Fig.~\ref{fig:GonH}. In general, there are two velocity peaks with a velocity separation of $\sim10\,{\rm km}\,{\rm s}^{-1}$, which is the relative speed of the GMC collision \citep[see][for the corresponding description of initial conditions]{Wu17}. We also examined the spectra of [C{\sc i}]~(1-0) and CO~(1-0) from the same slice, as well as the spectra of all aforementioned lines in a similar region in the diffuse cloud. We found that all those lines have simple, Gaussian-like shapes reflecting the turbulent motions of the gas. We note that the spectra of [C{\sc i}] and CO lines in the dense cloud have Gaussian-like shapes as well, since the colliding signature has disappeared in this MHD snapshot \citep{Bisb17b}.

Cosmic-rays with $\zeta_{\rm CR}{\lesssim}10^{-16}\,{\rm s}^{-1}$ do not have any effect on the collision signature in the [C{\sc ii}] line and only a minor effect in that from the [O{\sc i}] line. For $\zeta_{\rm CR}=10^{-15}\,{\rm s}^{-1}$, the [O{\sc i}] spectra does not show a significant change, but the [C{\sc ii}] line collision signature starts to diminish; this is reflected in the relative peaks in $T_{\rm B}$. Eventually, at $\zeta_{\rm CR}=10^{-14}\,{\rm s}^{-1}$, there are two interesting findings. The first one is that the two brightness temperature peaks in each of the [C{\sc ii}] and [O{\sc i}] lines differ so substantially ($\sim5$ times) that the bridge-effect signature appears to be severely diminished. Notably, the brightness temperature of [O{\sc i}]~$63\mu$m decreases in the first peak when $\zeta_{\rm CR}$ increases from $10^{-17}\,{\rm s}^{-1}$ to $10^{-15}\,{\rm s}^{-1}$ but increases rapidly when $\zeta_{\rm CR}=10^{-14}\,{\rm s}^{-1}$. This is because the increase in gas temperature resulting from the high cosmic-ray heating affects the line intensity at high densities, thus resulting in this high jump in brightness temperature, as explained in detail by \citet{Gold19}. The second is that at such high $\zeta_{\rm CR}$, both these lines trace the dense filament rather than the surrounding rarefied medium, as discussed in \S\ref{sec:emission}. This explains also the increase in the line broadening of the two fine-structure lines and especially that of the [O{\sc i}]~$63\mu$m. After examining further the [O{\sc i}] broadening, we found that it is identical to the CO~(1-0) one, thus originating from higher densities.

The second row of Fig.~\ref{fig:signatures} shows how the bridge-effect is affected as a function of $G_0$. As can be seen, the FUV photons do not change the shape of the [C{\sc ii}] line spectra. The [O{\sc i}] line is also not significantly affected for $G_0\lesssim100$. Increasing the FUV intensity can simply enhance the emission of these lines given that they originate from low optical depths, as described in \S\ref{ssec:emfuv}. In the case of $G_0=10^3$, [O{\sc i}] is emitted from higher densities (as with the case of high $\zeta_{\rm CR}$), therefore changing the relative peaks in $T_{\rm B}$. However, we find that the bridge-effect signature indicating a collision process, remains unaffected.

Finally, the bottom row of Fig.~\ref{fig:signatures} shows the dependency of the spectra on the changes in metallicity. For sub-solar metallicities, we find that the bridge effect in the [C{\sc ii}] line is diminishing. On the other hand, super-solar metallicities enhance it. This is because [C{\sc ii}] originates from higher number densities thus changing the linewidth in a similar way caused by high $\zeta_{\rm CR}$ as discussed above. The behaviour of the [O{\sc i}]~$63\mu$m line follows also the same concept as the [C{\sc ii}] line. Note, however, that the peaks of $T_{\rm B}$ alternate as metallicity changes; it is stronger for $v{\sim}0\,{\rm km/s}$ when compared to the $v{\sim}10\,{\rm km/s}$ peak, but as metallicity increases, the $T_{\rm B}$ at $v{\sim}10\,{\rm km/s}$ becomes stronger, as a result of the nature of [O{\sc i}] line becoming quickly optically thick. 

\begin{figure*}
    \centering
    \includegraphics[width=0.9\linewidth]{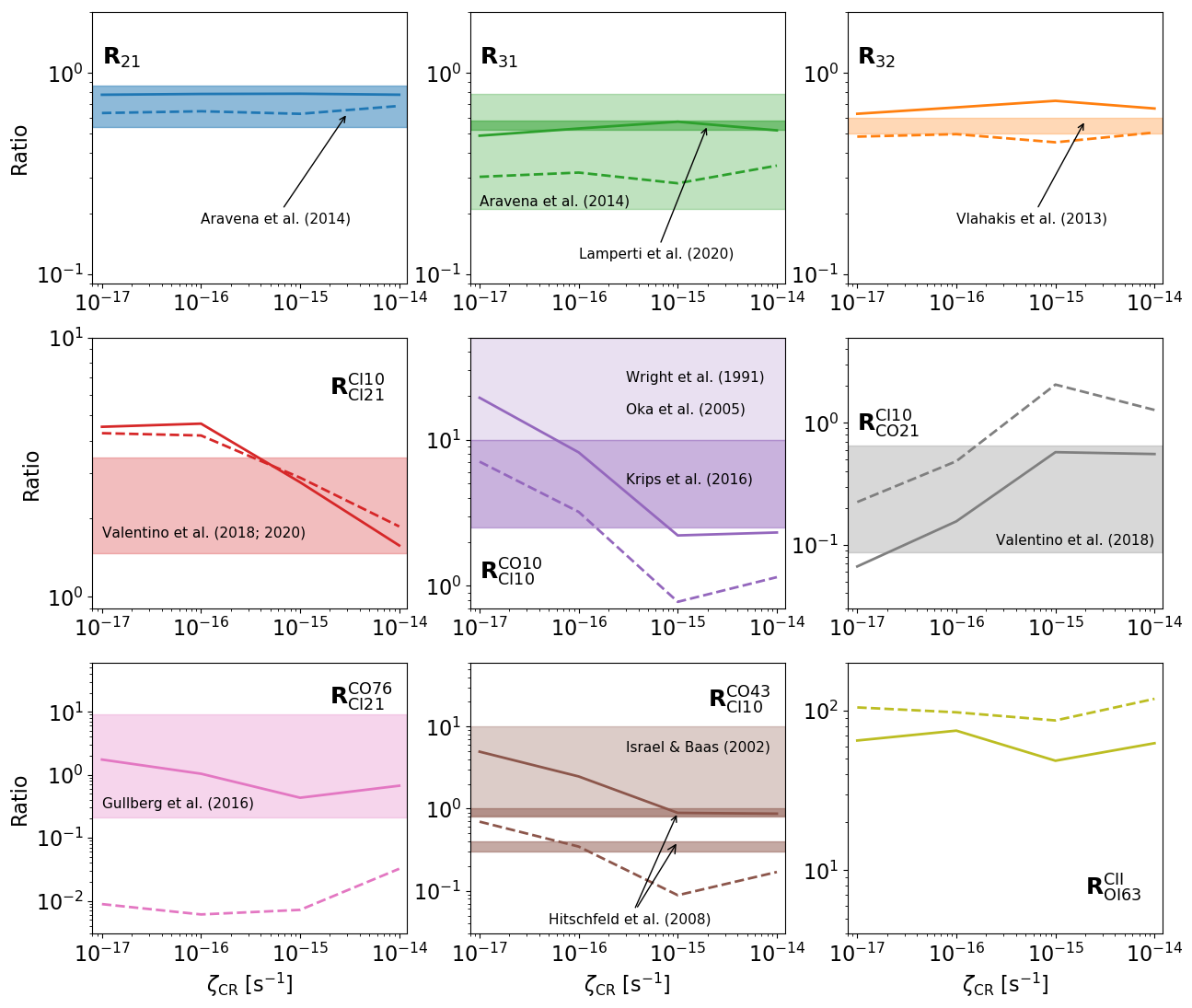}
    \caption{Ratios of various combinations between CO and [C{\sc i}] versus $\zeta_{\rm CR}$ and the FUV intensity for the dense (solid lines) and diffuse (dashed lines) clouds (see \S\ref{sec:ratios} for the corresponding nomenclature). The stripes correspond to observational estimates. In particular the stripes in the top left and top middle correspond to the \citet{Arav14} observations of the $R_{21}=0.7\pm0.14$ ratio and the $R_{31}=0.5\pm0.29$ ratio, respectively. The darker green stripe in the middle panel of the top row corresponds to the \citet{Lamp20} observations, giving $R_{31}=0.55\pm0.03$. The stripe on the top right panel is the $R_{32}=0.5-0.6$ ratio observed by \citet{Vlah13} for the whole M51 disk and its spiral arms. The stripe in the left panel of the middle row corresponds to the \citet{Vale18,Vale20} observations. In the central panel, the light-purple stripe corresponds to the \citet{Wrig91} and \citet{Oka05a} observations of the $R^{\rm CO10}_{\rm CI10}\gtrsim10$ ratio, while the darker purple stripe to the \citet{Krip16} $R^{\rm CO10}_{\rm CI10}=6.25\pm3.75$ observations. The light-gray stripe on the right panel of the middle row is the \citet{Vale18} observations of $R^{\rm CI10}_{\rm CO21}$. The light-pink stripe on bottom left panel corresponds to observations presented in \citet{Gull16}. The light-brown stripe in the bottom middle panel corresponds to the \citet{Isra02} observations and the darker-brown one to those of \citet{Hits08}. Overall we find very good agreement with observations.}
    \label{fig:ratios}
\end{figure*}

The above findings complement recent computational and observational works examining cloud formation in emission lines of the carbon cycle. \citet{Clar19} studied [C{\sc ii}], [C{\sc i}] and CO emission in dynamically evolved clouds, finding that the velocity dispersion of [C{\sc ii}] is larger than that of molecular line emission. This means that the $158\mu$m fine-structure line can be used as a diagnostic to identify ``bridging'' features for potential cloud-cloud collision activity. \citet{Lim20} studied the star cluster formation processes in Orion A using $^{13}$CO~(1-0) and [C{\sc ii}] data. The trends they observed, especially in [C{\sc ii}], provide evidence for an on-going cloud-cloud collision process which is a potential site for star formation activity. In a recent work, \citet{Beut20} studied the dynamical state of the infrared dark cloud G28 using [C{\sc i}] lines in addition to $^{13}$CO transitions. They identified a two-velocity component spectrum in the atomic lines, revealing a converging gas flow. They claim that this atomic emission originated from the diffuse, more extended part of the ISM surrounding the dense cloud. All the above are in good agreement with our results, e.g., as presented in Fig.~\ref{fig:signatures}, as well as with the results of \citet{Bisb17b,Bisb18}.

\section{Line ratios}
\label{sec:ratios}

\begin{figure*}
    \centering
    \includegraphics[width=0.9\linewidth]{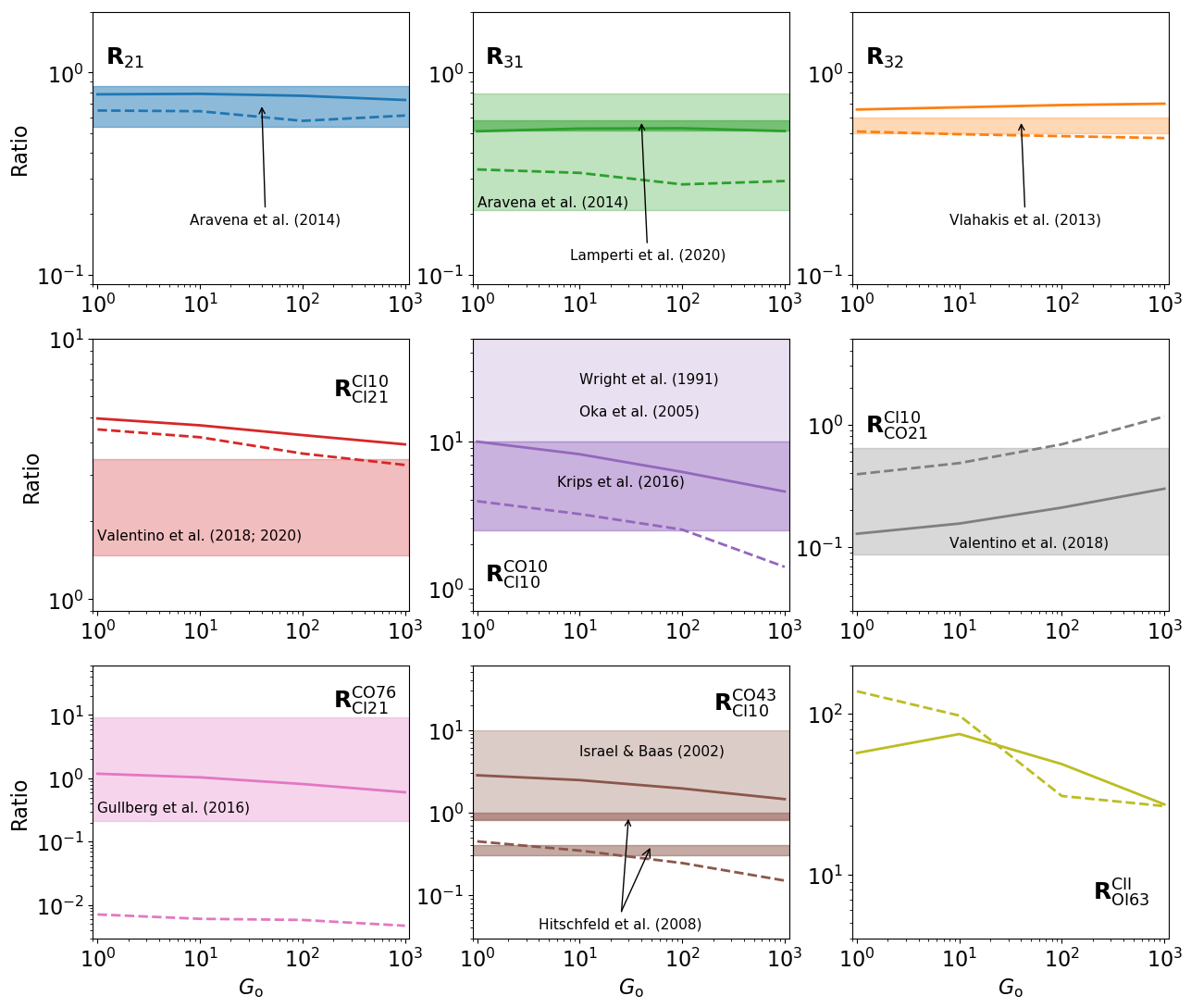}
    \caption{Same as Fig~\ref{fig:ratios} but for the FUV intensity as the free parameter.}
    \label{fig:ratiosFUV}
\end{figure*}

\begin{figure*}
    \centering
    \includegraphics[width=0.9\linewidth]{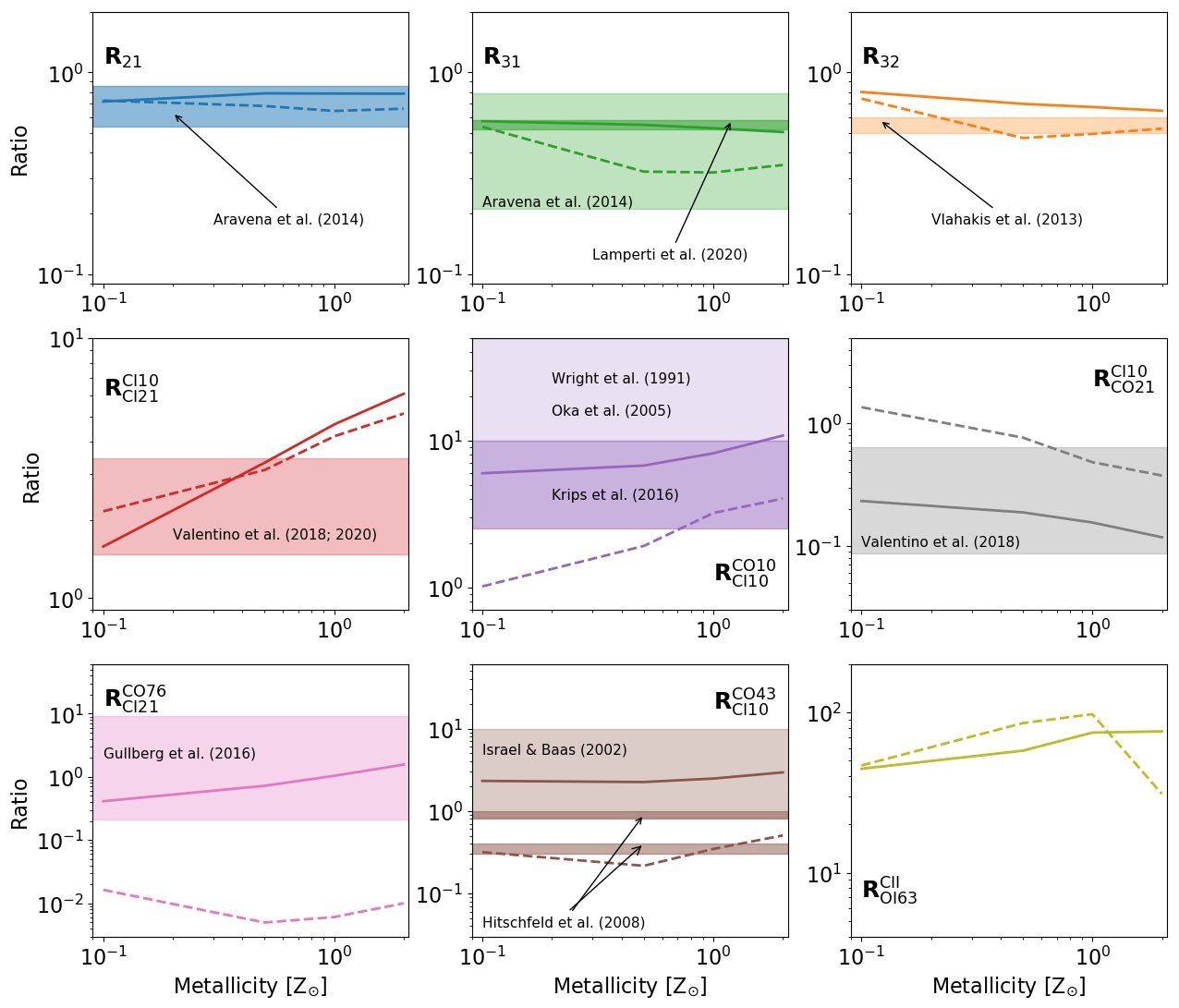}
    \caption{Same as Fig~\ref{fig:ratios} but for metallicity as the free parameter.}
    \label{fig:ratiosZ}
\end{figure*}

Line ratios are commonly used to study the physical state of the observed object, as well as to understand its molecular mass content. As we discuss in \S\ref{sec:conversion}, the emission of CO $J=1-0$ and, recently, of [C{\sc i}]~(1-0) lines are frequently used to infer to the column density of H$_2$. However, in the extragalactic context mid-$J$ CO lines are favoured since their higher frequencies offer better angular resolution for imaging. Moreover, the emission frequencies of CO $J=7-6$ and [C{\sc i}]~(2-1) transitions are so close that observing them simultaneously is entirely possible across any redshift. In a similar way, CO $J=4-3$ and [C{\sc i}]~(1-0) can be simultaneously observed with ALMA Bands 3-8. In addition, [C{\sc ii}] and [O{\sc i}]~$63\mu$m can be simultaneously observed using the SOFIA-upGREAT. It is therefore important to study how the different line ratios behave under the different ISM environmental parameters considered here. In doing so, we examine the ratios $R_{21}={\rm W_{ CO(2-1)}/W_{CO(1-0)}}$, $R_{31}={\rm W_{CO(3-2)}/W_{CO(1-0)}}$, $R_{32}={\rm W_{CO(3-2)}/W_{CO(2-1)}}$, as well as $R^{\rm CI10}_{\rm CI21}$, $R^{\rm CI10}_{\rm CO21}$, $R^{\rm CO10}_{\rm CI10}$, $R^{\rm CO43}_{\rm CI10}$, $R^{\rm CO76}_{\rm CI21}$, and $R^{\rm CII}_{\rm OI63}$, where the latter ratios follow the notation $R^X_Y={\rm W_{X}/W_Y}$.

Figure~\ref{fig:ratios} shows the results of the line ratios with the cosmic-ray ionization rate as the free ISM parameter compared with observed values. Figures~\ref{fig:ratiosFUV} and \ref{fig:ratiosZ} show the results when the FUV intensity and the metallicity are the free parameters, respectively. Solid lines correspond to the `dense cloud', while dashed lines correspond to the `diffuse cloud'. As can be seen in the top row, $R_{21}$, $R_{31}$ and $R_{32}$ show a weak dependency on the cosmic-ray ionization rate, since this affects the corresponding emission lines by approximately equal amounts. It is also found that both the FUV intensity and the metallicity do not significantly change these ratios. There is, however, a weak dependency on the column density distribution. For instance, the ratio of $R_{\rm 21,dense}/R_{\rm 21,diffuse}$ between the dense and diffuse clouds is ${\sim}1.5$. Similarly, for $R_{32}$ it is ${\sim1.5}$ as well, and for $R_{31}$ it is ${\sim}2.5$. \citet{Arav14} studied four massive star-forming galaxies at redshifts of $\sim1.5-2.2$ and reported average line brightness temperature ratios of $R_{21}=0.7\pm0.16$ and $R_{31}=0.50\pm0.29$. Recently, \citet{Lamp20} studied this ratio for a sample of 98 galaxies (Star Forming, AGNs, LIRGs) and found a mean ratio of $R_{31}\simeq0.55\pm0.03$, which agrees remarkably well with our dense cloud for $\zeta_{\rm CR}\ge10^{-16}\,{\rm s}^{-1}$. The top right panel shows the $R_{32}$ ratio. \citet{Vlah13} studied this ratio in the spiral galaxy M51 (NGC~5194) using the James Clerk Maxwell Telescope (JCMT). At a spatial resolution of ${\sim}600\,{\rm pc}$, they reported an average value of $R_{32}\sim0.5$ for the disk. Notably, they found $R_{32}\sim0.2$ in interarm regions and $R_{32}\sim0.6$ in arm and central regions. The simulated values of these ratios are in good agreement with observations in both cloud cases and for all ISM environmental parameters examined. In addition, our results are in agreement with the models of \citet{Pena18}, who examined the behaviour of $R_{21}$, $R_{31}$, and $R_{32}$ under similar ISM conditions.

In the left panel of the middle row, the results for the $R^{\rm CI10}_{\rm CI21}$ ratio between [C{\sc i}]~(1-0) and (2-1) are presented. An interesting feature is that this ratio is found to be very weakly depending on the density distribution for all ISM environmental parameters. For $\zeta_{\rm CR}>10^{-16}\,{\rm s}^{-1}$ it is found to depend on the cosmic-ray ionization rate as $R^{\rm CI10}_{\rm CI21}{\approx}3\times10^{-3}\zeta_{\rm CR}^{-0.2}$. Interestingly, this ratio depends weakly on the $G_0$ intensity but strongly on metallicity following a relationship of the form $R^{\rm CI10}_{\rm CI21}{\approx} 4Z^{0.37}$. This suggests that this ratio may be a promising new diagnostic for an estimation of $\zeta_{\rm CR}$ and $Z$, which is particularly interesting for the upcoming CCAT-prime telescope. Observations of various galaxies (red stripe) including Local IR bright galaxies, Main Sequence at a redshift of ${\sim}1-1.5$ and SMGs at a redshift of ${\sim}2.5-4$ by \citet{Vale18,Vale20} found a similar $R^{\rm CI10}_{\rm CI21}$ ratio throughout their sample, which matches with our simulations for an average $\zeta_{\rm CR}\gtrsim3\times10^{-15}\,{\rm s}^{-1}$. This is more than the average Galactic one (see \S\ref{ssec:3dpdr}), reaching to values as high as $\zeta_{\rm CR}\sim10^{-14}\,{\rm s}^{-1}$. However, because the sample contains galaxies that have much higher star formation rates (SFRs) than the Galactic one it is reasonable to expect such boosts of $\zeta_{\rm CR}$ \citep{Papa11}. The calculated $R^{\rm CI10}_{\rm CI21}$ ratio as a function of $G_0$ does not match with the observational data of \citet{Vale18,Vale20}. Note, however, that the trends of the corresponding panel in Fig.~\ref{fig:ratiosFUV} show that much higher FUV intensities may be also explaining these observations, although it is unlikely that such high $G_0$ can operate on a global scale in these systems. From Fig.~\ref{fig:ratiosZ} it can be seen that metallicities with $Z\lesssim0.5\,{\rm Z}_{\odot}$ can lower the $R^{\rm CI10}_{\rm CI21}$, matching the observations. Assuming solar metallicity for the aforementioned sample of galaxies, we propose that the cosmic-ray ionization rate is enhanced compared to the Milky Way average.

The central panel of the middle row shows the $R^{\rm CO10}_{\rm CI10}$ ratio between CO~(1-0) and [C{\sc i}]~(1-0). This ratio shows a strong dependency on the density distribution as it differs always by more than three times under all free parameters. In terms of the dependency on the free parameters, it is found that $R^{\rm CO10}_{\rm CI10}$ depends weakly on $G_0$ and $Z$ and more strongly on $\zeta_{\rm CR}$. Early Milky Way observations of that ratio by \citet{Wrig91} and more recently by \citet{Oka05a}\footnote{Both \citet{Wrig91} and \citet{Oka05a} report on the [C{\sc i}]~(1-0)/CO~(1-0) ratio which is the reverse of what is presented here; however we keep our nomenclature for consistency.}, show that in our Galaxy $R^{\rm CO10}_{\rm CI10}\gtrsim10$. These values are in good agreement with our calculations, particularly for the dense cloud and for $\zeta_{\rm CR}\lesssim10^{-16}\,{\rm s}^{-1}$, $G_0{\sim}1$ and $Z{\sim}1\,{\rm Z}_{\odot}$. We do not find a reasonable match of these observations with the diffuse cloud calculations. Recent observations by \citet{Krip16} in NGC~253, found a global ratio of the order of $R^{\rm CO10}_{\rm CI10}\sim2.5-10$, which agrees with our simulations for higher $\zeta_{\rm CR}$, higher $G_0$ and high metallicities, $Z>1\,{\rm Z}_{\odot}$ for which both clouds match the observations. This is in line with the \citet{Krip16} suggestion that in NGC~253, which is a starburst galaxy therefore having a different and more energetic ISM environment than the local one, shocks and their associated radiation and/or an enhancement of cosmic-ray energy densities may explain such $R^{\rm CO10}_{\rm CI10}$ values. It is, furthermore, in agreement with \citet{Gall08} who estimate a metallicity of ${\sim}2.19\,{\rm Z}_{\odot}$ for NGC~253. Observations by Dunne et al. (\textit{in preparation}) also report that $R^{\rm CO10}_{\rm CI10}=3.97-11.11$ in Main Sequence galaxies and $R^{\rm CO10}_{\rm CI10}=3.13-6.67$ in SMGs having high star forming rates, which according to our models suggest the presence of high cosmic-ray energy densities and high intensities of FUV radiation field.

The right panel of the middle row shows the $R^{\rm CI10}_{\rm CO21}$ ratio between [C{\sc i}]~(1-0) and CO~(2-1). This ratio varies with the density distribution (the dense cloud has overall a lower ratio), it increases with increasing $\zeta_{\rm CR}$ (but remains unchanged for $\zeta_{\rm CR}{\gtrsim}10^{-15}\,{\rm s}^{-1}$), slightly increases with increasing $G_0$ and slightly decreases with increasing $Z$. Observations of \citet{Vale18} of various galaxies including, Main Sequence, starbursts, Active Galactic Nuclei (all at $z\sim1.2$), and SPT SMGs ($z\sim4$) cover a similar range of the $R^{\rm CI10}_{\rm CO21}$ ratio. \citet{Vale18} also report that the particular SPT SMGs have the highest values of that ratio. 

The interesting ratio of CO~(7-6)/[C{\sc i}]~(2-1) is illustrated in the left panel of the bottom row in Figs.~\ref{fig:ratios}, \ref{fig:ratiosFUV} and \ref{fig:ratiosZ} for $\zeta_{\rm CR}$, $G_0$ and $Z$ as the free parameters, respectively. As can be seen, the $R^{\rm CO76}_{\rm CI21}$ ratio shows a strong dependency on the column density distribution as the differences between the two clouds can be between one and more-than-two orders of magnitude. In terms of the environmental parameters, both clouds show weak dependency on the $G_0$ intensity. The $R^{\rm CO76}_{\rm CI21}$ ratio of both clouds depend on the values of $\zeta_{\rm CR}$ and $Z$ but they follow different patterns. The pink stripe corresponds to observations presented in \citet{Gull16}. They found a low CO~(7-6)/[C{\sc i}]~(2-1) ratio ($R^{\rm CO76}_{\rm CI21}\sim0.2$) for the Spiderweb radio galaxy at a redshift of $\sim2.161$, which may be attributed to high cosmic-ray energy densities. However, given that we have agreement of this ratio for all free parameters explored, it may well indicate that the $R^{\rm CO76}_{\rm CI21}$ is not a good tracer of the physical conditions and that additional observations may be needed to understand the properties of the ISM environment of that galaxy.

The middle panel of the bottom row shows the $R^{\rm CO43}_{\rm CI10}$ ratio, which does not appear to have a monotonic relationship with $\zeta_{\rm CR}$. Instead, for $\zeta_{\rm CR}{\sim}10^{-15}$ it has a local minimum for the diffuse cloud. That particular ratio shows a strong dependence on the density distribution with a difference between the dense and diffuse clouds of approximately an order of magnitude (the ratio for the dense cloud is always higher). However, there is a very weak dependency on the intensity of FUV radiation field and on the metallicity. The light brown stripe corresponds to the observations of \citet{Isra02} in 13 spiral galaxies. The reported ratio of $R^{\rm CO43}_{\rm CI10}\sim0.8-10$ matches with the dense cloud results. Furthermore, they found that the ratio of CO~(4-3)/[C{\sc i}]~(1-0) decreases with increasing [C{\sc i}]~(1-0) luminosity. According to our results, this decrease of the $R^{\rm CO43}_{\rm CI10}$ corresponds to higher $\zeta_{\rm CR}$ due to the increase of [C{\sc i}]~(1-0) emission. In addition, \citet{Hits08} studied this ratio in the starburst galaxies NGC~4945 and Circinus and reported that $R^{\rm CO43}_{\rm CI10}\sim0.8-1.0$ and ${\sim}0.3-0.4$, respectively, matching with the diffuse cloud results.

Finally, in the bottom right panel of Figs.~\ref{fig:ratios}, \ref{fig:ratiosFUV} and \ref{fig:ratiosZ}, the [C{\sc ii}]/[O{\sc i}]~$63\mu$m ratio is plotted. As can be seen, there is a weak dependence on the density distribution. This ratio does not vary strongly with $\zeta_{\rm CR}$ and it decreases with increasing $G_0$. For the dense cloud, it increases with increasing $Z$, however for the diffuse cloud it peaks for $Z=1\,{\rm Z}_{\odot}$ but then decreases rapidly for $2\,{\rm Z}_{\odot}$. The line of [O{\sc i}]~$63\mu$m is much more optically thick than the line of [C{\sc ii}] and suffers from self-absorption which has a significant impact on the resulting $R^{\rm CII}_{\rm OI63}$ ratio. For instance, \citet{Schn18} report an $R^{\rm CII}_{\rm OI63}{\sim}0.03-0.2$ in the massive star-forming region S106. This ratio is much lower than what we find for any ISM condition explored. However, \citet{Schn18} find that a radiation field with $G_0>10^4$ is needed to model this ratio. Although we do not explore such high FUV intensities, it can be seen from the corresponding panel of Fig.~\ref{fig:ratiosFUV} that it is in agreement with the trend we find against $G_0$. More recently, \citet{Kram20} reported an $R^{\rm CII}_{\rm OI63}=0.2-20$ in M33 with a mean of $4.5\pm2.6$ which matches with the \citet{Kauf06} one-dimensional uniform density PDR models of $n_{\rm H}\sim2\times10^2-10^4\,{\rm cm}^{-3}$ interacting with $G_0\sim1.5-60$. After experimenting with such simple PDR models, we also confirm the observed ratio of \citet{Kram20} using {\sc 3d-pdr} in one-dimensional slabs but we are unable to reproduce this ratio with our full three-dimensional models. We therefore conclude that a ratio between these two lines is rather challenging to use as a diagnostic \citep[see also][]{Kram98}.

\section{Conversion factors}
\label{sec:conversion}

As discussed in \S\ref{sec:intro}, although being the most abundant molecule in the ISM, H$_2$ does not emit significant radiation from most molecular clouds, as its lowest transition energy difference at $\Delta E/k_{\rm B}{\sim}510\,{\rm K}$ above ground is much higher than the gas temperature typically found in the ISM at high column densities. The next most abundant molecule, CO, is frequently used to trace the H$_2$-rich gas via the velocity integrated emission of its $J=1-0$ transition as it corresponds to an energy difference of $\Delta E/k_{\rm B}{\sim}5\,{\rm K}$ above ground. This emission is converted to a molecular column density via the $X_{\rm CO}$--factor \citep{Stro96,Dame01,Bell07,Bola13,Szuc16}, which is taken to be constant. Other tracers have been considered as alternatives to the CO $J=1-0$ emission with the [C{\sc i}]~$609\mu$m \citep{Papa04} to become one of the most popular as it is bright, present in H$_2$-rich regions optically thinner than CO $J=1-0$ and its lowest transition energy difference corresponds to a temperature of $\Delta E/k_{\rm B}{\sim}24\,{\rm K}$. In this Section, it is examined how the $X_{\rm CO}$-- and $X_{\rm CI}$--factors depend on the ISM parameters we consider. In all cases the conversion factor, $X_{\rm C^{\star}}$ (denoting either CO or C{\sc i}), is defined as:
\begin{eqnarray}
X_{\rm C^{\star}}=\frac{\int N_{\rm H_2} {\rm d}S}{\int W_{\rm C^{\star}}{\rm d}S}.
\end{eqnarray}

In general we find that both conversion factors depend on both the free ISM parameters and the density distribution. The best-fit functions we find obey the formula $X_{\rm C^*}=10^b ({\rm FP})^k\,[{\rm cm}^{-2}\,{\rm K}^{-1}\,{\rm km}^{-1}\,{\rm s}]$, where FP is the ISM free parameter ($\zeta_{\rm CR}$, $G_0$ or $Z$).

\subsection{Dependency on the cosmic-ray ionization rate}
\label{ssec:Xcr}

\begin{figure}
    \centering
    \includegraphics[width=\linewidth]{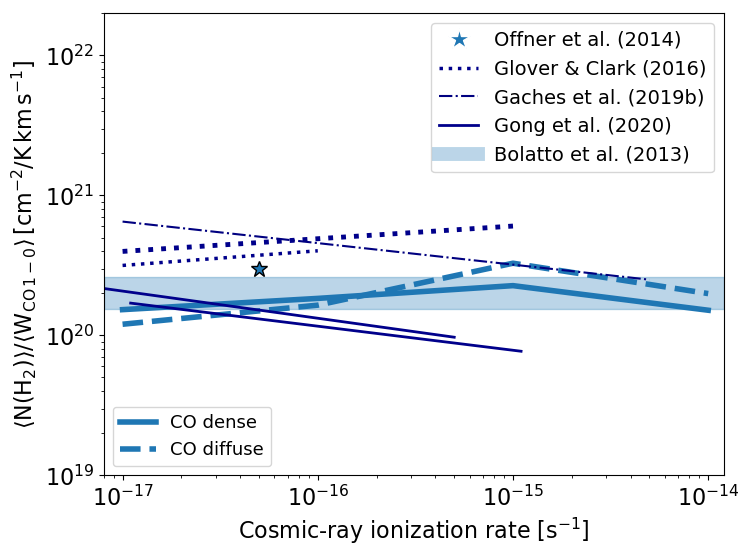}
    \includegraphics[width=\linewidth]{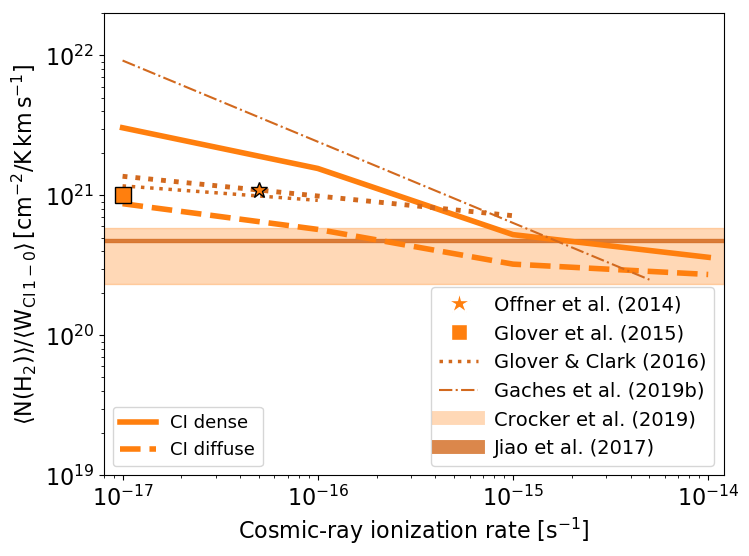}
    \caption{Conversion factors based on CO~(1-0) (top panel) and [C{\sc i}]~(1-0) (bottom panel) for the dense (thick solid lines) and diffuse (thick dashed lines) clouds as a function of $\zeta_{\rm CR}$. The recommended value of $X_{\rm CO}=2\times10^{20}\,{\rm cm}^{-2}\,{\rm K}^{-1}\,{\rm km}^{-1}\,{\rm s}$ $\pm30\%$ spread \citep{Bola13} is shown in blueish stripe in the top panel. The values of $X_{\rm CI}$ reported by \citet{Croc19} and \citet{Jiao17} are shown in light and darker orange stripes in the bottom panel, respectively. Simulation results from \citet{Gach19b}, \citet{Offn14}, \citet{Glov15}, \citet{Glov16} and \citet{Gong20} are also shown. In particular, the thin dotted lines in \citet{Glov16} correspond to $\chi_0=10$, while the thick dotted to $\chi_0=100$. The upper dark blue solid line of the \citet{Gong20} simulations correspond to their `R4' model, while the lower one to their `R2' model (Milky Way conditions at a radius of 4kpc and 2kpc, respectively). We find that $X_{\rm CI}$ is more strongly depending on both $\zeta_{\rm CR}$ and column density distributions than $X_{\rm CO}$ and that the observational data of \citet{Jiao17} and \citet{Croc19} may be connected with galaxies with enhanced cosmic-ray energy densities.}
    \label{fig:xco}
\end{figure}

The top panel of Fig.~\ref{fig:xco} shows the resultant conversion factors for CO~(1-0) versus $\zeta_{\rm CR}$. Solid lines correspond to the dense cloud and dashed to the diffuse cloud. As discussed in \S\ref{ssec:densities} and also in \citet{Bisb17a}, the column density of CO is severely decreased with increasing $\zeta_{\rm CR}$ while $N({\rm H_2})$ remains almost unaffected. Intuitively, this would imply a different $X_{\rm CO}$--factor as $W({\rm CO}~1-0)$ may be affected. However, as discussed in \S\ref{ssec:growth}, the increase in gas temperature increases the brightness temperature to high values even for low CO column densities. This, in turn, keeps the emission of CO~(1-0) strong, leading to small changes in the $X_{\rm CO}$--factor. As can be seen in this panel, $X_{\rm CO}$ increases by only a factor of ${\sim}2$ between $\zeta_{\rm CR}=10^{-17}\,{\rm s^{-1}}$ and $10^{-15}\,{\rm s}^{-1}$, at which point it has a local maximum. In particular, for $\zeta_{\rm CR}\lesssim10^{-15}\,{\rm s}^{-1}$ we find $b{\simeq}22.70$ and $k{\simeq}0.15$. For $\zeta_{\rm CR}=10^{-14}\,{\rm s}^{-1}$, $X_{\rm CO}$ decreases by a factor of $\sim1.5$. This is because CO has been destroyed in the low-density medium allowing the strong line of CO~(1-0) to originate from the denser and warmer (due to cosmic-ray heating) gas. In this regime of $10^{-15}\lesssim\zeta_{\rm CR}\lesssim10^{-14}\,{\rm s}^{-1}$, we find that $b{\simeq}17.49$ and $k{\simeq}-0.20$. The dependence on the density distribution is minor; both dense and diffuse cloud appear to require a similar $X_{\rm CO}$--factor to infer to the molecular mass content. Overall, $X_{\rm CO}$ varies from ${\sim}1.2-3.3\times10^{20}\,{\rm cm}^{-2}\,{\rm K}^{-1}\,{\rm km}^{-1}\,{\rm s}$ across the different $\zeta_{\rm CR}$. 

The blue stripe is the recommended \citet{Bola13} value of $X_{\rm CO}=2\times10^{20}\,{\rm cm}^{-2}\,{\rm K}^{-1}\,{\rm km}^{-1}\,{\rm s}$ with the recommended $\pm30\%$ variation. Our results are in broad agreement with this value, particularly for $\zeta_{\rm CR}\sim10^{-16}\,{\rm s}^{-1}$. \citet{Offn14} studied a turbulent $600\,{M}_{\odot}$ Milky Way star-forming cloud \citep[presented in][]{Offn13} under two different interstellar radiation fields of $\chi_0=1,~10$ \citep[normalized according to][]{Drai78}, the `full' UMIST2012 chemical network of 215 species and a cosmic-ray ionization rate of $\zeta_{\rm CR}=5\times10^{-17}\,{\rm s}^{-1}$. They found $X_{\rm CO}=3\times10^{20}\,{\rm cm}^{-2}\,{\rm K}^{-1}\,{\rm km}^{-1}\,{\rm s}$ which is approximately two times higher than our calculated for both clouds. \citet{Glov16} modelled a $10^4\,{\rm M}_{\odot}$ cloud under several ISM environmental parameters. The two dotted lines are the simulations for $\chi_0=10$ (thin dotted) and $\chi_0=100$ (thick dotted). As can be seen, both derived $X_{\rm CO}-$factors increase with increasing $\zeta_{\rm CR}$ with a slope very similar to our modelled clouds. However, their highest cosmic-ray ionization rate is $\zeta_{\rm CR}=10^{-15}\,{\rm s}^{-1}$, a value above which our results indicate a declining trend of $X_{\rm CO}$. 
\citet{Gach19a} modified {\sc 3d-pdr} to account for cosmic-ray attenuation as a function of the total column density to study star-forming clouds including embedded protostellar clusters. This approach was used in a subsequent work \citep{Gach19b} to examine the impact of cosmic-rays on both $X_{\rm CO}$ and $X_{\rm CI}$ (see below) conversion factors. For a Voyager spectrum of cosmic-rays, they found that $X_{\rm CO}$ decreases with increasing $\zeta_{\rm CR}$ as opposed to our simulations. In particular, for $\zeta_{\rm CR}\lesssim10^{-16}\,{\rm s}^{-1}$ they found an $X_{\rm CO}\gtrsim5\times10^{20}\,{\rm cm}^{-2}\,{\rm K}^{-1}\,{\rm km}^{-1}\,{\rm s}$, which is higher than the \citet{Bola13} recommended value. 
Recently, \citet{Gong20} presented simulations of three clouds representing the Milky Way ISM and studied the dependency of $X_{\rm CO}-$factor in different environmental parameters. Contrary to our findings, they report a trend of decreasing $X_{\rm CO}$ with increasing cosmic-ray ionization rate even for $\zeta_{\rm CR}{\gtrsim}5\times10^{-18}\,{\rm s}^{-1}$. We argue that this difference may arise from a number of different factors such as the density distribution they modelled (512~pc scale clouds at a resolution of 2~pc), the attenuation of cosmic-rays as a function of the column density they assumed, and the chemical network they adopted \citep{Gong18}.

On the other hand, the $X_{\rm CI}$--factor depends stronger on the value of cosmic-ray ionization rate and varies by approximately one order of magnitude between the two $\zeta_{\rm CR}$ extrema. In addition, $X_{\rm CI}$ is higher for the dense cloud by a factor of $\sim2-3$. We find that the dense cloud is best-fitted with $b{\simeq}15.95$ and $k{\simeq}-0.32$ while the diffuse one with $b{\simeq}17.92$ and $k{\simeq}-0.18$. This strong decrease of $X_{\rm CI}$ with increasing $\zeta_{\rm CR}$ is due to the increase of the [C{\sc i}]~(1-0) emission, given also the fact that it is more optically thin than CO~(1-0), as described in \S\ref{ssec:emcr} and \S\ref{ssec:growth}. Various computational efforts have been made to compute the value of $X_{\rm CI}$--factor. 
\citet{Offn14} found that $X_{\rm CI}=1.1\times10^{21}\,{\rm cm}^{-2}\,{\rm K}^{-1}\,{\rm km}^{-1}\,{\rm s}$ (represented with the orange star in Fig.~\ref{fig:xco}), a value confirmed observationally by \citet{Lo14}. The value of this conversion factor is in-between our modelled clouds. Similarly, simulations of a $10^4\,{\rm M}_{\odot}$ cloud with $\langle n_{\rm H}\rangle\sim276\,{\rm cm}^{-3}$ by \citet{Glov15} under a $\chi_0=1$ isotropic radiation field and a $\zeta_{\rm CR}=10^{-17}\,{\rm s}^{-1}$ found an $X_{\rm CI}\sim1.1\times10^{21}\,{\rm cm}^{-2}\,{\rm K}^{-1}\,{\rm km}^{-1}\,{\rm s}$. The \citet{Glov15} setup is very similar to our diffuse cloud and for that specified $\zeta_{\rm CR}$ we find good agreement. In the follow-up paper, \citet{Glov16} found also a decreasing $X_{\rm CI}$ with increasing $\zeta_{\rm CR}$ which is in line with our findings. For instance, for $\zeta_{\rm CR}=10^{-15}\,{\rm s}^{-1}$ they find an $X_{\rm CI}=6.04\times10^{20}\,{\rm cm}^{-2}\,{\rm K}^{-1}\,{\rm km}^{-1}\,{\rm s}$ which is ${\sim}1.1$ times higher than the one for the dense cloud and ${\sim}1.9$ times higher than for the diffuse one. The trends of \citet{Gach19b} are also in accordance with our simulations, although in most cases its value surpasses the corresponding one of our dense cloud model.

\citet{Jiao17} observed 71 galaxies with {\it Herschel} SPIRE-FTS, out of which 62 where LIRGs and 9 ULIRGs. They report a line ratio of $R_{\rm CI10}^{\rm CO10}{\sim}2-10$, which matches with that of \citet{Krip16}, and an average C{\sc i}-to-H$_2$ conversion factor of $\alpha_{\rm [CI]~(1-0)}=7.6\,{\rm M}_{\odot}\,{\rm pc}^{-2}\,{\rm K}^{-1}\,{\rm km}^{-1}\,{\rm s}$ or $X_{\rm CI}=4.8\times10^{20}\,{\rm cm}^{-2}\,{\rm K}^{-1}\,{\rm km}^{-1}\,{\rm s}$ without accounting for Helium contribution. This value agrees with our models for values of $\zeta_{\rm CR}$ higher than the average Milky Way value; for a density distribution corresponding to the diffuse cloud, we estimate a $\zeta_{\rm CR}{\gtrsim}5\times10^{-16}\,{\rm s}^{-1}$. More recently, \citet{Croc19} studied 18 nearby galaxies also observed with {\it Herschel} SPIRE-FTS. They report a similar conversion factor with that of \citet{Jiao17}, of $X_{\rm CI}=3.7\times10^{20}\,{\rm cm}^{-2}\,{\rm K}^{-1}\,{\rm km}^{-1}\,{\rm s}$ (including He contribution), thus matching with our models for high $\zeta_{\rm CR}$. Such cosmic-ray ionization rates, higher than the Milky Way average, are further supported by the line ratios $R_{\rm CI21}^{\rm CI10}{\sim}0.2-1.7$, $R_{\rm CI10}^{\rm CO43}{\sim}0.7-2$ and $R_{\rm CI21}^{\rm CO76}{\sim}0.2-0.6$ they report (see \S\ref{sec:ratios}). In all above cases, we argue that the observed galaxies have enhanced cosmic-ray energy densities.

\subsection{Dependency on the intensity of the FUV radiation field}

\begin{figure}
    \centering
    \includegraphics[width=\linewidth]{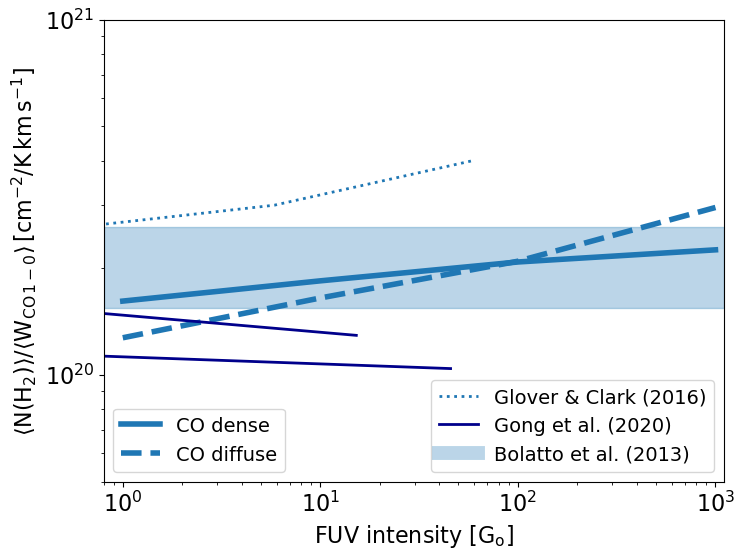}
    \includegraphics[width=\linewidth]{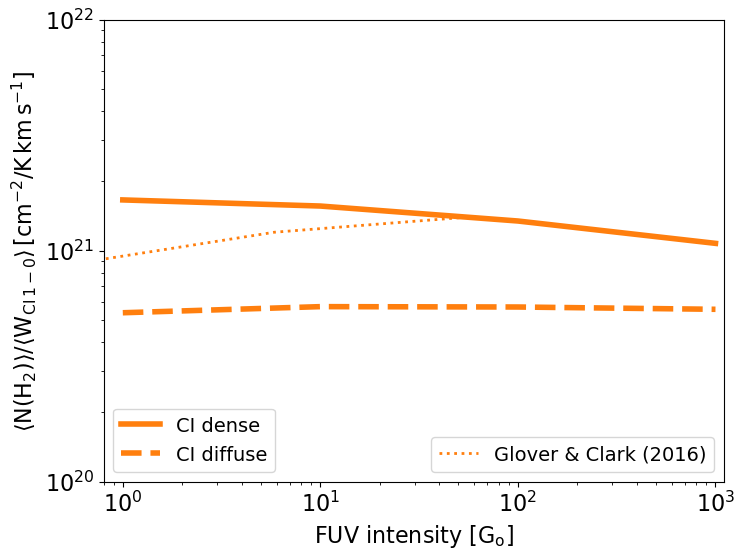}
    \caption{As in Fig.~\ref{fig:xco} but for the FUV intensity as the free ISM parameter. We find that both $X_{\rm CO}$ and $X_{\rm CI}$ factors remain relatively unchanged as a function of $G_0$. The results of \citet{Glov16} (for both tracers) and \citet{Gong20} (for CO~(1-0); upper/lower dark blue solid lines are their `R4'/`R2' models, respectively) are illustrated for comparison.}
    \label{fig:xco2}
\end{figure}

Figure~\ref{fig:xco2} shows how the conversion factors depend on the intensity of FUV radiation field. We find that this dependency is a weak function of $G_0$. For $X_{\rm CO}$, we further find that it depends weakly on the density distribution.

As can be seen from the upper panel, our results agree well with the \citet{Bola13} average for all $G_0$ intensities. There is a slight increase of $X_{\rm CO}$ as $G_0$ increases. We find that the CO-to-H$_2$ conversion factor for both clouds can be fitted with $b{\simeq}20.15$ and $k{\simeq}0.08$. The slope is similar to the one found by \citet{Glov16}, whose results are plotted with dotted lines (their $\chi_0=10$ simulation). However, the best-fit slope of \citet{Gong20} shows a decreasing $X_{\rm CO}$ as $G_0$ increases, although their slope of ${\sim}-0.03$ indicates a weak dependency on the FUV intensity as well. Similarly, \citet{Gach18} report a weak dependency of $X_{\rm CO}$ on $G_0$ when studying the effect of radiation field on ISM gas chemistry from embedded protostars.

The lower panel of Fig.~\ref{fig:xco2} shows the $X_{\rm CI}$ dependency on $G_0$. Here, both conversion factors remain relatively constant under all $G_0$ intensities, however there is a stronger dependency on the density distribution. For the diffuse cloud, $X_{\rm CI}{\simeq}5.3\times10^{20}\,{\rm cm}^{-2}\,{\rm K}^{-1}\,{\rm km}^{-1}\,{\rm s}$ (independent on $G_0$) and for the dense cloud it is fitted with $b{\simeq}21.23$ and $k{\simeq}-0.06$. \citet{Glov16} find also similar values of $X_{\rm CI}$, although their results indicate an increasing factor with $\chi_0$.

\subsection{Dependency on the metallicity}

\begin{figure}
    \centering
    \includegraphics[width=\linewidth]{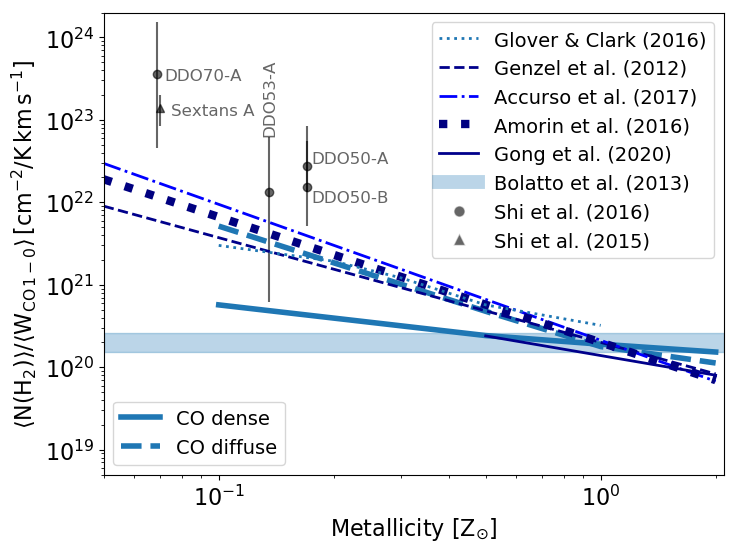}
    \includegraphics[width=\linewidth]{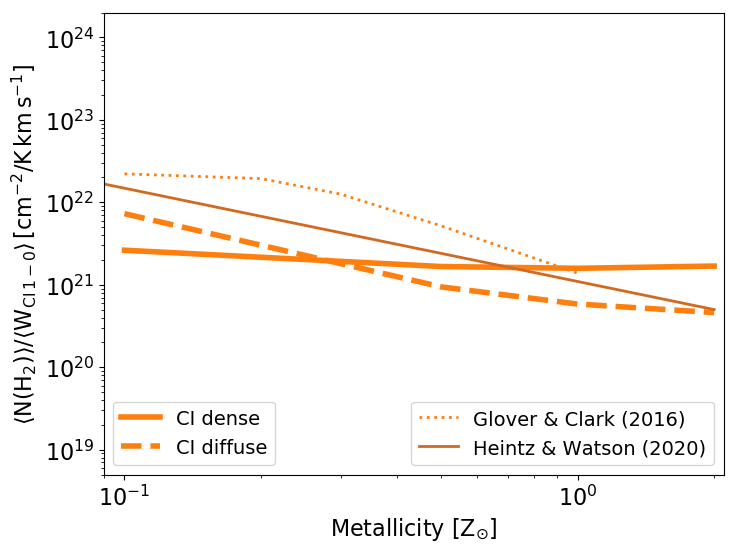}
    \caption{As in Fig.~\ref{fig:xco} but for the metallicity as the free ISM parameter. We find that both factors increase with decreasing $Z$ and they both depend on the column density distribution. For comparison, we plot the numerical results of \citet{Glov16} for both tracers (thin dotted lines), for the [C{\sc i}]-to-H$_2$ factor the observations of \citet{Hein20} (thin solid dark-orange line) and for the CO-to-H$_2$ factor the best-fits of \citet{Genz12} (dark blue dashed line), \citet{Accu17b} (light blue dash-dotted line), \citet{Amor16} (dark blue thick dotted line) and \citet{Gong20} (dark blue solid line). We also plot individual data points of four DDO galaxies \citep{Shi16} and Sextans A \citep{Shi15}.}
    \label{fig:xco3}
\end{figure}

Figure~\ref{fig:xco3} shows the correlation of the conversion factors with metallicity which is arguably of great interest in studying high-redshift galaxies. The $X_{\rm CO}-$factor depends strongly on metallicity, while $X_{\rm CI}-$ depends much more weakly on $Z$, which is in agreement with numerous other observational \citep{Isra97, Lero11, Genz12, Schr12, Shi15, Shi16, Amor16, Hein20} and theoretical \citep{Wolf10, Glov11, Nara12, Feld12, Sand13, Glov16, Accu17b, Gong20, Madd20} works. 

We find that at low metallicities, the $X_{\rm CO}-$factor additionally depends strongly on the density distribution. The diffuse cloud is best-fitted with with $b{\simeq}20.34$ and $k{\simeq}-1.32$. As shown in the top panel of Fig.~\ref{fig:xco3}, our results for the diffuse cloud are in agreement with the observations of \citet{Genz12} and \citet{Amor16} as well as with the theoretical works of \citet{Glov16} (their $G_0=10$ simulation is plotted here) and \citet{Accu17b}. In this panel, it is further plotted the observations of \citet{Shi15,Shi16} on Sextans~A and four DDO objects (70-A, 53-A, 50-A, 50-B), which are all very metal poor galaxies. All these galaxies provide evidence of stronger $Z-X_{\rm CO}$ correlations than we find here. Considering that the slope of $X_{\rm CO}$ between the two modelled clouds correlates remarkably strongly with the density distribution, we argue that the observed emission of CO~(1-0) in the \citet{Shi15,Shi16} galaxies may be connected with their diffuse and extended ISM. Stronger $Z-X_{\rm CO}$ correlations have been also reported by other studies \citep[e.g.][]{Isra97, Wolf10, Glov11, Schr12, Madd20}. The connection of the column density distribution with the slope of $X_{\rm CO}$ is further supported with the models of \citet{Glov16}, who compare the resultant conversion factor in different dynamical times and report such a dependency.

The dense cloud is best-fitted with $b{\simeq}20.29$ and $k{\simeq}-0.45$. This weaker dependency on $Z$ (when compared to the diffuse cloud) is due to the ability of the cloud to remain molecular and bright in CO~(1-0), as the column densities can provide enough shielding against the FUV photons, even for the $0.1\,{\rm Z}_{\odot}$ case. This is better illustrated in the corresponding column of Fig.~\ref{fig:satur} which shows how $W(\rm CO\,1-0)$ builds versus $N({\rm H}_2)$ (see \S\ref{ssec:growth}) and also from the SLED presented in the bottom right panel of Fig.~\ref{fig:sleds}. In the upper panel of Fig.~\ref{fig:xco3} we plot the results of \citet{Gong20} (for $Z=0.5-1\,{\rm Z}_{\odot}$), which agrees better with our results for the dense cloud, as can be seen. Weaker correlations of $Z-X_{\rm CO}$ have been reported in other works, including \citet{Nara12,Feld12,Sand13}.

The lower panel of Fig.~\ref{fig:xco3} shows the correlation of the C{\sc i}-to-H$_2$ factor with metallicity. Overall there is a weaker dependency with $Z$ as well as with the density distribution. The dense cloud is best-fitted with $b{\simeq}21.22$ and $k{\simeq}=-0.16$; thus it is nearly independent on metallicity. On the other hand, the diffuse cloud is best-fitted with $b{\simeq}20.83$ and $k{\simeq}-0.96$. This is in agreement with the correlation obtained by the models of \citet{Glov16} and the recent absorption-derived $X_{\rm CI}$ observations of high-redshift star-forming galaxies by \citet{Hein20}.

\section{Conclusions}
\label{sec:conclusions}

In this paper we explored a range of different ISM environments by varying the cosmic-ray ionization rate spanning thee orders of magnitude ($\zeta_{\rm CR}=10^{-17}-10^{-14}\,{\rm s}^{-1}$), the FUV intensity spanning three orders of magnitude ($G_0=1-10^3$) and the metallicity from $Z=0.1\,{\rm Z}_{\odot}$ to $2\,{\rm Z}_{\odot}$ in two snapshots of simulations examining the evolution of magnetised, turbulent, self-gravitating molecular clouds. This paper continues the studies of \citet{Bisb15,Bisb17a} and of \citet{Wu15,Wu17,Bisb17b,Bisb18} in examining how the observables and the dynamical diagnostics change under different ISM environment, respectively. For our analysis we used {\sc 3d-pdr} to determine the abundances of species and level populations of the coolants explored in full thermal-balance calculations and a radiative transfer algorithm to estimate the velocity integrated emission of the most important coolants. Our results are summarized below.

\subsection{Effect of cosmic-rays}

Increasing the cosmic-ray ionization rate changes the carbon phases more rapidly than the H{\sc i}-to-H$_2$ transition. This results in a decrease in CO and an increase in C{\sc i} and C{\sc ii} column densities, as well as in O{\sc i}. The gas temperature increases with $\zeta_{\rm CR}$ as a result of cosmic-ray heating and number densities ${\gtrsim}10^5\,{\rm cm^{-3}}$ can be as warm as ${\sim}40-50\,{\rm K}$. The emission of all fine-structure lines ([C{\sc ii}]~$158\mu$m, [C{\sc i}]~(1-0), (2-1), [O{\sc i}]~$63\mu$m, $146\mu$m), brightens with increasing $\zeta_{\rm CR}$. The line of [C{\sc ii}] is tightly connected with the distribution of H$_2$ column densities for $\zeta_{\rm CR}\sim10^{-14}\,{\rm s}^{-1}$ and its emission at these high cosmic-ray ionization rates becomes very bright. In our simulations, the isotope of [$^{13}$C{\sc ii}] becomes bright only under ISM conditions of high $\zeta_{\rm CR}$ values for which the optical depth of C{\sc ii} is also high. Similarly, both [C{\sc i}] lines become bright for high $\zeta_{\rm CR}$. On the other hand, both [O{\sc i}] lines, although brightening with increasing $\zeta_{\rm CR}$, remain much fainter as they are quite optically thick and suffer from self-absorption. For low values of $\zeta_{\rm CR}$, low-$J$ CO lines are more extended when compared to high $\zeta_{\rm CR}$ models. This may cause extended ISM regions to appear more clumpy than they really are. Higher-$J$ CO lines are emitted from high column densities and they become brighter as $\zeta_{\rm CR}$ increases. The growth of CO~(1-0) emission with H$_2$ column density is weakly dependent on the cosmic-ray ionization rate.  On the other hand the line of [C{\sc i}]~(1-0) increases with $\zeta_{\rm CR}$ and its optical depth decreases. This makes this line an excellent alternative H$_2$ gas tracer. The line of [C{\sc ii}] at $158\mu$m shows no correlation with H$_2$ column density for $\zeta_{\rm CR}\le10^{-15}\,{\rm s}^{-1}$. However, for $\zeta_{\rm CR}10^{-14}\,{\rm s}^{-1}$ it shows an almost linear correlation with $N(\rm H_2)$ up to ${\sim}3\times10^{22}\,{\rm cm}^{-2}$, also making it a potential alternative H$_2$ gas tracer in cosmic-ray dominated regions. Cosmic-ray heating, especially when $\zeta_{\rm CR}\gtrsim10^{-15}\,{\rm s}^{-1}$, may excite high-$J$ CO lines and thus change the shape of SLEDs. We compared our results with the \citet{Pon16} observations of three different IRDCs and found that, as an alternative to shock heating, cosmic-ray heating may also explain the observed high-$J$ CO lines in such objects. When it comes to such CO SLED fitting, we found that the column density distribution also plays an important role and more exploration of different distributions is needed. The fine-structure line of [C{\sc ii}] remains a good diagnostic for cloud-cloud collisions for intermediate-to-low cosmic-ray ionization rates. For high $\zeta_{\rm CR}$, the `bridge-effect' feature that signals the cloud-cloud collision is diminished. Similarly to [C{\sc ii}], the [O{\sc i}]~$63\mu$m line may be used as such a diagnostic, although its brightness temperature is much fainter. 
The increase of cosmic-rays decreases the line ratios of $R^{\rm CI10}_{\rm CI21}$, $R^{\rm CO10}_{\rm CI10}$ and $R^{\rm CO43}_{\rm CI10}$, while they increase the $R^{\rm CI10}_{\rm CO21}$ line ratio. When the value of metallicity of an object is known, the particular ratio of the two atomic carbon lines appears to be a new promising diagnostic of cosmic-ray dominated regions, especially as it depends only weakly on the density distribution. The CO-to-H$_2$ conversion factor does not depend strongly on the cosmic-ray ionization rate or the column density distribution. Its value is in broad agreement with the \citet{Bola13} recommended value under all $\zeta_{\rm CR}$ rates and we further find a local maximum of $X_{\rm CO}$ for $\zeta_{\rm CR}=10^{-15}\,{\rm s}^{-1}$. On the other hand, the C{\sc i}-to-H$_2$ factor decreases with increasing $\zeta_{\rm CR}$ and depends more strongly on the density distribution.

\subsection{Effect of intensity of FUV radiation field}

Increasing the intensity of the FUV radiation field boosts the photodissociation of CO and creates a surplus of C{\sc i} (in more compact and shielded regions) and C{\sc ii}, as well as an increase in O{\sc i} abundances and therefore their column densities. The H{\sc i}-to-H$_2$ transition shifts to higher column densities as a result of the photodissociation of H$_2$. The gas temperature remains as low as ${\sim}10\,{\rm K}$ at high column densities, even for $G_0=10^3$, since the FUV photons attenuate rapidly as a function of visual extinction. However, the gas temperature increases at the outer parts of both clouds as $G_0$ increases. As a consequence of the abundance and gas temperature changes, the emissions of [C{\sc ii}], [C{\sc i}] and [O{\sc i}] become stronger as $G_0$ increases. However, they always arise from the gas surrounding the higher density medium. The relation of W(CO~1-0) and W(C{\sc i}~1-0) with $N(\rm H_2)$ does not substantially change as a function of $G_0$. However, W(C{\sc ii}) and W(O{\sc i}~63$\mu$m) increase with $G_0$, but they do not scale as a function of the H$_2$ column. In regards to CO SLEDs, it is found that in diffuse gas, $G_0$ affects significantly only the low-$J$ transitions; high $G_0$ leads to a suppression of these transitions. Since the origin of the [C{\sc ii}] and [O{\sc i}] emission is always connected with the gas surrounding the filamentary structures in the dense cloud, we find that any potential signature for cloud-cloud collisions remains relatively unchanged, even for $G_0=10^3$. We also find that the $R^{\rm CII}_{\rm OI63}$ line ratio depends strongly on the value of the FUV intensity and that this ratio is rather challenging to use as a diagnostic in ISM studies. Strong $G_0$ may affect line ratios including atomic carbon by decreasing the $R^{\rm CO10}_{\rm CI10}$ ratio and increasing the $R^{\rm CI10}_{\rm CO21}$ ratio. Finally, both $X_{\rm CO}$ and $X_{\rm CI}$ conversion factors remain relatively constant as a function of the FUV intensity.

\subsection{Effect of metallicity}

Low metallicities decrease the CO and H$_2$ shielding against FUV, as well as the H$_2$ formation rate on dust grains, therefore impacting the carbon cycle in relation to the H{\sc i}-to-H$_2$ transition and increasing the overall gas temperature. We find that for $Z=0.1\,{\rm Z}_{\odot}$, the dense cloud remains molecular and becomes equally rich in C{\sc ii} and C{\sc i} abundances as it is in CO abundance. The diffuse cloud becomes H{\sc i}-dominated with the carbon abundance to be almost entirely in C{\sc ii} form. For super-solar metallcities, both clouds are entirely molecular CO-dominated. Consequently, the emission of all cooling lines modelled, change accordingly. For $Z=0.1\,{\rm Z}_{\odot}$ it is particularly interesting to note that the emission of [C{\sc ii}] originates from higher column densities corresponding to H$_2$-rich gas, making this line a promising tracer for molecular gas in low-metallicity environments \citep[see also][]{Madd20}. As a result of the CO photodissociation in metal-poor ISM, low $Z$ affect the mid- and low-$J$ transitions of the CO line and hence the corresponding SLEDs. In regards to the bridge-effect signature, we find that its shape depends only weakly on metallicity for $Z\geq0.5\,{\rm Z}_{\odot}$, but it diminishes for ${\sim}0.1\,{\rm Z}_{\odot}$ as these fine structure lines originate from higher column densities that may not carry this collision information any longer. We find that as metallicity increases, the line ratios of $R^{\rm CI10}_{\rm CI21}$ and $R^{\rm CO10}_{\rm CI10}$ increase, while $R^{\rm CI10}_{\rm CO21}$ decrease. Finally, we find that both $X_{\rm CO}$ and $X_{\rm CI}$ factors are depending strongly on metallicity in agreement with several observations, as well as on the density distribution of the ISM.

\vspace{0.2cm}
As a final conclusion, future studies using ALMA, SOFIA and the forthcoming CCAT-prime telescope of ISM environments that differ from the local one, such as the Galactic Centre, low-metallicity galaxies and galaxies with high star-formation rates, can provide data to help understand the ISM environment in the high-redshift Universe, particularly at $z \sim 2-3$ marking the `cosmic noon' of galaxy assembly. The models we have presented will be helpful in the interpretation of such data.

\section*{Data availability}

The data underlying this article will be shared on reasonable request to the corresponding author.

\section*{Acknowledgements}

The authors thank the anonymous referee for their comments and the overall detailed and timely review which improved the clarity of the work. The authors also thank Zhi-Yu Zhang, Daniel Seifried, Stefanie Walch, Brandt Gaches, Loretta Dunne, Francesco Valentino and Munan Gong for the useful discussions.
TGB acknowledges support from Deutsche Forschungsgemeinschaft (DFG) grant No. 424563772. 
JCT acknowledges support from the Chalmers Foundation, NSF grant AST-2009674 and ERC Advanced Grant 788829 (MSTAR).
KEIT acknowledges support from NAOJ ALMA Scientific Research grant No. 2017-05A, and JSPS KAKENHI grant Nos. JP19H05080, JP19K14760. 

TGB dedicates this work to the memory of Professor John-Hugh Seiradakis.

\bsp	
\label{lastpage}

\appendix

\section{Additional emission maps and optical depth diagrams}
\label{app:plots}

\begin{figure*}
    \centering
    \includegraphics[width=\linewidth]{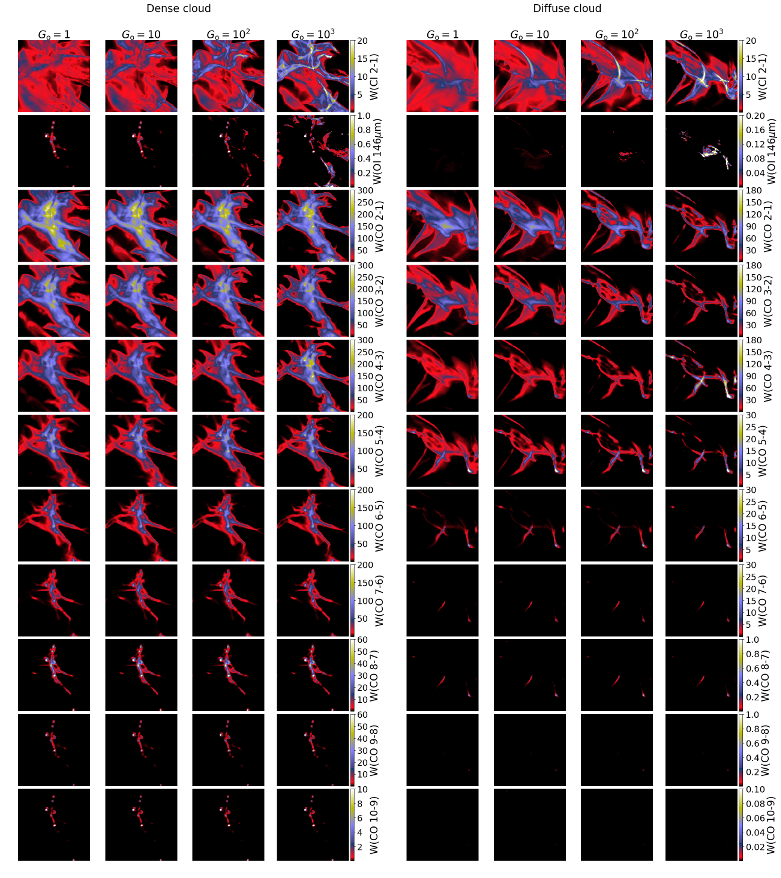}
	\caption{The rest of emission lines following Fig.~\ref{fig:Ga_striking} for the FUV intensity as the free parameter. From top-to-bottom, the lines of [C{\sc i}]~(2-1), [O{\sc i}]~$146\mu$m and then CO $J=2-1$ to $J=10-9$ are presented.}
    \label{fig:Gapp}
\end{figure*}

\begin{figure*}
    \centering
    \includegraphics[width=\linewidth]{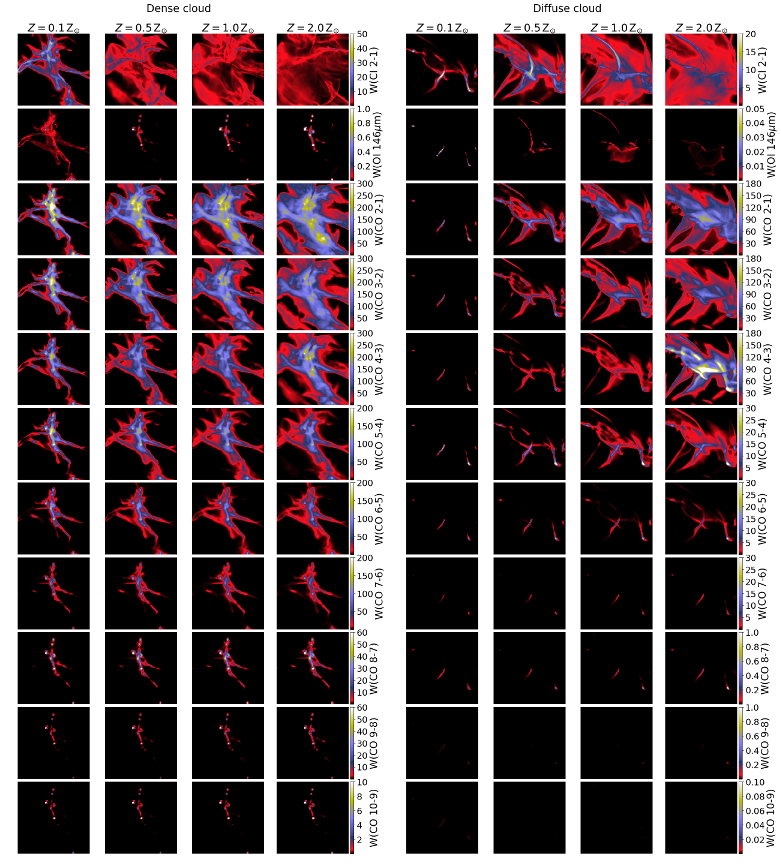}
	\caption{The rest of emission lines following Fig.~\ref{fig:Za_striking} for the metallicity as the free parameter. From top-to-bottom, the lines of [C{\sc i}]~(2-1), [O{\sc i}]~$146\mu$m and then CO $J=2-1$ to $J=10-9$ are presented.}
    \label{fig:Zapp}
\end{figure*}

Figures~\ref{fig:Gapp} and \ref{fig:Zapp} present emission maps for the lines of [C{\sc i}]~(2-1), [O{\sc i}]~$146\mu$m and CO with $J=2-1$ up to $J=10-9$ for the FUV intensity and the metallicity as the free ISM parameters, respectively.

\begin{figure*}
    \centering
    \includegraphics[width=0.9\linewidth]{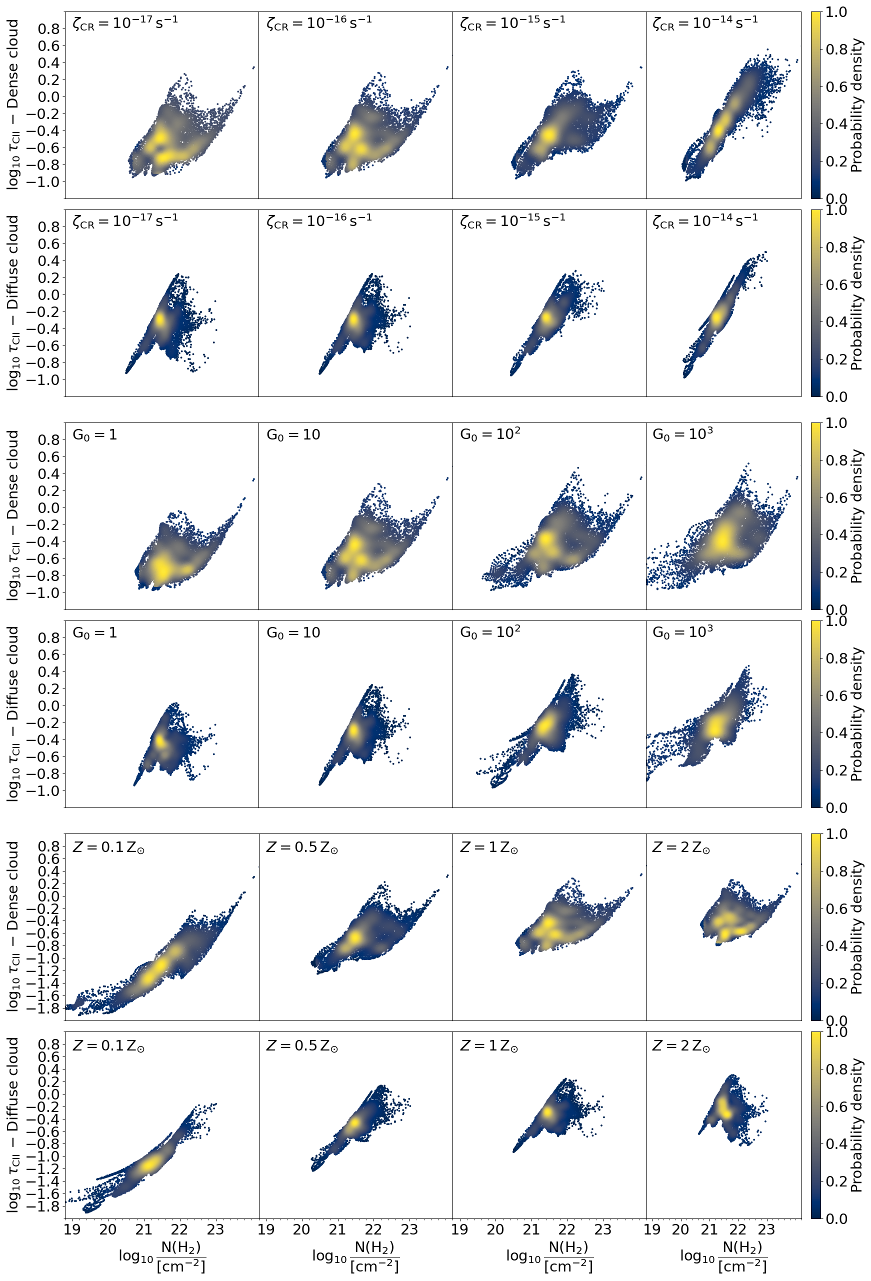}
	\caption{Optical depth of the [C{\sc ii}]~$158\mu$m line versus N(H$_2$) for all conditions explored.}
    \label{fig:taucii}
\end{figure*}

\begin{figure*}
    \centering
    \includegraphics[width=0.9\linewidth]{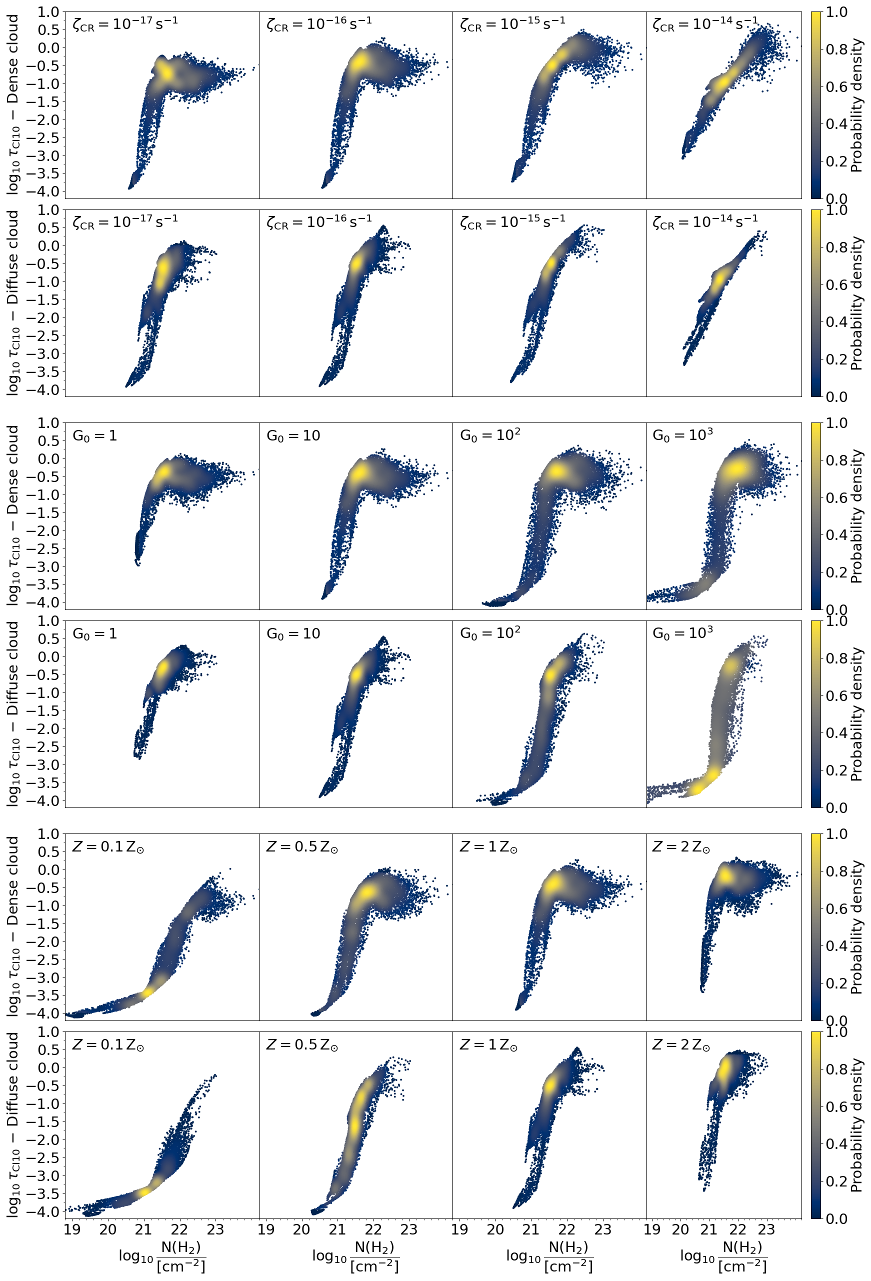}
	\caption{Optical depth of the [C{\sc i}]~(1-0) line versus N(H$_2$) for all conditions explored.}
    \label{fig:tauci}
\end{figure*}

\begin{figure*}
    \centering
    \includegraphics[width=0.9\linewidth]{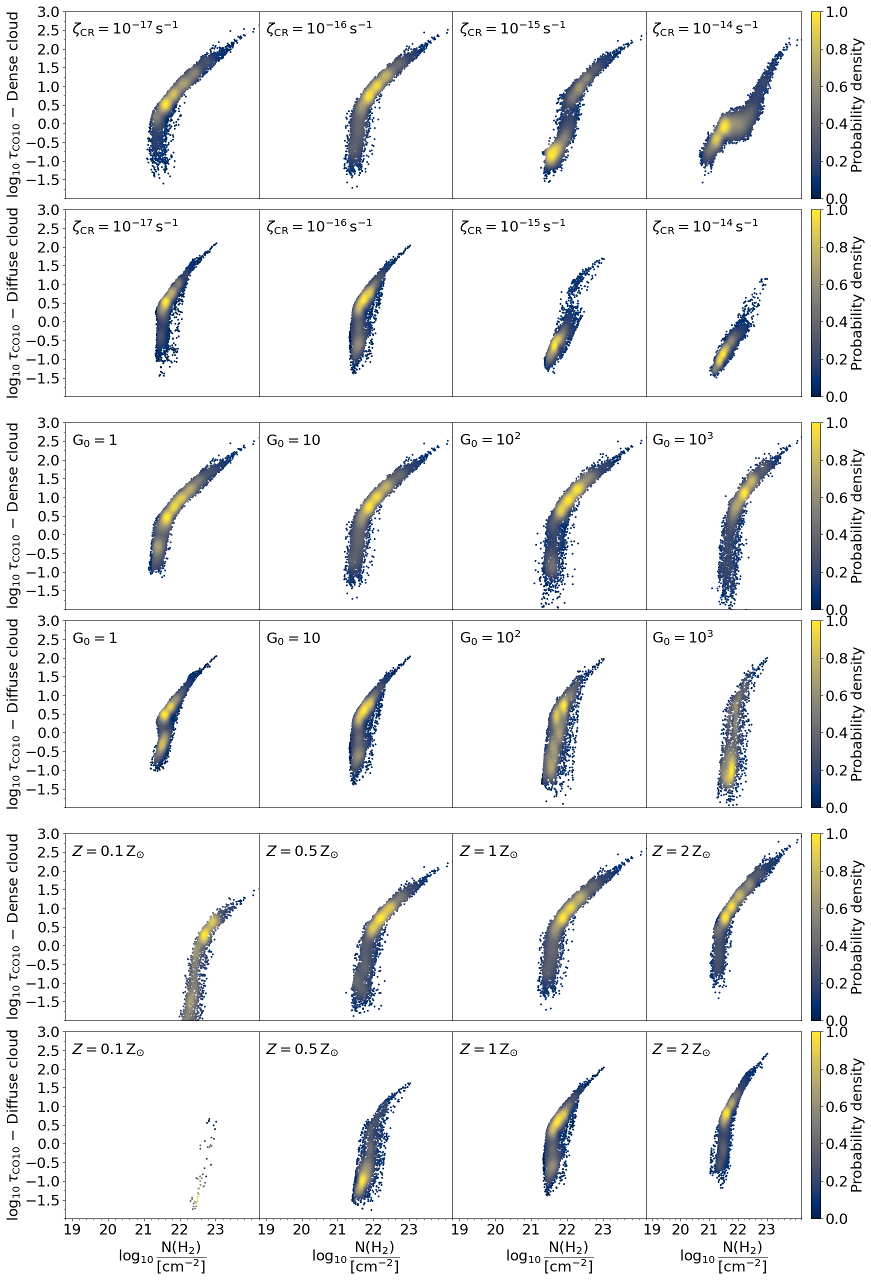}
	\caption{Optical depth of the CO $J=1-0$ line versus N(H$_2$) for all conditions explored.}
    \label{fig:tauco}
\end{figure*}

\begin{figure*}
    \centering
    \includegraphics[width=0.9\linewidth]{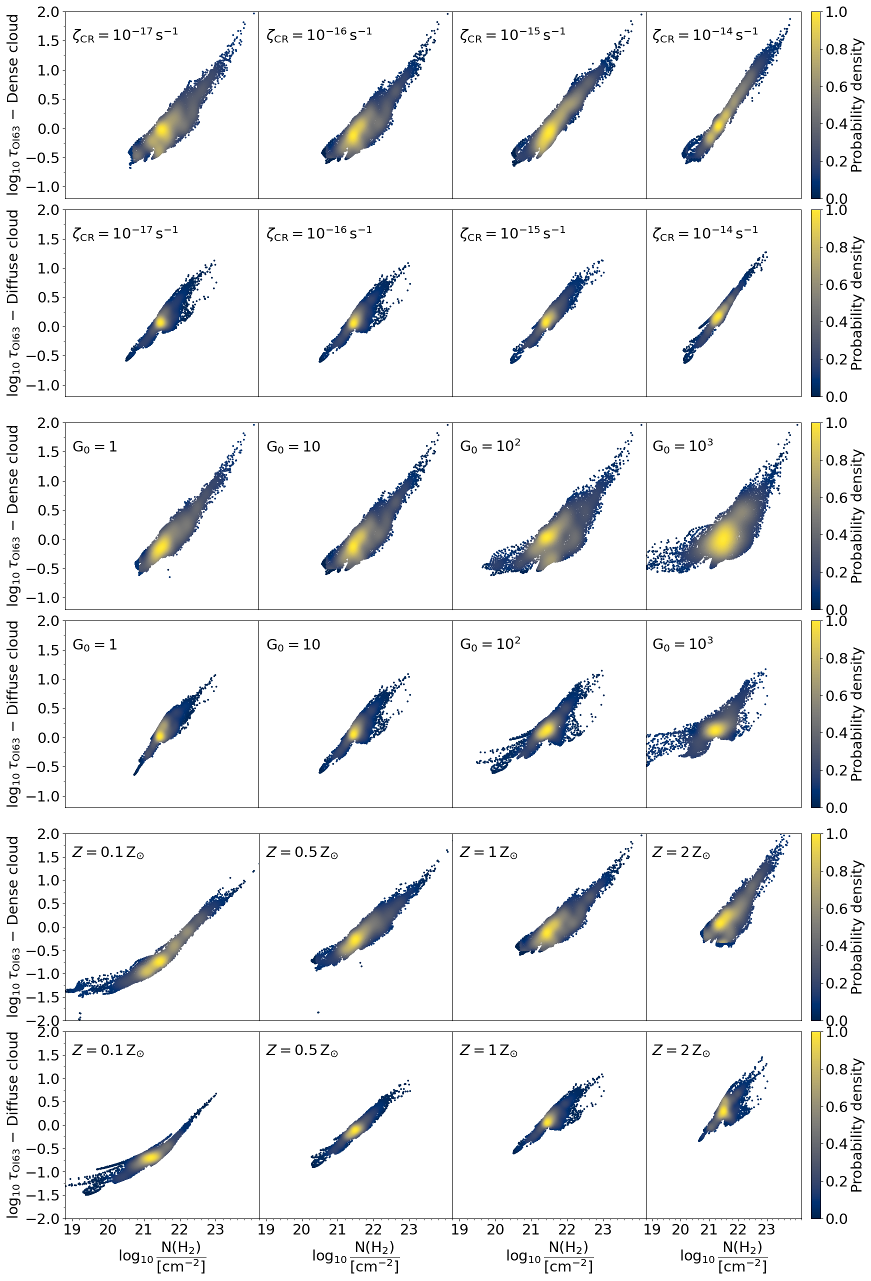}
	\caption{Optical depth of the [O{\sc i}]~$63\mu$m line versus N(H$_2$) for all conditions explored.}
    \label{fig:tauoi}
\end{figure*}

Figures~\ref{fig:taucii} -- \ref{fig:tauoi} show the average optical depth ($\tau$) weighted with the brightness temperature for the lines of [C{\sc ii}]~$158\mu$m, [C{\sc i}]~(1-0), CO $J=1-0$ and [O{\sc i}]~$63\mu$m versus the column density of H$_2$, respectively. The aforementioned optical depth is given by the expression:
\begin{eqnarray}
	\tau = \frac{\int_{\varv_{\rm min}}^{\varv_{\rm max}} \tau_{\varv} T_{\rm A}d\varv}{\int_{\varv_{\rm min}}^{\varv_{\rm max}} T_{\rm A} d\varv},
\end{eqnarray}
where $\tau_{\varv}$ is the optical depth of a given velocity channel and the integration is over velocities as discussed in \S\ref{ssec:rt}. In each panel we show a scatter plot of $\tau$, coloured according to the probability density. High values of the probability density gives the dominant value of $\tau$ under all ISM conditions explored.

Note that the probability density of the $\tau_{\rm CII}$ in the dense cloud appears to contain four different `groups'. This occurs for the cases of the fiducial model, of $\zeta_{\rm CR}=10^{-17}\,{\rm s}^{-1}$ and of $Z=2\,{\rm Z}_{\odot}$, and it is a result of edge effects in the boundaries of the map i.e. where the isotropic radiation impinges from.
Furthermore, in the particular case of $\tau_{\rm CO10}$ in the diffuse cloud and at $Z=0.1\,{\rm Z}_{\odot}$, there are only a few scatter points as a result of the CO depletion and photodissociation.

\end{document}